%% file: 4C.tex
\documentclass[journal=jctcce,manuscript=article]{achemso}
\usepackage{amssymb,amsmath,amsfonts,multicol,multirow,longtable,array,mathpazo}
\usepackage{lscape}
\usepackage[version=3]{mhchem} % Formula subscripts using \ce{}
\usepackage{appendix}
\usepackage{soul} %text highlighting
\usepackage[usenames]{color}
\usepackage{makecell}
\usepackage{enumerate}
\usepackage{xr}
\usepackage[T1]{fontenc} % Use modern font encodings
\usepackage{booktabs}
\usepackage{tikz}
\usetikzlibrary{shapes.geometric}
\usepackage{threeparttable}
\usepackage{graphicx}
\usepackage{subfigure}
\usepackage{colortbl}
\usepackage{framed}
\usepackage{verbatim}
\usepackage{amsbsy}
\usepackage{bm}% bold math
\usepackage{bbm}
\usepackage[normalem]{ulem}
\usepackage{float}
\usepackage[linesnumbered,boxed]{algorithm2e}
\usepackage{algorithm2e}
\usepackage{float}
\usepackage{subfigure}
\usepackage{simplewick} % time-ordered contraction
\newcounter{xscheme}

\newcommand{\ii}{\mathbbm{i}}

\newcommand{\bsigma}{\boldsymbol{\sigma}}
\newcommand{\balpha}{\boldsymbol{\alpha}}
\newcommand{\sgn}{ {\rm{sgn}} }

\newfloat{Algorithm}{htbp}{alg}
\floatname{Algorithm}{Algorithm}

\newcounter{exe}[figure]
\newcommand{\iexe}{\refstepcounter{exe}\the\value{exe}:}

\setkeys{acs}{maxauthors = 0} % references of many authors

\author{Ning Zhang\footnote{Current address: Division of Chemistry and Chemical Engineering, California Institute of Technology, Pasadena, CA 91125, USA}}
\author{Wenjian Liu}\email{liuwj@sdu.edu.cn}
\affiliation{Qingdao Institute for Theoretical and Computational Sciences, School of Chemistry and Chemical Engineering,
	Shandong University, Qingdao, Shandong 266237, China}

\title{Unified Implementation of Relativistic Wave Function Methods: 4C-iCIPT2 as a Showcase}

\setkeys{acs}{articletitle=true}

\begin{document}
	
\begin{abstract}
In parallel to the unified construction of relativistic Hamiltonians based solely on physical arguments [J. Chem. Phys. 160, 084111 (2024)],
a unified implementation of relativistic wave function methods is achieved here via programming techniques (e.g., template metaprogramming and polymorphism in C++).
That is, once the code for constructing the Hamiltonian matrix is made ready, all the rest can be generated automatically from existing
templates used for the nonrelativistic counterparts.
This is facilitated by breaking a second-quantized relativistic Hamiltonian down to diagrams that are
topologically the same as those required for computing the basic coupling coefficients between spin-free
configuration state functions (CSF). Moreover, both time reversal and binary double point group symmetries can readily be incorporated
into molecular integrals and Hamiltonian matrix elements. The latter can first be evaluated
in the space of (randomly selected) spin-dependent determinants and then transformed to that of
spin-dependent CSFs, thanks to simple relations in between. As a showcase, we consider here
the no-pair four-component relativistic iterative configuration interaction with selection and perturbation correction (4C-iCIPT2),
which is a natural extension of the spin-free iCIPT2
[J. Chem. Theory Comput. 17, 949 (2021)], and can provide near-exact numerical results within the manifold of positive energy states (PES),
as demonstrated by numerical examples.
%However, the results are still dependent on how the PESs are generated. It turns out that
%such dependence can largely be removed by invoking a simple pair correction (PC) that is rooted in quantum electrodynamics.
%The efficacy of 4C-iCIPT2-PC is demonstrated by fine-structure calculations of heavy elements.
\end{abstract}

\maketitle

\clearpage
\newpage

\section{INTRODUCTION}
Accurate descriptions of the electronic structure of systems containing heavy elements
should in principle treat relativistic, correlation, and even quantum electrodynamics (QED) effects
simultaneously\cite{LiuWIRES2023}, so as to match experimental measurements without relying on error compensations\cite{IP-QED2017,SchwerdtfegerAuPRL2017,SaueeQED2022}.
Both relativistic and leading QED effects can be built into the Hamiltonian\cite{eQED,LiuPhysRep,IJQCrelH,IJQCeQED,X2C2016,LiuPerspective2020,LiuSciChina2020},
whereas correlation is carried on by the wave function parameterized by a particular ansatz.
Given the complete `Hamiltonian ladder'\cite{IJQCrelH,LiuPhysRep}, from which one can pick up the right Hamiltonian
according to the target problem and accuracy, the task remains
to search for suitable parameterizations of the wave function,
so as to describe electron correlation as accurately as possible.
Since many-electron relativistic Hamiltonians can only be formulated properly in Fock space
\cite{eQED,LiuPhysRep,IJQCrelH,IJQCeQED,X2C2016,LiuPerspective2020,LiuSciChina2020,Kutzelnigg2012,PCCPNES}, explicitly correlated
wave function methods are excluded from the outset, unless special care is taken\cite{PCCPNES,RelR12,cutting-projector,punching-projector,hollosy2024one}.
In contrast, any orbital product-based wave function ansatz can be combined with
any second-quantized relativistic Hamiltonian. The available relativistic wave function methods
can be classified into two categories, with spinors or scalar orbitals as building units.
The former can further be classified into two categories, four-component (4C) or two-component (2C).
Since the two sets are indistinguishable in the correlation step after integral transformations,
they can be regarded as equivalent from the computational point of view.
Essentially all variants of wave function methods developed in the nonrelativistic regime have been transferred to the 4C/2C domain
under the no-pair approximation (NPA), including
multiconfiguration self-consistent field (SCF)
\cite{4C-MCSCF1996,2C-CASSCF1996,2C-CASSCF2003,2C-CASSCF2013,4C-MCSCF2008,
4C-CASSCF2015,4C-CASSCF2018,LixiaosongX2CCASSCF},
many-body perturbation theory\cite{4C-MP21994,4C-MRPT21999,4C-CASPT22006,2C-MRPT22014,4C-icMRCI-CASPT2-NEVPT22015,X2C-MRPT20222,4C-DSRG-MRPT2-2024}, coupled-cluster\cite{4C-CCSD1995,4C-CCSD1996,2C-CC2001,2C-CC2005,2C-CC2007,4C-IHFSCC2000,4C-IHFSCC2001,4C-FSCC2001,4C-CC2007,
4C-CC2010,2C4C-CC2011,4C-CC2016,2C-CC2017,Cheng2C-CC2018,Saue4C-CC2016,Saue4C-CC2018,2C-EOM-CCSD2019,Cheng2C-CCrev,CVS-EOM-CCSD2021},
configuration interaction (CI)\cite{4C-CISD1993,2C-CI1996,2C-MRCI2001,4C-MRCI2002,4C-GASCI2003,
4C-CI-MCSCF2006,4C-CI2008,2C-CICC2012,Fleig2012,2C-MRCI2020,4C2C-HCI2023},
density-matrix renormalization group\cite{4C-DMRG2014,4C-DMRG2018,4C-DMRG2020}, and full configuration interaction (FCI) quantum Monte Carlo\cite{4C-FCIQMC}.
It deserves to be mentioned that the correlation contribution of negative energy states (NES) can readily be accounted for
in both the 4C and 2C frameworks, so as to go beyond the NPA\cite{eQED,LiuPhysRep,eQEDBook2017}.

Although methods that start with scalar orbitals and treat SOC in the correlation step are sometimes also termed two-component,
we prefer to call them one-component (1C) to distinguish them from those working with two-component spinors.
The former involve spin-separated Hamiltonians, whereas the latter do not. Such methods can also be classified into two categories,
one- or two-step\cite{Marian2001}. The former type of approaches\cite{DGCIa,SpinUGA1,DGCIb,DGCIc,DetSOCI1997,CI-SOC1998,SPOCK2006,WF-CC2008,WF-CC2011,CAS-SOC2013,sf-X2C-EOM-SOC2017,HBCISOC2017,SOC-CCSDT-GPU2021,SO-MRCI2023,SOiCI}
aims to treat spin-orbit and electron-electron interactions on an equal footing,  whereas the latter type of approaches
\cite{Hess1982,CIPSO1983,CI-SOC1997,SOCsingle-1,SOCsingle-2,SOCsingle-3,MRCI-SOC2000,Teichteil2000,DFTMRCI-SOC2002,
CASPT2-SOC2004,SI-SOC2006,EOMIPSO2008,mai2014perturbational,DMRGSO2015,DMRGSO2016,nosiDMRGSO2016,cheng2018perturbative,
ChengEOMCCSO2020,WangMP2020,Suo-SOC2021,vanWullen2021,MPSSI2021}
amounts to treating SOC after correlation, by
constructing and diagonalizing an effective spin-orbit Hamiltonian matrix over a set of close-lying correlated scalar states.
It is obvious that such two-step-1C approaches
work well only when SOC and correlation are roughly additive. Of course, they become
identical with the one-step-1C approaches if all correlated scalar states from the space
spanned by individual configurations are included. It is also true that the one-step-1C approaches
can approach the 4C/2C ones  in the no-pair FCI limit, particularly
when spin orbital relaxations induced by SOC are further accounted for by variational optimizations\cite{SOCASSCF2013,SOiCISCF}.
 However, in practice, great care has to be taken of the choice of
active (scalar) orbitals\cite{SOiCI} and even of basis sets\cite{SOiCISCF} to avoid spurious results,
especially for spin-orbit splittings of $np$ ($n\ge 5$) orbitals.

Notwithstanding the above developments, relativistic wave function methods that are highly accurate for
strongly correlated, especially open-shell, systems containing heavy elements are still highly desired.
Considering that relativity (especially SOC) and correlation reside in different regions of the Hilbert space
due to their very different physical origins
(primarily one-body magnetic vs. primarily two-body Coulomb interactions), such methods must be self-adaptive and balanced,
based on some selection procedure. As an attempt to this end, we consider here the 4C variant of
iCIPT2 (iterative configuration interaction (iCI)\cite{iCI} with selection and second-order perturbation correction)\cite{iCIPT2,iCIPT2New},
which has proved to be one of the most accurate and efficient methods\cite{Blindtest}, and is applicable to arbitrary open-shell systems
thanks to use of configuration state functions (CSF) as the many-electron basis functions (MBF).
Both time reversal and double group symmetries ought to be incorporated into 4C-iCIPT2, so as to facilitate not only
the computation but also the state assignment (see Appendices \ref{SecSymmERI} and \ref{doublegroup2}).
It turns out that the implementation of 4C-iCIPT2 can be simplified greatly by leveraging the power of meta-programming furnished by C++:
once the code for constructing the relativistic Hamiltonian matrix is made ready, all the rest can be generated automatically
from the existing templates used for iCITP2, without the need to distinguish between complex and real algebras. In particular,
sophisticated techniques available in iCIPT2, e.g., memory allocation, loop unrolling, data sorting, and intermediates recycling, etc.,
can directly be transferred to 4C-iCIPT2. Therefore, a unified implementation of relativistic and nonrelativistic wave function methods
has been achieved, in parallel to the unified construction of relativistic Hamiltonians\cite{UnifiedH}.
This is made possible only by invoking a diagrammatical representation of the relativistic Hamiltonian
that is topologically the same as that employed in the unitary group approach\cite{Paldus1980}
for determining the basic coupling coefficients (BCC) between spin-free CSFs (cf. Sec. \ref{Hdecomposition}).

The rest of the paper is organized as follows. Sec. \ref{SecTheory} is devoted to the theoretical aspects, including a brief discussion of
relativistic Hamiltonians in Sec. \ref{Hmat}, general structure of the Hamiltonian matrix in Sec. \ref{SecHstructure}, and special treatment
of Hamiltonian matrix elements in Sec. \ref{SecMat}. The spin-free iCIPT2\cite{iCIPT2,iCIPT2New} is then recapitulated in Sec. \ref{SDS},
where the implementation of 4C-iCIPT2 is also presented.
Some benchmark calculations are provided in Sec. \ref{pilot_app}.
The paper is finally closed with concluding remarks in Sec. \ref{Conclusion}.

The Einstein summation convention over repeated indices is always employed.
%Several quantum chemistry packages are also available to perform these relativistic electronic structure calculations\cite{pyscf1,pyscf2,BDF2020,CFOUR,BAGEL,BERTHA}.

\section{THEORY}\label{SecTheory}
\subsection{Relativistic Hamiltonians}\label{Hmat}
As emphasized repeatedly\cite{Kutzelnigg2012,PCCPNES,eQED,LiuPhysRep,X2C2016,LiuPerspective2020}, many-body
relativistic Hamiltonians can only be formulated properly in Fock space,
for all configuration space-based, first-quantized relativistic Hamiltonians suffer\cite{PCCPNES,eQED} from unphysical contaminations of negative energy states (NES) of the Dirac equation. As a matter of fact, by interpreting the solutions $\{\psi_p\}$ of the effective one-body Dirac equation
\begin{align}
h\psi_p&=\epsilon_p\psi_p,\quad p \in \mbox{ PES, NES},\label{DEQ0}\\
h&=D + U(\mathbf{r}), \label{hop}\\
D&=c\balpha  \cdot\boldsymbol{p}+\beta c^2+V_{\mathrm{nuc}},\label{Dop}\\
\balpha  &=\begin{pmatrix}0&\bsigma \\
\bsigma &0\end{pmatrix},\quad \beta=\begin{pmatrix}1&0\\
0&-1\end{pmatrix}, \quad\boldsymbol{p}=-\ii \hbar\boldsymbol{\partial}
%\sigma_x&=\begin{pmatrix}0&1 \\  1&0 \end{pmatrix},\quad
%\sigma_y =\begin{pmatrix}0&-\ii\\  \ii&0 \end{pmatrix},\quad
%\sigma_z =\begin{pmatrix}1&0 \\  0&-1\end{pmatrix},\\
%V(\boldsymbol{r})&=-\sum_{A}\frac{Z_A}{|\boldsymbol{r}-\boldsymbol{R}_A|}.
\end{align}
as classical fields and second-quantizing them properly\cite{LiuPerspective2020}, %i.e.,
%\begin{align}
%\hat{\phi}(\mathbf{r})=a_p\psi_p(\mathbf{r}),\quad a_p|\mathrm{vac}\rangle=0,\quad p\in\mbox{ PES, NES},
%\end{align}
one can readily obtain the normal-ordered many-body relativistic quantum electrodynamics (QED) Hamiltonian
\begin{align}
H_n&=(h_{pq}+Q_{pq}) \{E_{pq}\}_n+\frac{1}{2}g_{pq,rs}\{e_{pq,rs}\}_n,\quad p,r,r,s\in\mbox{ PES, NES},\label{HeQED}\\
h_{pq}&=\langle\psi_p|h|\psi_q\rangle,\quad Q_{pq}=-\frac{1}{2}\bar{g}_{pq,ss}\sgn(\epsilon_s),\label{QDef}\\
g_{pq,rs}&=(pq|V(1,2)|rs),\quad \bar{g}_{pq,rs}=g_{pq,rs}-g_{ps,rq},\\
E_{pq}&=a_p^\dag a_q,\quad
e_{pq,rs}=a^{pr}_{qs}=E_{pq} E_{rs}-\delta_{qr}E_{ps}=e_{rs,pq}=-e_{ps,rq}=-e_{rq,ps},
\end{align}
provided that the charge-conjugated contraction\cite{eQED}
\begin{align}
\acontraction[0.5ex]{}{a^p}{}{a_q}a^pa_q&=-\acontraction[0.5ex]{}{a_q}{}{a^p}a_qa^p=\langle 0;\tilde{N}|\frac{1}{2}[a^p, a_q]|0;\tilde{N}\rangle
=-\frac{1}{2}\delta^p_q \sgn(\epsilon_q), \quad p, q \in \mbox{PES, NES},\label{CCC}
\end{align}
is employed when normal ordering
with respect to $|0;\tilde{N}\rangle$, a reference that is built up
with the NES of Eq. \eqref{DEQ0} and represents the filled Dirac sea when the number ($\tilde{N}$) of negative energy electrons
approaches infinity. The so-obtained effective potential $Q$ \eqref{QDef} describes vacuum polarization and electron self-energy,
the genuine QED effect\cite{eQED}. It has recently been shown decisively\cite{CommentQED,ReCommentQED} that
the Hamiltonian \eqref{HeQED} is the only correct and complete QED Hamiltonian (in the form of eigenvalue problem),
at variance with other proposals\cite{WrongQED,RespCommentQED}.

Although relativistic, correlation, and QED effects can in principle be treated on an equal footing\cite{LiuWIRES2023},
for the time being we ignore QED effects and instead consider the no-pair approximation.
%focus on the proper combination
%of the no-pair four-component relativistic Hamiltonian with a highly accurate method for electron correlation.
That is, the QED Hamiltonian \eqref{HeQED} is to be simplified to
\begin{align}
H_+&=h_{pq}E_{pq}+\frac{1}{2}g_{pq,rs}e_{pq,rs},\quad p, q, r, s \in\mbox{ PES},\label{np-H}
\end{align}
which accounts only for electron correlation within the manifold of positive energy states (PES), mediated by
the instantaneous electron-electron interaction
\begin{align}
V(1,2)&=\frac{1}{r_{12}}+\mathrm{c_g}\frac{\balpha_1\cdot\balpha_2}{r_{12}}+\mathrm{c_b}\balpha_1\cdot
\frac{\boldsymbol{r}_{12}\otimes\boldsymbol{r}_{12}}{r_{12}^3}\cdot\balpha_2,\label{V12full}
\end{align}
which reduces to the Coulomb (C), Coulomb-Gaunt (CG), and Coulomb-Breit (CB) interactions by setting $(\mathrm{c_g},\mathrm{c_b})$ to $(0,0)$, $(-1,0)$, and $(-1/2, -1/2)$, respectively. To avoid over notation, we will denote the Coulomb and Gaunt/Breit electron repulsion integrals (ERI)
simply as $(pq|rs)$ and $(p\boldsymbol{\alpha}q|r\boldsymbol{\alpha}s)$, respectively.
If wanted, the no-pair four-component Hamiltonian \eqref{np-H} can further be transformed to the quasi-four-component (Q4C)\cite{Q4C,Q4CX2C}
and exact two-component (X2C)\cite{X2CName,X2C2005,X2C2009} variants that can be constructed in a unified manner,
based solely on physical arguments\cite{UnifiedH}.

%One point to be emphasized lies in that the no-pair energy is always dependent on how the spinors are generated. However,
%such dependence can largely be removed by introducing the simple correction that is rooted in quantum electrodynamics\cite{eQED}
%\begin{align}
%E^{(2)}_{\mathrm{pc}}&=-\frac{U_{\tilde{i}s} U_{t\tilde{i}}\gamma_{ts}}{\epsilon_{\tilde{i}}-\epsilon_t}
%+\frac{g_{pr,\tilde{i}s}U_{t\tilde{i}}\Gamma_{pr,ts}}{\epsilon_{\tilde{i}}-\epsilon_t}
%+\frac{U_{\tilde{i}t}g_{pr,s\tilde{i}}\Gamma_{pr,st}}{\epsilon_r+\epsilon_{\tilde{i}}-\epsilon_p-\epsilon_s}, \label{EPC}\\
%\gamma_{pq}&=\langle\Psi|E_{pq}|\Psi\rangle,\quad \Gamma_{pq,rs}=\langle\Psi|e_{pq,rs}|\Psi\rangle,\quad p,q,r,s\in\mbox{ PES},
%\end{align}
%where the index $\tilde{i}$ runs over all NES, whereas all other indices are confined to PES. Note that Eq. \eqref{EPC}
%is a generalization  (see the second terms
%in Eqs. (79)--(81) in Ref. \citenum{eQED})
%of the previous formulation\cite{QED1999} that is valid only for closed-shell systems.
%In the case of Q4C or X2C, the NES can be generated in the end of self-consistent field calculations\cite{UnifiedH}.

In the absence of magnetic interactions, the Hamiltonians \eqref{hop} and \eqref{np-H} commute
with the time reversal operator $\mathcal{K}$ ($=-\ii \mathbf{I}_2\otimes \sigma_y K_0$).
The spinors can therefore be grouped into Kramers pairs $\{\psi_p, \psi_{\bar p}=\mathcal{K}\psi_p\}_{p=1}^{K_P}$. As shown in Appendix \ref{StorageIntPermutation},
compared to the brute-force storage,
the storage requirement for the ERIs can be reduced by a factor of 16 by making proper
use of time reversal and permutation symmetries. Point group symmetries
inherent in the ERIs are further revealed in Sec. \ref{doublegroup1}.

\subsection{General Structure of Hamiltonian Matrix}\label{SecHstructure}
\begin{figure}
	\centering
	{\resizebox{0.7\textwidth}{!}{\includegraphics{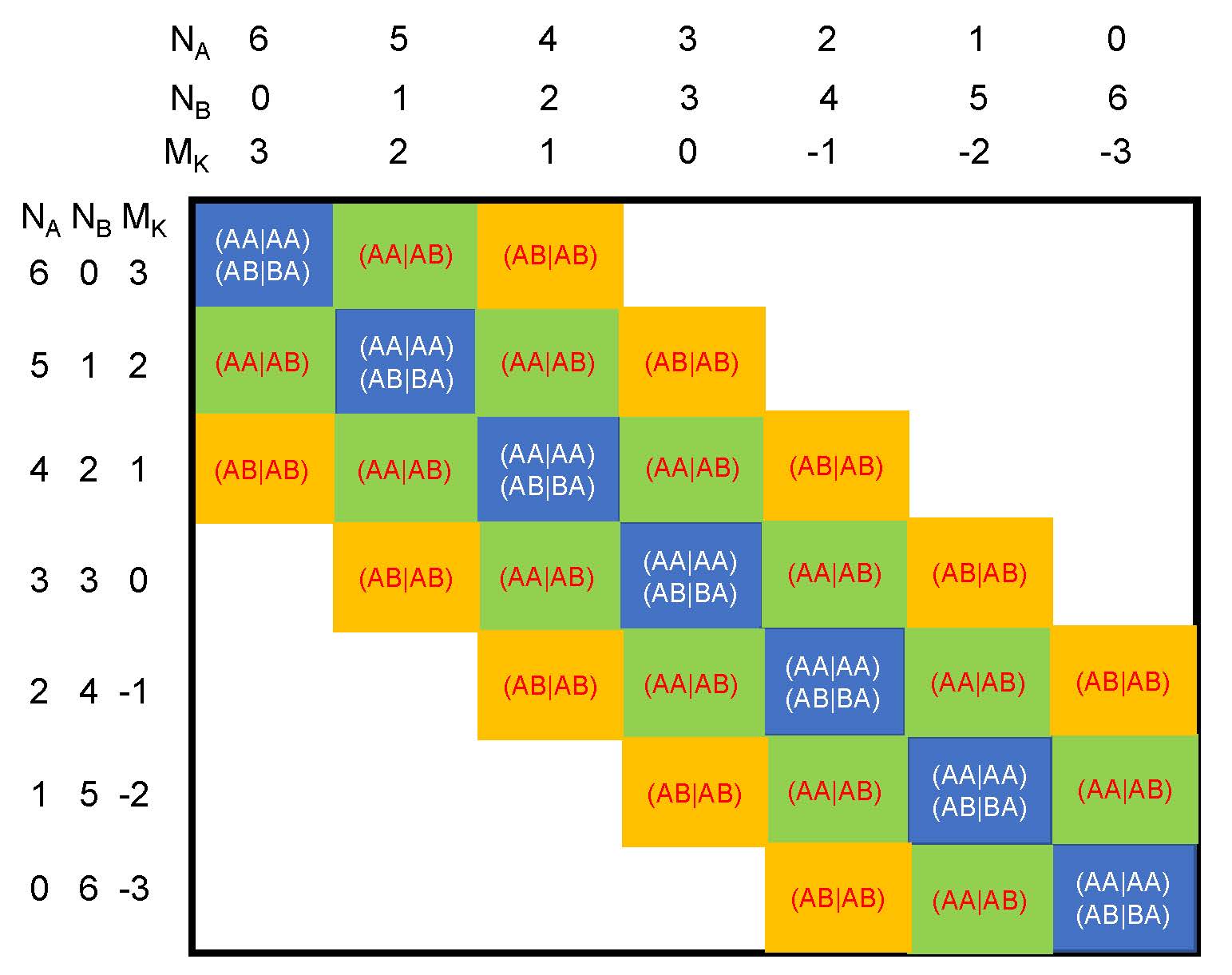}}}
	\caption{Block pentadiagonal structure of relativistic Hamiltonian matrix over $M_K$-ordered determinants.
The $(AA|AB)$ type of ERIs exist only for the $C_1^\ast$ and $C_i^\ast$ groups,
and all the ERIs become real-valued for the $C_{2v}^\ast$, $D_2^\ast$, and $D_{2h}^\ast$ groups (cf. Appendix \ref{StorageIntPermutation}). }
	\label{Hstructure}
\end{figure}

Consider all possible Slater determinants (SD) that can be constructed by distributing $N$ electrons in a given set of $K_P$ Kramers pairs $\{\psi_p, \psi_{\bar p}=\mathcal{K}\psi_p\}_{p=1}^{K_P}$.
Every SD can be characterized by the pseudo-quantum number $M_K=(N_A-N_B)/2$, with $N_A$ and $N_B$ ($=N-N_A$)
 being the numbers of unbarred (A) and barred (B) spinors.
If SDs of the same $M_K$ are grouped into one set (of size $C_{K_P}^{N_A}C_{K_P}^{N_B}$) and different sets are ordered in a descendent order of $M_K$,
the Hamiltonian matrix will become block pentadiagonal\cite{RQC_BOOK_Dyall}  (see Fig. \ref{Hstructure}), for the matrix elements are zero when
two sets of SDs differ in their $M_K$ values by more than two and are hence connected by more than double excitations.
For the main diagonal blocks with $|M_K(L)-M_K(M)|=0$, the matrix elements $H_{LM}$ consist of one-electron integrals of the type $\langle A|h|A\rangle$
and ERIs of the types $(AA|AA)$ and $(AB|BA)$ (see Sec. \ref{StorageIntPermutation}).
For the first off-diagonal blocks with $|M_K(L)-M_K(M)|=1$, the matrix elements $H_{LM}$ consist of one-electron integrals of the type $\langle A|h|B\rangle$
and ERIs of the type $(AA|AB)$, whereas
for the second off-diagonal blocks with $|M_K(L)-M_K(M)|=2$, the matrix elements $H_{LM}$ consist of
ERIs of the type $(AB|AB)$. Moreover, the following relations
\begin{align}
|\bar{M}\rangle&=\hat{\mathcal{K}}|M\rangle,\quad \hat{\mathcal{K}}|\bar{M}\rangle  =(-1)^N|M\rangle,\\
|\bar{M}\rangle&=|M\rangle \mbox{ (closed-shell)},\\
\langle\bar{L}|H_+|\bar{M}\rangle&=\langle L|H_+| M\rangle^*=\langle M|H_+|L\rangle, \\
\langle \bar{L}|H_+|M\rangle&=(-1)^N \langle L|H_+|\bar{M}\rangle^*=(-1)^N \langle\bar{M}|H_+|L\rangle
\end{align}
dictate that the Hamiltonian matrix has the following structure
 \begin{equation}
 \begin{pmatrix}\langle L|H_+| M\rangle&\langle L|H_+|\bar{M}\rangle\\
 \langle\bar{L}|H_+|M\rangle&\langle\bar{L}|H_+|\bar{M}\rangle\end{pmatrix}=
 	\begin{bmatrix}
 		A_{LM} & B_{LM}            \\
 		(-1)^N B_{LM}^* & A_{LM}^* \\
 	\end{bmatrix},\quad \mathbf{A}=\mathbf{A}^\dag,\quad \mathbf{B}=\mathbf{B}^T,\label{Mat_TR}
 \end{equation}
which shows that the number of unique matrix elements can be reduced by a factor of two.
For the case of even $N$, it is even possible to construct
orthonormal, time-even (real-valued) scalar MBFs\cite{RQC_BOOK_Dyall}
\begin{equation}
	\left|M^s\right\rangle=\frac{i^{(1-s) / 2}}{\sqrt{2\left(1+\delta_{M, \bar{M}}\right)}}(|M\rangle+s|\bar{M}\rangle), \quad s=\pm 1, \label{even_basis}
\end{equation}
over which the Hamiltonian matrix become real and symmetric, viz.,
\begin{align}
&\begin{pmatrix}\langle L^+|H_+| M^+\rangle&\langle L^+|H_+|M^-\rangle\\
 \langle L^-|H_+|M^+\rangle&\langle L^-|H_+|M^+\rangle\end{pmatrix}=
 	\begin{bmatrix}
 		H^{++}_{LM} & H^{+-}_{LM}            \\
 		H^{-+}_{LM} & H^{--}_{LM}  \\
 	\end{bmatrix}, \label{even_hmat}
\end{align}
where
\begin{align}
H^{++}_{LM}&=\operatorname{Re}\left(H_{L M}+H_{L \bar{M}}\right) / G_{L M}=H^{++}_{ML}, \\
H^{+-}_{LM}&=-\operatorname{Im}\left(H_{L M}-H_{L \bar{M}}\right) / G_{L M}=H^{-+}_{ML}, \\
H^{-+}_{LM}&=\operatorname{Im}\left(H_{L M}+H_{L \bar{M}}\right) / G_{L M}=H^{+-}_{ML}, \\
H^{--}_{LM}&=\operatorname{Re}\left(H_{L M}-H_{L \bar{M}}\right) / G_{L M}=H^{--}_{ML},\\
G_{L M}&=\sqrt{\left(1+\delta_{L, \bar{L}}\right)\left(1+\delta_{M, \bar{M}}\right)}.
\end{align}

As shown in Appendix \ref{doublegroup2}, a second pseudo-quantum number, $\tilde{M}_K=2M_K\mod 4$,
can be invoked to classify the MBFs according to binary double point group symmetries. As a result,
 for both odd and even $N$, the combined use of time reversal and point group
symmetries reduces the computational cost by the order of the given group
(i.e., 2, 4, 4, 4, 8, 8, 8, and 16 for
the $C_1^\ast$, $C_i$, $C_2^\ast$, $C_s^\ast$, $C_{2h}^\ast$, $C_{2v}^\ast$, $D_2^\ast$, and $D_{2h}^\ast$ point groups, respectively).

\subsection{Breakdown of Relativistic Hamiltonian}\label{Hdecomposition}

It has been shown\cite{iCIPT2} that the spin-free, second-quantized Hamiltonian can be broken down to terms
that have one-to-one correspondence with the diagrams employed in the unitary group approach\cite{Paldus1980} for
determining the basic coupling coefficients (BCC) between CSFs.
In the first place, the breakdown classifies the Hamiltonian matrix elements $\langle I\mu | \hat{\mathcal{H}} | J\nu\rangle$ into distinct groups
based on the excitation level (0, 1, or 2) from the ket orbital configuration (oCFG) $|J\rangle$ to the bra  $|I\rangle$.
Secondly, the common zero or doubly occupied orbitals in the bra and ket oCFGs
can be deleted, ending up with reduced occupation tables (ROT) characterizing oCFG pairs.
All oCFG pairs sharing the same ROT have the same BCCs, thereby avoiding repeated evaluations of the BCCs.
Thirdly, a closed-shell reference oCFG can be introduced to simplify the evaluation of
the Hamiltonian matrix elements over zero or singly excited oCFGs.
It is shown here that
such techniques can also be extended to the relativistic domain working with SDs though,
by defining the occupation number (PON) of a Kramers pair $\mathcal{P}_p=(p,\bar{p})$ as $\tilde{n}_p=n_p+n_{\bar{p}}$.
Different SDs $|J\nu\rangle$ generated from the same oCFG $|J\rangle$ share the same PONs but
differ in their spinor occupations (SON).

The general rule of thumb for the breakdown of the Hamiltonian $H_+$ \eqref{np-H} is to reserve
the creation and annihilation characters of indices $(p, r)$ and $(q, s)$, respectively,
when recasting the unrestricted summation into various restricted summations. For instance,
the first, one-body term of $H_+$ \eqref{np-H} can be decomposed as
\begin{eqnarray}
H_1&=&H_1^0+H_1^1,\\
H_1^0&=&\sum_p h_{pp} E_{pp}, \label{H1-0}\\
H_1^1&=&\sum_{p< q} h_{pq}E_{pq}+\sum_{p>q}h_{pq}E_{pq}.\label{H1-1}
\end{eqnarray}
The superscripts 0, 1, and 2 in $H_i$ ($i=1,2$) indicate that the terms would
contribute to the Hamiltonian matrix elements over two oCFGs that are related by zero, single and double excitations, respectively.
To breakdown the second, two-body term of $H_+$ \eqref{np-H}, we notice that
\begin{align}
	\sum_{p,q,r,s}&=\sum_{\{p,r\}\cap\{q,s\}\ne \emptyset}+\sum_{\{p,r\}\cap\{q,s\}= \emptyset},\label{H2split}
\end{align}
where
\begin{align}
	\sum_{\{p,r\}\cap\{q,s\}\ne\emptyset}&=\left(\sum_{p=q=r=s}+\sum_{p=q\ne r=s}+\sum_{p=s\ne r=q}\right)\nonumber\\
    &+\left(\sum_{p=q=r\ne s}+\sum_{p=q=s\ne r}+\sum_{r=s=p\ne q}+\sum_{r=s=q\ne p}\right)\nonumber\\
	&+\left(\sum_{p=q\ne r\ne s}+\sum_{p=s\ne r\ne q}+\sum_{r=q\ne p\ne s}+\sum_{r=s\ne p\ne q}\right),\label{H2a}\\
	\sum_{\{p,r\}\cap\{q,s\}=\emptyset}&=\sum_{p\ne r\ne q\ne s}+\sum_{p=r\ne q\ne s}+\sum_{q=s\ne p\ne r}+\sum_{p=r\ne q=s}.\label{H2b}
\end{align}
By making further use of the relation $e_{pq,rs}=e_{rs,pq}$, the first summation of Eq. \eqref{H2split} can be written as
\begin{align}
	H_2^0+H_2^1&=\frac{1}{2}\sum_{\{p,r\}\cap\{q,s\}\ne \emptyset}g_{pq,rs}e_{pq,rs},\label{H201def}\\
	H_2^0&=\frac{1}{2}\sum_{p}g_{pp,pp}e_{pp,pp}+\frac{1}{2}\sum_{p\ne q}\left[g_{pp,qq}e_{pp,qq}+g_{pq,qp}e_{pq,qp}\right],\label{H2-0a}\\
	H_2^1&=\sum_{p\ne q}\left[g_{pq,pp}e_{pq,pp}+g_{pq,qq}e_{pq,qq}\right]
          +\sum_{p\ne q\ne r}\left[g_{pq,rr}e_{pq,rr}+g_{pr,rq}e_{pr,rq}\right].\label{H2-1a}
\end{align}
The second summation of $H_2^1$ \eqref{H2-1a} reads more explicitly
\begin{eqnarray}
	\sum_{p\ne q\ne r}&=&\left(\sum_{p<q<r}+\sum_{q<p<r}+\sum_{r<p<q}+\sum_{r<q<p}\right)+\left(\sum_{p<r<q}+\sum_{q<r<p}\right).\label{ijksum}
\end{eqnarray}
Similarly, the second summation of Eq. \eqref{H2split} can be written as
\begin{align}
	H_2^2&=\frac{1}{2}\sum_{\{p,r\}\cap\{q,s\}=\emptyset}g_{pq,rs}e_{pq,rs}, \\
	&=\frac{1}{2}\sum_{p\ne r\ne q\ne s}g_{pq,rs}e_{pq,rs}+\frac{1}{2}\sum_{p\ne q\ne r}g_{pq,rq}e_{pq,rq}\nonumber\\
    &+\frac{1}{2}\sum_{p\ne q\ne s}g_{pq,ps}e_{pq,ps}+\frac{1}{2}\sum_{p\ne q}g_{pq,pq}e_{pq,pq}.\label{H2-2}
\end{align}
The first term of $H_2^2$ \eqref{H2-2} can further be written as
\begin{align}
	\frac{1}{2}\sum_{p\ne r\ne q\ne s}g_{pq,rs}e_{pq,rs}=\sum_{p<r}\sum_{q<s}^{\prime}\left[g_{pq,rs}e_{pq,rs}+g_{ps,rq}e_{ps,rq}\right],
\end{align}
where the summation takes the following form
\begin{eqnarray}
	\sum_{p<r}\sum_{q<s}^{\prime}=\sum_{p<r<q<s}+\sum_{q<s<p<r}+\sum_{p<q<r<s}+\sum_{p<q<s<r}+\sum_{q<p<r<s}+\sum_{q<p<s<r}.\label{H2sum-a}
\end{eqnarray}
%It is obvious that interchanging $j$ ($i$) and $l$ ($k$) in the second (third) term gives rise to identical result, thereby leading to
%\begin{eqnarray}
%\frac{1}{2}\sum_{i\ne j\ne l}(ij|il)e_{ij,il}&=&\sum_i\sum_{j<l}^\prime(ij|il)e_{ij,il},\\
%\frac{1}{2}\sum_{i\ne j\ne k}(ij|kj)e_{ij,kj}&=&\sum_j\sum_{i<k}^\prime(ij|kj)e_{ij,kj},
%\end{eqnarray}
%where the summations read
%\begin{eqnarray}
%\sum_i\sum_{j<l}^\prime&=&\sum_{i<j<l}+\sum_{j<i<l}+\sum_{j<l<i},\\
%\sum_j\sum_{i<k}^\prime&=&\sum_{i<k<j}+\sum_{i<j<k}+\sum_{j<i<k}.
%\end{eqnarray}
The second term of Eq. \eqref{H2-2} can further be written as
\begin{align}
	\frac{1}{2}\sum_{p\ne q\ne r}g_{pq,rq}e_{pq,rq}&=\sum_q\sum_{p<r}^\prime g_{pq,rq}e_{pq,rq},\label{H2-2nd}\\
	\sum_q\sum_{p<r}^\prime&=\sum_{p<r<q}+\sum_{p<q<r}+\sum_{q<p<r},\label{H2sum-b}
\end{align}
by observing that interchanging $p$ and $r$ on the left hand side of Eq. \eqref{H2-2nd} gives rise to an identical result.
Similarly, the third term of Eq. \eqref{H2-2} can further be written as
\begin{align}
	\frac{1}{2}\sum_{p\ne q\ne s}g_{pq,ps}e_{pq,ps}&=\sum_p\sum_{q<s}^\prime g_{pq,ps}e_{pq,ps},\label{H2-3rd}\\
	\sum_p\sum_{q<s}^\prime&=\sum_{q<s<p}+\sum_{q<p<s}+\sum_{p<q<s}.\label{H2sum-c}
\end{align}
In sum, $H_2^2$ reads
\begin{align}
	H_2^2&=\sum_{p<r}\sum_{q<s}^{\prime}\left[g_{pq,rs}e_{pq,rs}+g_{ps,rq}e_{ps,rq}\right]+\sum_j\sum_{p<r}^\prime g_{pq,rq}e_{pq,rq}\nonumber\\
	&+\sum_i\sum_{q<s}^\prime g_{pq,ps}e_{pq,ps}+\frac{1}{2}\sum_{p\ne q} g_{pq,pq}e_{pq,pq},\label{H2-2f}
\end{align}
in conjunction with Eqs. \eqref{H2sum-a}, \eqref{H2sum-b}, and \eqref{H2sum-c} for the first three summations, respectively.
Alternatively, $H_2^2$ \eqref{H2-2f} can be written as
\begin{eqnarray}
H_2^2=\sum_{p\le r}\sum_{q\le s}^\prime \left[ 2^{-\delta_{pr}\delta_{qs}} g_{pq,rs}e_{pq,rs}+(1-\delta_{pr})(1-\delta_{qs})g_{ps,rq}e_{ps,rq}\right],\label{EqnDiff2}
\end{eqnarray}
where the prime indicates that $\{ p, r\}\cap\{q, s\} =\emptyset$.

The individual terms of $H_+$ \eqref{np-H} are summarized in the Table \ref{H-diagram}. The corresponding diagrams are shown in Figs. \ref{Diagrams-0} to \ref{Diagrams-2},
which are drawn with the following conventions:
(1) The enumeration of spinor levels starts with zero and increases from bottom to top.
(2) The left and right vertices (represented by filled diamonds)
indicate creation and annihilation operators, respectively, which form a single generator when connected by a non-vertical line.
(3) Products of single generators should always be understood as normal ordered.
For instances, Fig. \ref{Diagrams-1}(a) and 2(b) mean $E_{pq}$ ($p<q$) and $E_{pq}$ ($p>q$), respectively.
The former is a raising generator (characterized
by a generator line going upward from left to right),
whereas the latter is a lowering generator (characterized by
a generator line going downward from left to right);
Fig. \ref{Diagrams-1}(m) means $\{E_{pr}E_{rq}\}=E_{pr}E_{rq}-E_{pq}=e_{pr, rq}$ ($p<q<r$),
which is the exchange counterpart
of the direct generator $\{E_{pq}E_{rr} \}=\{E_{rr}E_{pq}\}=e_{pq, rr}$ ($p<q<r$) shown in Fig. \ref{Diagrams-1}(g).
Such diagrams for the BCCs between spin-dependent SDs
are topologically the same as those for the BCCs between spin-free CSFs\cite{iCIPT2,Paldus1980}.
Therefore, the same tabulated unitary group approach (TUGA)\cite{iCIPT2} for the
calculation and reutilization of the BCCs between randomly selected spin-free CSFs
can also be applied here.  However, it should be kept in mind that every vertex of the diagrams [i.e., every term in the algebraic expressions
$H_1^0$ \eqref{H1-0}, $H_2^0$ \eqref{H2-0a}, $H_1^1$ \eqref{H1-1}, $H_2^1$ \eqref{H2-1a}, and $H_2^2$ \eqref{H2-2f}/\eqref{EqnDiff2}]
is composed of both A- and B-spinors (e.g., the first term of $H_2^0$ \eqref{H2-0a} is composed of
16 terms!).

Compared to the straightforward use of the Slater-Condon rules for the BCCs of individual pairs of SDs,
the present scheme (i.e., TUGA) takes oCFGs as the organization units and avoids completely
redundant calculations of the BCCs by invoking a sorting procedure. Since the number of oCFGs is much smaller than that of SDs
(NB: a single oCFG with $N_o$ singly occupied Kramers pairs
gives rise to $2^{No}$ SDs), the high efficiency of TUGA is obvious.

\begin{table}[!htp]
	\centering
	\small
	\caption{Diagrammatic representation of the relativistic Hamiltonian $H_+$ \eqref{np-H}. }
	\begin{threeparttable}
		\begin{tabular}{lllll}\toprule\toprule \multicolumn{1}{l}{Hamiltonian}&\multicolumn{1}{l}{expression$^a$}&\multicolumn{1}{l}{range}&\multicolumn{1}{l}{diagram$^b$}&\\\toprule
			$H_1^0$ \eqref{H1-0} &$h_{pp}E_{pp}$                     &         &w1\\
			$H_2^0$ \eqref{H2-0a}&$\frac{1}{2} g_{pp,pp}e_{pp,pp}$      &         &w2 \\
			&$g_{pp,qq}e_{pp,qq}$                 &$p<q$    &w3 \\
			&$g_{pq,qp}e_{pq,qp} $                &$p<q$    &b7=c7\\\midrule
			$H_1^1$ \eqref{H1-1} &$h_{pq} E_{pq}$                    &$p<q$    & s1\\
			&                                   &$p>q$    & s2\\
			$H_2^1$ \eqref{H2-1a} &$g_{pq,pp}e_{pq,pp}$                &$p<q$    & s3\\
			&                                   &$p>q$    & s4 \\
			&$g_{pq,qq}e_{pq,qq}$               &$p<q$    & s5 \\
			&                                   &$p>q$    & s6 \\
			&$g_{pq,rr}e_{pq,rr}$               &$p<q<r$  & s7 \\
			&                                   &$q<p<r$  & s8\\
			&                                   &$r<p<q$  & s9\\
			&                                   &$r<q<p$  & s10\\
			&                                   &$p<r<q$  & s11\\
			&                                   &$q<r<p$  & s12\\
			&$g_{pr,rq}e_{pr,rq}$               &$p<q<r$  & b6\\
			&                                   &$q<p<r$  & c6\\
			&                                   &$r<p<q$  & b4\\
			&                                   &$r<q<p$  & c4\\
			&                                   &$p<r<q$  & a2\\
			&                                   &$q<r<p$  & d2\\\midrule
			$H_2^2$ \eqref{H2-2f}/\eqref{EqnDiff2}&\multirow{6}{*}{$g_{pq,rs}e_{pq,rs}+g_{ps,rq}e_{ps,rq}$}&$p<r<q<s$&a3 + a5&\\
			&                                   &$q<s<p<r$&d3 + d5\\
			&                                   &$p<q<r<s$&a1 + b5\\
			&                                   &$p<q<s<r$&c1 + c3\\
			&                                   &$q<p<r<s$&b1 + b3\\
			&                                   &$q<p<s<r$&d1 + c5\\\cline{2-5}
			&                                   &$p<r<q=s$  &a6&\\
			&$g_{pq,rq}e_{pq,rq}$               &$p<q=s<r$  &c2\\
			&                                   &$q=s<p<r$  &d4\\\cline{2-5}
			&                                   &$q<s<p=r$  &d6&\\
			&$g_{pq,ps}e_{pq,ps}$               &$q<p=r<s$  &b2\\
			&                                   &$p=r<s<q$  &a4\\\cline{2-5}
			&\multirow{2}{*}{$\frac{1}{2}g_{pq,pq}e_{pq,pq}$}      &$p=r<q=s$    &a7&\\
			&                                                    &$q=s<p=r$    &d7\\
			\bottomrule\bottomrule
		\end{tabular}
		\begin{tablenotes}
         \item[a] Every index should be assigned freely to A- or B-spinor (e.g., $h_{pp}$ is composed of
         $h_{pq}$, $h_{\bar{p}q}$, $h_{p\bar{q}}$, and $h_{\bar{p}\bar{q}}$).
		 \item[b] See Figs. \ref{Diagrams-0}--\ref{Diagrams-2}.
		\end{tablenotes}
	\end{threeparttable}\label{H-diagram}
\end{table}

%\section{Diagrammatic representation of the spin-free, second-quantized Hamiltonian}\label{SecDiagram}
\clearpage
%\newpage
%\input{Diagram1}
%\input{Diagram2}
%\input{Diagram3}
\input{Diagram}

\subsection{Explicit expressions for Hamiltonian matrix elements}\label{SecMat}
As discussed in Appendix \ref{doublegroup2}, spin-dependent CSFs (time reversal and point group symmetry-adapted
MBFs) are related to the SDs in a very simple way. Specifically, a single SD is
already a CSFs in the case of odd $N$, whereas a CSF is merely a simple combination of two SDs (see Eq. \eqref{even_basis})
in the case of even $N$. Moreover, each CSF has a simple structure in terms of binary double point group symmetries.
As such, the symmetrized Hamiltonian matrix can first be constructed in the space of
(randomly selected) SDs and then transformed to that of CSFs. The matrix elements of the Coulomb interaction
are documented here, whereas those of the Gaunt/Breit interaction are given in Appendix \ref{GauntME}.

\subsubsection{Zero Excitations}\label{ZeroEDiff}
For matrix elements between SDs of the same oCFG $|I\rangle$, it is the $H_1^0$ \eqref{H1-0} and $H_2^0$ \eqref{H2-0a} parts of $H_+$ \eqref{np-H} that should be adopted, i.e.,
\begin{align}
\langle I\mu|H_1^0+H_2^0|I\nu\rangle&=H_{\mathrm{D}}^0\delta_{\mu\nu}+\langle I\mu|H_{\mathrm{O}}^0|I\nu\rangle.
\end{align}
As shown in Appendix \ref{SecC1}, introducing a closed-shell reference oCFG $\{\tilde{\omega}_p=\omega_p+\omega_{\bar p}=2 \mbox{ or } 0\}$
permits a more efficient evaluation of the diagonal term, viz.,
\begin{align}
	H^0_{\text{D}}-E_{\mathrm{ref}} 	& =
	  \left\{\sum_{p} [(\bar{p}\bar{p}|pp) - (\bar{p}p|p\bar{p})]\tilde{\Delta}_p\delta_{\tilde{\omega}_p,2}
	+ \sum_{p} [(\bar{p}\bar{p}|pp) - (\bar{p}p|p\bar{p})][\delta_{\tilde{\Delta}_p,2} +  \delta_{\tilde{\Delta}_p,-2}]\right\}   \nonumber \\
	& +    \left\{ \sum_q \left [ h_{qq} +  \sum_{p\neq q} [2(pp|qq)-(pq|qp)-(\bar{p}q|q\bar{p})]\delta_{\tilde{\omega}_p,2}\right]
	\tilde{\Delta}_q + \sum_{p<q} (pp|qq) \tilde{\Delta}_p\tilde{\Delta}_q   \right\}                                               \nonumber \\
	& -   \sum_{p\neq q}  [(pq|qp)+(\bar{p}q|q\bar{p})] \tilde{\Delta}_p\delta_{\tilde{\Delta}_q,2}
	  +   \sum_{p\neq q}  [(pq|qp)+(\bar{p}q|q\bar{p})] \tilde{\Delta}_p\delta_{\tilde{\Delta}_q,-2}  \\
	& -   \left\{\sum_{p<q}^\prime (pq|qp)\left[\Delta_p\Delta_q+\Delta_{\bar{p}}\Delta_{\bar{q}}\right]
	+ (\bar{p}q|q\bar{p})\left[\Delta_{\bar{p}}\Delta_q + \Delta_{p}\Delta_{\bar{q}}\right]\right\}, \label{Dg_3}\\
	E_{\mathrm{ref}}&=\sum_p [2 h_{pp} + (pp|\bar{p}\bar{p}) - (p\bar{p}|\bar{p}p)]\delta_{\tilde{\omega}_p,2}  \nonumber\\
	&+\sum_{p<q} 2[2(pp|qq)-(pq|qp)-(\bar{p}q|q\bar{p})] \delta_{\tilde{\omega}_p,2}\delta_{\tilde{\omega}_q,2}.\label{Dg_Const_1}
\end{align}
Here, the summations run over Kramers pairs (instead of spinors) and $\sum^\prime$ means that both Kramers pairs $\mathcal{P}_p$ and $\mathcal{P}_q$ are singly occupied.
$E_{\mathrm{ref}}$ \eqref{Dg_Const_1} is the energy of
the closed-shell reference (i.e., common zero point energy for all SDs).
The second and third terms of Eq. \eqref{Dg_3} depend only on
the differential occupations $\Delta_{p}=n_p-\omega_p$,
$\Delta_{\bar{p}}=n_{\bar{p}}-\omega_{\bar{p}}$, and $\tilde{\Delta}_{p}=\Delta_{p}+\Delta_{\bar{p}}$,
and are therefore the same for all diagonal elements for the same oCFG $|I\rangle$.

 %\textcolor[rgb]{1.00,0.00,0.00}{The following should be put into equations in Eq. (92) Appendix C.1, such that
 %Eq. (47) is the final expression}
%As for the last term of Eq. \eqref{Dg_3}, it is important to note that if one of the Kramers pairs, say $\mathcal{P}_q$,
%satisfies $\tilde{\Delta}_q=\pm2$ (i.e., $\Delta_q=\Delta_{\bar{q}}=\pm 1$), then both $\Delta_p\Delta_q+\Delta_{\bar{p}}\Delta_{\bar{q}}$ and $\Delta_{\bar{p}}\Delta_q + \Delta_{p}\Delta_{\bar{q}}$ are equal to $\pm\tilde{\Delta}_p$. In this case, this term can be merged with
%the second term. As such, this term is to be evaluated only when $\tilde{n}_p=\tilde{n}_q=1$.

The off-diagonal matrix elements read
\begin{align}
\langle I\mu|H_{\mathrm{O}}^0|I\nu\rangle
	& = \sum_{p<q} \left\{
	[(\bar{p}p|\bar{q}q) - (\bar{p}q|\bar{q}p)]\Gamma^{I\mu I\nu}_{\bar{p} p \bar{q}q}+
	[(\bar{p}p|q\bar{q}) - (\bar{p}\bar{q}|qp)]\Gamma^{I\mu I\nu}_{\bar{p} p q\bar{q}}\right.  \nonumber  \\
	& + \left.
	[(p\bar{p}|\bar{q}q)-(pq|\bar{q}\bar{p})]\Gamma^{I\mu I\nu}_{p\bar{p} \bar{q}q}+
	[(p\bar{p}|q\bar{q}) - (p\bar{q}|q\bar{p})]\Gamma^{I\mu I\nu}_{p\bar{p}q\bar{q}} \right\}\nonumber \\
	& - \sum_{p<q} \left\{
	 (\bar{p}q|qp) (n_q-n_{\bar{q}})\gamma^{I\mu I\nu}_{\bar{p}p}
	+ (p\bar{q}|qp) (n_p-n_{\bar{p}})\gamma^{I\mu I\nu}_{q\bar{q}}  \right. \nonumber \\
	& \left.
	+ (pq|\bar{q}p)(n_p-n_{\bar{p}})\gamma^{I\mu I\nu}_{\bar{q}q}
	+ (pq|q\bar{p})(n_q-n_{\bar{q}})\gamma^{I\mu I\nu}_{p\bar{p}}  \right\},\label{H0off}\\
\gamma^{I\mu J\nu}_{p_{\sigma}q_{\tau}}&=\langle I\mu|E_{p_{\sigma}q_{\tau}}|J\nu\rangle,\quad \sigma,\tau\in A,B;\quad \bar{\sigma},\bar{\tau}\in B, A,\label{gamma1}\\
 \Gamma^{I\mu J\nu}_{p_{\sigma}q_{\tau},r_{\lambda}s_{\gamma}}&=\langle I\mu|e_{p_{\sigma}q_{\tau},r_{\lambda}s_{\delta}}|J\nu\rangle. \label{gamma2}
\end{align}
Again, the summations in Eq. \eqref{H0off} run over Kramers pairs.
Every term therein is nonzero only for singly occupied Kramers pairs.
The Greek subscripts in Eqs. \eqref{gamma1} and \eqref{gamma2} indicate the type of spinors, e.g.,
if $p_{\sigma}$ is an A/B-spinor $p$/$\bar{p}$, $p_{\bar{\sigma}}$ would be
the opposite, $\bar{p}$/$p$.

\subsubsection{Single Excitations}\label{OneEDiff}
When an oCFG $|I\rangle$ can be obtained from $|J\rangle$ by exciting a single electron from spinor $q_\tau$ to $p_\sigma$ ($\ne q_\tau$), we have
\begin{align}
	\langle I\mu|H_1^1+H_2^1|J\nu\rangle
	&=\left\{f_{p_\sigma q_\tau} + (p_\sigma q_\tau|p_{\bar{\sigma}}p_{\bar{\sigma}})n_{p_{\bar{\sigma}}}^J
	+(p_\sigma q_\tau|q_{\bar{\tau}}q_{\bar{\tau}}) n_{q_{\bar{\tau}}}^J \right.\nonumber\\
	&\left.-\sum_\lambda\sum_{r}^\prime
	(p_\sigma r_\lambda|r_{\lambda}q_\tau) \delta_{\tilde{n}_r^J,1} n_{r_\lambda}^J\right\} \gamma^{I\mu J\nu}_{p_\sigma q_\tau} \nonumber\\
	&+\sum_\lambda\sum_{r}^\prime
	(p_\sigma r_\lambda|r_{\bar{\lambda}}q_\tau)
	\Gamma_{p_\sigma r_\lambda, r_{\bar{\lambda}}q_\tau}^{I\mu J\nu},\label{SingleME} \\
	f_{p_\sigma q_\tau} & = f^{\mathrm{c}}_{p_\sigma q_\tau}
	+ \sum^\prime_{r} (p_\sigma q_\tau |rr)\tilde{\Delta}_r^J
	-\sum_\lambda\sum_{r}^\prime (p_\sigma r_\lambda|r_\lambda q_\tau)\delta_{\tilde{n}_r^J,2}, \label{f1}
\end{align}
where the summation $\sum_r^\prime$ over Kramers pairs excludes $r= p$ or $q$.
All the terms need to be calculated only for Kramers pairs with different PONs or
those that are commonly singly occupied.
It should also be clear from Table \ref{H-diagram} and Fig. \ref{Diagrams-1} that only half of the diagrams (with $p<q$) need to be evaluated,
for the other half can simply be obtained by hermitian conjugation.

\subsubsection{Double Excitations}\label{TwoEDiff}
When an oCFG $|I\rangle$ can be obtained from oCFG $|J\rangle$ by exciting two electrons
from $q_\tau,s_\delta$ to $p_\sigma,r_\lambda$ (subject to ${p_\sigma \leq r_\lambda, q_\tau \leq s_\delta,
\{p_\sigma,r_\lambda\}\cap\{q_\tau,s_\delta\}=\emptyset})$,
the Hamiltonian matrix elements read (cf. Eq. \eqref{EqnDiff2})
\begin{align}
\langle I\mu|H_2^2|J\nu\rangle=
[2^{-\delta_{p_\sigma r_\lambda}\delta_{q_\tau s_\delta}}(p_\sigma q_\tau|r_\lambda s_\delta) - (1-\delta_{p_\sigma r_\lambda})(1-\delta_{q_\tau s_\delta})(p_\sigma s_\delta|r_{\lambda}q_{\tau})]
\Gamma_{p_\sigma q_\tau, r_\lambda s_\delta}^{I\mu J\nu}.
\end{align}
As can be seen from Table \ref{H-diagram} and Fig. \ref{Diagrams-2}, there are in total 20 diagrams
(i.e., 20 cases for the conditions ${p \leq r, q \leq s, \{p,r\}\cap\{q,s\}=\emptyset})$). However,
since diagrams c1+c3 and b1+b3, c5+d1 and b5+a1, d3+d5 and a3+a5, c2 and b2, d4 and a4, d6 and a6, as well as
d7 and a7 are conjugate pairs, respectively,
only the cx and dx types of diagrams need to be evaluated explicitly. All the rest
can simply be obtained by hermitian conjugation.

\section{4C-iCIPT2}\label{SDS}

Having discussed the basic elements that are common to all relativistic quantum chemical methods, we now present
a specific realization of the ideas by implementing 4C-iCPT2 on top of the C++ templates employed for
the spin-free iCIPT2\cite{iCIPT2,iCIPT2New}, which stems from the (restricted) static-dynamic-static (SDS) ansatz\cite{SDS}
for constructing many-electron wave functions. The ansatz can best be understood as follows. Without loss of generality,
the $N_P$ FCI solutions $\{|\Psi_I\rangle\}_{I=1}^{N_p}$ of a second-quantized Hamiltonian can be written as
\begin{align}
|\Psi_I\rangle&=\sum_{\mu=1}^{N_T} |\Phi_{\mu}\rangle C_{\mu I},\quad C_{\mu I}=\sum_{k=1}^{N_T} \tilde{C}_{\mu k}\bar{C}_{kI}\label{SVD}\\
&=\sum_{k=1}^{N_T} |\tilde{\Phi}_k\rangle \bar{C}_{kI},\quad |\tilde{\Phi}_k\rangle=\sum_{\mu=1}^{N_T}|\Phi_{\mu}\rangle\tilde{C}_{\mu k}\label{SVD2}\\
&\approx\sum_{k=1}^{\tilde{N}}|\tilde{\Phi}_k\rangle \bar{C}_{kI},\label{SVD3}
\end{align}
where $\{\Phi_{\mu}\}_{\mu=1}^{N_T}$ represent all possible MBFs,
while $\{\tilde{\Phi}\}_{k=1}^{N_T}$ are the contracted ones. Note that
going from Eq. \eqref{SVD} to Eq. \eqref{SVD2} does not introduce any error, but renders
the Hamiltonian matrix approximately diagonal (exactly diagonal if $\tilde{\mathbf{C}}=\mathbf{C}$).
It is immediately clear that, as long as the contraction coefficients $\tilde{C}_{\mu k}$ are chosen properly,
only a small number of contracted MBFs $\{\tilde{\Phi}_k\}_{k=1}^{\tilde{N}}$ should be enough!
It is of interest to see that the FCI solutions $\{|\Psi_I\rangle\}_{I=1}^{N_P}$ in the form of
Eq. \eqref{SVD3} are analogs of the Hartree-Fock (HF) solutions: $\{\Phi_{\mu}\}_{\mu=1}^{N_T}$,
$\{|\tilde{\Phi}_k\rangle\}_{k=1}^{N_T}$, and  $\{|\tilde{\Phi}_k\rangle\}_{k=1}^{\tilde{N}}$
correspond to the pre-chosen primitive, generally contracted but untruncated, and generally contracted
and truncated one-particle basis functions, respectively, in typical HF calculations.
The very first realization of Eq. \eqref{SVD3} is the particular choice of $\tilde{N}=3N_P$ for $N_P$ states, viz.,
\begin{align}
|\Psi_I\rangle&=\sum_{k=1}^{N_P}|\Psi_k^{(0)}\rangle\bar{C}_{kI}+\sum_{k=1}^{N_P}|\Xi_k^{(1)}\rangle\bar{C}_{(k+N_P)I}
+\sum_{k=1}^{N_P}|\Theta_k\rangle\bar{C}_{(k+2N_P)I},\label{ixc1}
\end{align}
where the $N_P$ zeroth-order states $\{|\Psi_k^{(0)}\rangle\}$ (which are usually generated by
CASSCF (complete active space self-consistent field) calculations) describe the primary static correlation,
the $N_P$ first-order corrections $\{|\Xi_k^{(1)}\rangle\}$ (which can be generated by any perturbation theory)
describe the primary dynamic correlation, whereas the $N_P$ secondary states $\{|\Theta_k\rangle\}\}$ (which can be generated
in a number of ways\cite{SDS}) describe the secondary static correlation (i.e.,
relaxation of the zeroth-order coefficients in the presence of dynamic correlation). Both theoretical analysis and numerical
evidence show\cite{SDS} that $|\Psi_k^{(0)}\rangle$, $|\Xi_k^{(1)}\rangle$, and $|\Theta_k\rangle$
have decreasing weights in the wave function $|\Psi_I\rangle$ \eqref{ixc1}, thereby justifying the ''static–
dynamic–static'' characterization. Obviously, Eq. \eqref{ixc1} is nothing but
a minimal multi-state MRCI model and has been dubbed SDSCI\cite{SDS}. Numerous examples
have revealed\cite{SDSRev} that SDSCI is very close in accuracy to
internally contracted MRCI with singles and doubles (ic-MRCISD), yet at a computational cost
only of one iteration of the latter. Taking SDSCI as a seed, a family of methods
can be derived\cite{LiuWIRES2023}, among which iCI (iterative configuration interaction)\cite{iCI} as an iterative version of SDSCI
is an exact FCI solver, which converges quickly and monotonically from above by diagonalizing a $3N_p$-by-$3N_P$ matrix
at each iteration. Further combined with an iterative selection of configurations for a compact variational space
and the Epstein-Nesbet second-order
perturbation (ENPT2)\cite{Epstein,Nesbet} for the residual dynamic correlation gives rise to iCIPT2\cite{iCIPT2,iCIPT2New}, which is
one of the most efficient near-exact methods\cite{Blindtest} and can handle arbitrary open-shell systems with full molecular symmetry.

As explained in Secs. \ref{Hdecomposition} and \ref{SecMat},
the essential ingredients of iCIPT2, including the diagrammatical representation of the Hamiltonian, the TUGA for the
calculation and reutilization of the BCCs between
randomly selected configurations, and the special treatment of the Hamiltonian matrix elements
can directly be transferred to 4C-iCPT2. Moreover, the criteria for configuration selection\cite{iCIPT2New}, the
algorithms for the storage and sorting of selected configurations\cite{iCIPT2New}, and the
constraint-based algorithm\cite{ASCI2018PT2} for the ENPT2 correction all remain unchanged.
As such, what is really needed to do is to make use of the template meta-programming and polymorphism furnished by C++, by
writing a code for constructing the relativistic Hamiltonian matrix and taking it as input
for the existing code for the configuration selection and ENPT2 correction, with no need to distinguish between complex and real algebras.
It also deserves to be mentioned that nothing needs to be done on
memory allocation, loop unrolling, and parallelization, etc.
This way, the implementation task of 4C-iCIPT2 (and any other relativistic wave function method) is greatly reduced.

\section{BENCHMARK CALCULATIONS}\label{pilot_app}
The 4C-iCIPT2 method introduced here was implemented in the BDF program package\cite{BDF2020}, and first applied to
investigate the spin-orbit splittings (SOS) of the ground states of the second- to fifth-row elements of groups 13 to 17.
Since 4C-CASSCF is out of our disposal, we started with Kramers-restricted Dirac-Hartree-Fock (DHF) calculations on the
$X^+$, $X^{2+}$, $X^{3+}$, $X^{2-}$, and $X^{-}$ closed-shell configurations (in conjunction with the finite Gaussian nuclear model\cite{VISSCHER1997207})
for groups 13 to 17, respectively,
by using the PySCF program package\cite{pyscf1,pyscf2}.
The spinors were then symmetrized to $D_{2h}^\ast$ symmetry.
%The states tested in this work correspond to the lowest 6, 10, 14, 10, 6 states for groups 13 to 17 respectively.
The selection of important configurations in 4C-iCIPT2 was controlled by a single parameter, $C_{\mathrm{min}}$. That is,
those CSFs of coefficients smaller in absolute value than $C_{\mathrm{min}}$ were pruned away in each iterative selection of
configurations (for more details, see Ref. \citenum{iCIPT2}). Upon convergence, the ENPT2 corrections ($E_{\mathrm{c}}^{(2)}$) were performed for each state individually.
In valence-only calculations, $C_{\mathrm{min}}$ was set to be $5\times10^{-7}$, which gives essentially convergent results.
%With perturbation correction, the energy difference between the degenerate state is within $1\mu E_h$.
%As the perturbation correction for each state is up to $0.2m E_h$, the total energy $E_{\text{tot}}$ ($=E_{\text{var}} + E_{\mathrm{c}}^{(2)}$) is thought to be converged.
When inner-shell electrons were further correlated, the energies ($E_{\mathrm{tot}}$) were calculated with five values for $C_{\mathrm{min}}$, i.e., $\{30,10,9,7,5\}\times10^{-6}$. Noticeably, the so-calculated energies are not strictly identical for a degenerate manifold of states
(e.g., the $^2P_{3/2,3/2}$ and $^2P_{3/2,1/2}$ components of $^2P_{3/2}$). However,
the splittings are at most 5 cm$^{-1}$ even for the heaviest elements considered here.
Therefore, the energies were simply averaged ($E_{\mathrm{tot}}^{\mathrm{av}}$) and then extrapolated to the zero limit of $C_{\mathrm{min}}$
by a linear fit of $E_{\mathrm{tot}}^{\mathrm{av}}$ vs  $|E_{\mathrm{c}}^{(2)}|$.

The Br atom was first taken as a showcase to reveal possible basis set and inner-shell electron effects, for correlation consistent basis sets up to quintuple-zeta quality
are not available for heavier atoms. As shown in Tables \ref{DC_Basis_Set} and \ref{DCB_Basis_Set}, when only valence electrons are correlated,
the uncontracted cc-pVTZ basis sets\cite{Dunning1999} already yield converged SOS for the $^2P$ state of Br,
for both the DC and DCB Hamiltonians.
In contrast, the contracted cc-pVXZ-DK basis sets\cite{Dixon2001} not only have sizable errors but also have no hint of convergence.
To see if this stems from the underlying scalar relativistic contractions\cite{Dixon2001} of cc-pVXZ\cite{Dunning1999}, we repeated the calculations
with the DHF spinors as the bases. They were obtained with the uncontracted cc-pVXZ\cite{Dunning1999}, but with the energetically highest spinors
removed to end up with the same numbers of spinors as those
by the scalar relativistic contractions\cite{Dixon2001}. As can be seen from Tables \ref{DC_Basis_Set} and \ref{DCB_Basis_Set}
(and also Tables S1 to S4), such spinor-based contractions do work very well.

It can also be seen from Tables \ref{DC_Basis_Set} and \ref{DCB_Basis_Set} that the fully occupied
$3d$ spinors have a relatively small contribution (less than 1\%)
to the SOS of the $^2P$ state of Br, for both the DC and DCB Hamiltonians in conjunction with
various basis sets. This is different from calculations where scalar relativistic orbitals are taken as the basis
and spin-orbit coupling is accounted for only at the correlation step. For instance, in the one-step-1C SOiCI calculations\cite{SOiCI}, the $3d$-shell
contributes as large as 367 cm$^{-1}$ (or 11\%) to the SOS of the $^2P$ state of Br
[3745 cm$^{-1}$ for $3d^{10}4s^24p^5$ vs 3378 cm$^{-1}$ for $4s^24p^5$]. This is of course not surprising: compared to the
pure correlation energies in 4C/2C approaches, a larger amount of
spin-orbit-correlation energies have to be accounted for in one-component approaches, so as to compensate for the large portion of spin-orbit
energies missed in the mean-field step\cite{2C-CICC2012,SOiCISCF,Cheng-MP-2023}.
Note, however, the fully occupied $(n-1)d$ spinors have large contributions to the SOSs of the fourth- and fifth-row atoms of groups 13 to 15 in a relative sense,
simply because the SOSs of these atoms are themselves relatively small (see Tables S3 and S4).

\begin{table}[htbp]
	\small
	\centering
	\caption{Spin-orbit splitting (SOS in cm$^{-1}$) of the $^2\text{P}$ of Br
with contracted and uncontracted basis sets in conjunction with the DC Hamiltonian}
\begin{threeparttable}
	\begin{tabular}{ccccccccc} \toprule
	\multirow{2}[0]{*}{basis} & \multicolumn{2}{c}{uncontracted} & \multicolumn{2}{c}{contracted} &
	\multicolumn{2}{c}{atomic spinor\tnote{c}} & \multirow{2}[0]{*}{$\Delta_1$\tnote{d,f}}  & \multirow{2}[0]{*}{$\Delta_2$\tnote{d,g}} \\ \cline{2-7}
	& SOS\tnote{d}   & error\tnote{d,e} & SOS\tnote{d}   & error\tnote{d,e}
	& SOS\tnote{d}   & error\tnote{d,e} \\ \toprule
	cc-pVDZ-DK\tnote{a} & 3665[3680]  & -0.6[-0.1]  & 3417[3407]  & -7.3[-7.5]  & 3641[3638] & -1.2[-1.3] & -247[-273] & -24[-42]\\
	cc-pVTZ-DK\tnote{a} & 3617[3630]  & -1.8[-1.5]  & 3470[3462]  & -5.8[-6.1]  & 3612[3621] & -2.0[-1.7] & -147[-168] & -5[-9]  \\
	cc-pVQZ-DK\tnote{a} & 3618[3644]  & -1.8[-1.1]  & 3497[3483]  & -5.1[-5.5]  & 3617[3644] & -1.8[-1.1] & -121[-161] & -1[0]   \\
	cc-pV5Z-DK\tnote{a} & 3621[3609]  & -1.7[-2.1]  & 3643[3675]  & -1.1[-0.3]  & 3620[3589] & -1.8[-2.6] &  22[66]    & -1[-20] \\
	ANO-RCC\tnote{b}    & 3616[3665]  & -1.9[-0.6]  & 3689[3701]  &  0.1[0.4]   & 3616[3577] & -1.9[-2.9] &  73[37]    & 0[-88]  \\\bottomrule
\end{tabular}%
	\begin{tablenotes}
\item[a]Refs. \citenum{Dunning1999,Dixon2001}
\item[b]Ref. \citenum{ANO2}
\item[c]DHF spinors obtained with uncontracted basis but truncated to the same number as the original scalar relativistic contraction.
\item[d]$4s^24p^5$[$3d^{10}4s^24p^5$].
\item[e]Percentage error compared to the experimental value of 3685 cm$^{-1}$\cite{NIST1}.
\item[f]Difference between contracted and uncontracted basis.
\item[g]Difference between atomic spinors and uncontracted basis.
	\end{tablenotes}
\end{threeparttable}
	\label{DC_Basis_Set}
\end{table}%

\begin{table}[htbp]
	\small
	\centering
	\caption{Spin-orbit splitting (SOS in cm$^{-1}$) of the $^2\text{P}$ of Br
with contracted and uncontracted basis sets in conjunction with the DCB Hamiltonian}
\begin{threeparttable}
	\begin{tabular}{ccccccccc} \toprule
	\multirow{2}[0]{*}{basis} & \multicolumn{2}{c}{uncontracted} & \multicolumn{2}{c}{contracted} &
	\multicolumn{2}{c}{atomic spinor\tnote{c}} & \multirow{2}[0]{*}{$\Delta_1$\tnote{d,f}}  & \multirow{2}[0]{*}{$\Delta_2$\tnote{d,g}} \\ \cline{2-7}
	& SOS\tnote{d}   & error\tnote{d,e} & SOS\tnote{d}   & error\tnote{d,e}
	& SOS\tnote{d}   & error\tnote{d,e} \\ \toprule
         cc-pVDZ-DK\tnote{a} &  3595[3611] & -2.4[-2.0] &3375[3367]  & -8.4[-8.6]  & 3572[3569] & -3.1[-3.1] & -221[-244]  & -23[-42]\\
		 cc-pVTZ-DK\tnote{a} &  3548[3575] & -3.7[-3.0] &3425[3421]  & -7.1[-7.2]  & 3543[3552] & -3.8[-3.6] & -124[-154]  & -5[-23] \\
         cc-pVQZ-DK\tnote{a} &  3550[3573] & -3.7[-3.0] &3453[3442]  & -6.3[-6.6]  & 3549[3568] & -3.7[-3.2] & -96[-131]   & -1[-5]  \\
		 cc-pV5Z-DK\tnote{a} &  3550[3545] & -3.7[-3.8] &3589[3596]  & -2.6[-2.4]  & 3552[3513] & -3.6[-4.7] &  39[50]     & 1[-32]  \\
		 ANO-RCC\tnote{b}    &  3548[3564] & -3.7[-3.3] &3625[3596]  & -1.6[-2.4]  & 3547[3511] & -3.7[-4.7] &  78[31]     & -1[-54] \\\bottomrule
	\end{tabular}%
	\begin{tablenotes}
\item[a]Refs. \citenum{Dunning1999,Dixon2001}
\item[b]Ref. \citenum{ANO2}
\item[c]DHF spinors obtained with uncontracted basis but truncated to the same number as the original scalar relativistic contraction.
\item[d]$4s^24p^5$[$3d^{10}4s^24p^5$].
\item[e]Percentage error compared to the experimental value of 3685 cm$^{-1}$\cite{NIST1}.
\item[f]Difference between contracted and uncontracted basis.
\item[g]Difference between atomic spinors and uncontracted basis.
	\end{tablenotes}
\end{threeparttable}
	\label{DCB_Basis_Set}

\end{table}%

The SOSs for the second- to fifth-row $p$-block atoms calculated with DC/DCB-4C-iCIPT2
and various basis sets are documented in Tables S1 to S4, respectively. They can be taken as reference data for other methods. 
%The mean absolute percentage errors (MAPE)
%from the experimental values\cite{NIST1} are further plotted in Fig. \ref{fig_cmpr_basis}.
Among others, we just mention that, compared to the DC Hamiltonian,
the Breit interaction tends to reduce the SOSs for the second- and third-row atoms (cf. Tables S1 and S2),
a well known tendency\cite{marian2012SOCISC}.
This can be understood from a theoretical argument.
Since the spin-dependent part of the gauge term of the Breit interaction
does not contribute to the energy at $\mathcal{O}(c^{-2})$\cite{MC-DPT1}, it is clear that
it is the spin-other-orbit interaction from the Gaunt-exchange term that is
responsible for such reduction of the SOSs resulting from the Coulomb interaction.
Since it amounts to twice the spin-same-orbit interaction from the Coulomb-exchange term at $\mathcal{O}(c^{-2})$\cite{X2CSOC2},
its importance for the SOSs of light elements can be expected from the outset. It is of particular interest
to see that the DC Hamiltonian even predicts
a wrong ordering for the $^2D_{5/2}$ and $^2D_{3/2}$ states of \ce{N}, in contrast to the DCB Hamiltonian (see Table S1).
The situation is somewhat different for
heavier atoms, where the Gaunt/Breit term contributes merely a few percents to the total SOSs,
thereby justifying an approximate treatment of them\cite{SOMF1}.

%\begin{figure}
%	\includegraphics[width=0.9\textwidth]{PertErrDistributionBasisSet.png}
%	\caption{Distribution of percentage errors of 4C-iCIPT2 for spin-orbit splittings of $p$-block atoms compared to experimental results\cite{NIST1} (see Tables S1--S4).
%The DCB and DC Hamiltonians were used for the second/third- and fourth/fifth-row atoms, respectively.
%		Probability density was estimated via kernel density estimation using Gaussian kernel.
%		%The $^4\text{S}_{3/2}\rightarrow$$^2\text{D}_{5/2}$/$^2\text{D}_{3/2}$ transitions of \ce{N}, \ce{P}, \ce{As} and \ce{Sb} were excluded.
%		%For unc-cc-pVDZ and unc-cc-pV5Z basis sets, the fifth-row atoms were excluded due to the lack of basis sets.
%		%The SOS of \ce{Te} were also excluded as the ordering of the states are still incorrect even with the largest basis set tested
%		Exclusions: $^4\text{S}_{3/2}\rightarrow$$^2\text{D}_{5/2}$/$^2\text{D}_{3/2}$ transitions of \ce{N}, \ce{P}, \ce{As}, \ce{Sb}; fifth-row atoms for unc-cc-pVDZ and unc-cc-pV5Z basis sets; \ce{Te} (due to incorrect ordering)
%	}
%	\label{fig_cmpr_basis}
%\end{figure}

\begin{figure}
	\includegraphics[width=0.9\textwidth]{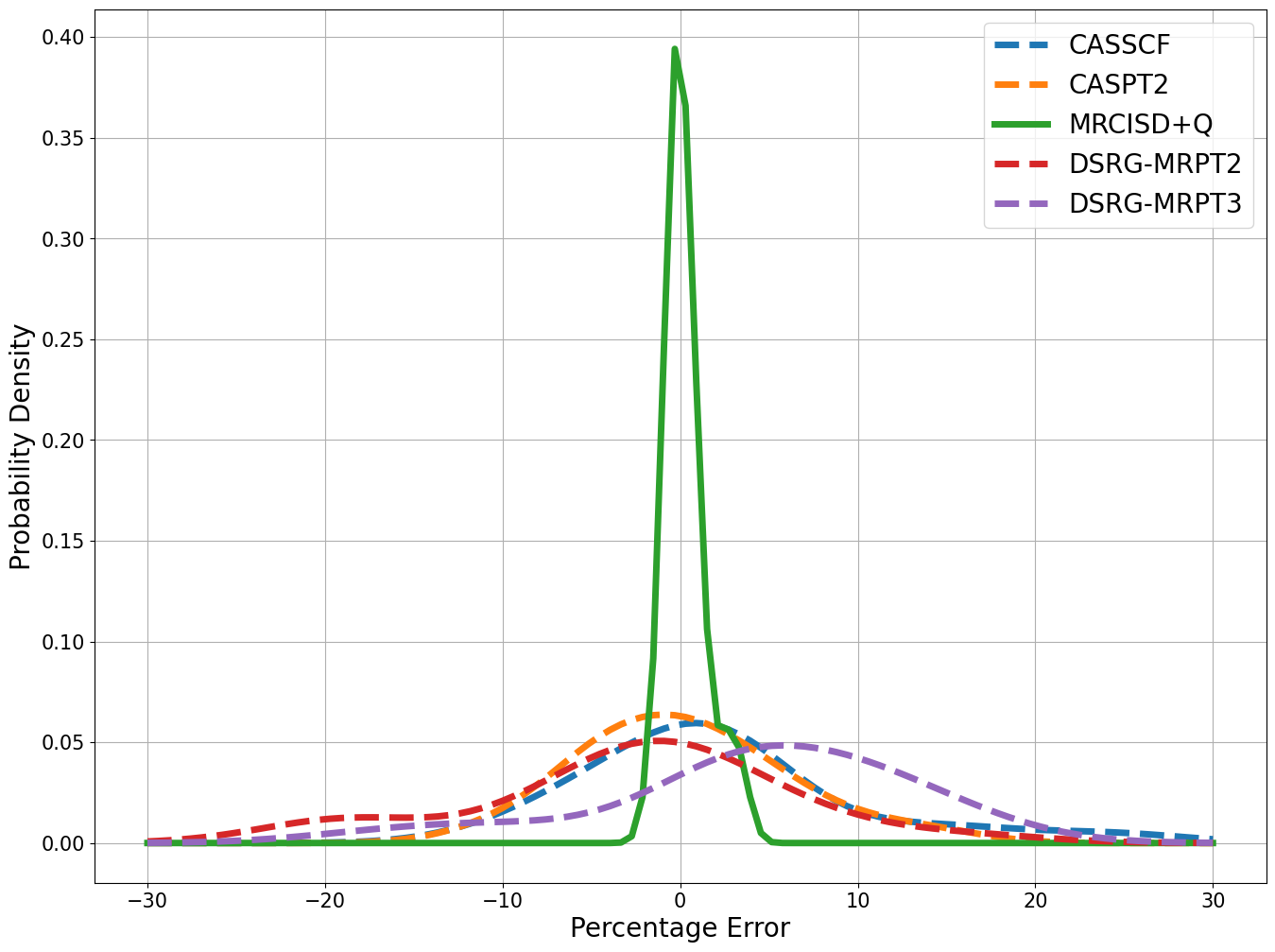}
	\caption{Distribution of percentage errors  of spin-orbit splittings of selected $p$-block atoms
with the DCB Hamiltonian and uncontracted cc-pVTZ basis sets compared to 4C-iCIPT2 results (see Table \ref{comparison}).
	Probability density was estimated via kernel density estimation using Gaussian kernel.
	The $^4\text{S}_{3/2}\rightarrow$$^2\text{D}_{5/2}$/$^2\text{D}_{3/2}$ transitions of \ce{N}, \ce{P}, and \ce{As} were excluded.
		}
	\label{fig_cmpr_ici}
\end{figure}

\begin{figure}
	\includegraphics[width=0.9\textwidth]{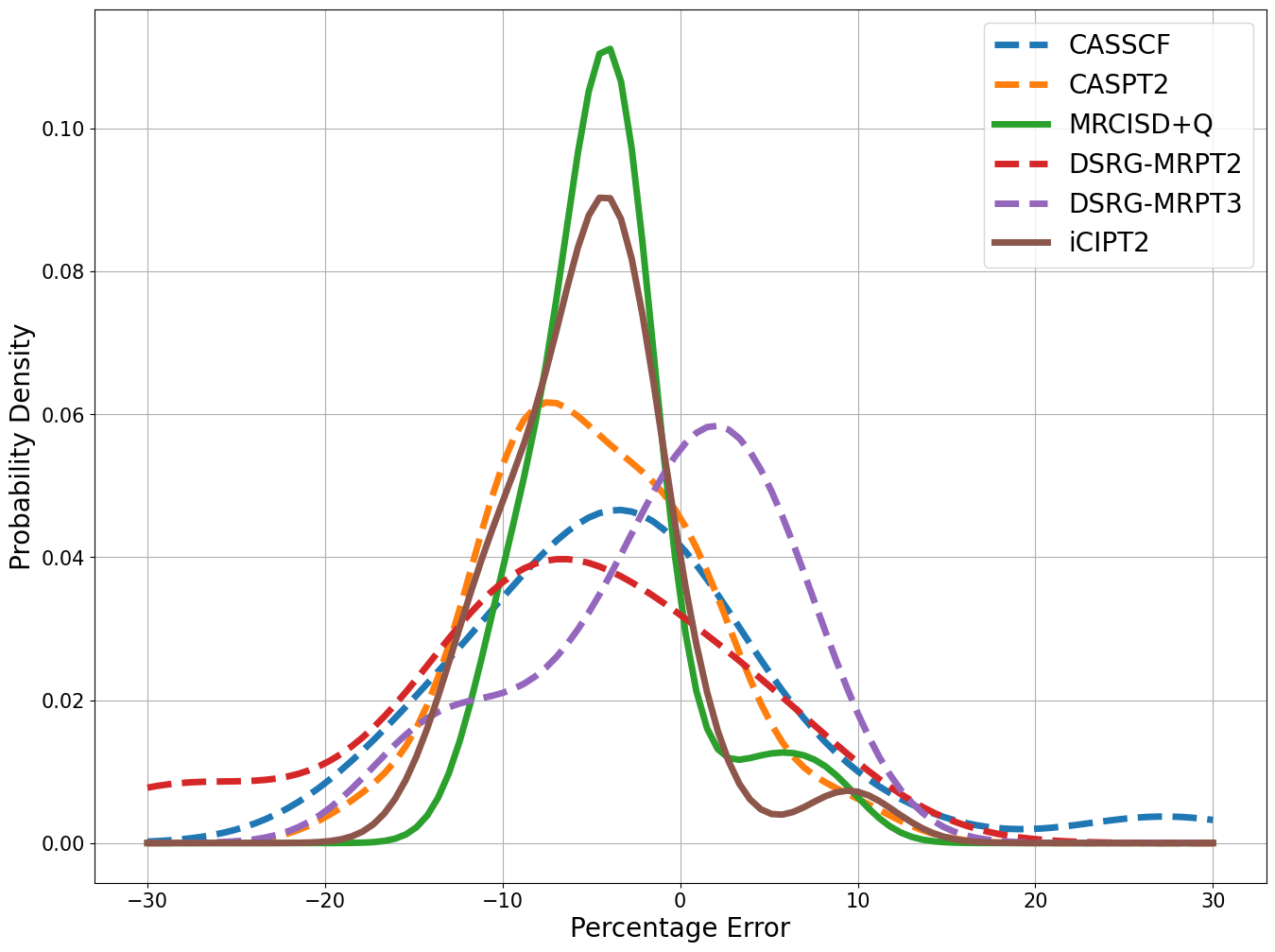}
	\caption{Distribution of percentage errors of spin-orbit splittings of selected $p$-block atoms
with the DCB Hamiltonian and uncontracted cc-pVTZ basis sets compared to experimental results\cite{NIST1} (see Table \ref{comparison}).
	Probability density was estimated via kernel density estimation using Gaussian kernel.
	The $^4\text{S}_{3/2}\rightarrow$$^2\text{D}_{5/2}$/$^2\text{D}_{3/2}$ transitions of \ce{N}, \ce{P}, and \ce{As} were excluded.
	}
\label{fig_cmpr_exp}
\end{figure}

In the present account, it is of more interest to make a close comparison between different methods. To this end, we used the same setup
(all-electron correlation with the DCB Hamiltonian and uncontracted cc-pVTZ basis sets\cite{Dunning1989,Dunning1993,Dunning1999})
as in the previous calculations on the selected $p$-block atoms (see Table \ref{comparison}),
including 4C-CASSCF, 4C-CASPT2, and 4C-MRCISD+Q (internally contracted multireference
configuration interaction with singles and doubles as well as Davidson corrections) in Ref. \citenum{4C-MR2018}, as well as
the driven similarity renormalization group based second- and third-order multireference perturbation theories
(DSRG-MRPT2/MRPT3) in Ref. \citenum{4C-DSRG-MRPT2-2024}. Since the present 4C-iCIPT2 is nearly exact,
it should be taken as benchmark to calibrate the other methods. As can be seen from Table \ref{comparison} and Fig. \ref{fig_cmpr_ici},
4C-MRCISD+Q is closest to 4C-iCIPT2. Yet, DSRG-MRPT3 appears to be closest to experiment (cf. Fig. \ref{fig_cmpr_exp}).
However, this should not be taken seriously, for the cc-pVTZ basis sets are not sufficiently large, even though uncontracted.
As a matter of fact, it was already noticed long ago\cite{RelaBasisCorr1} that some tight $p$-functions
have to be added to the cc-pVTZ basis sets for more accurate descriptions of the SOSs, especially of the second-row atoms.
We therefore repeated the calculations on \ce{B} to \ce{F} by using the tight $p$-functions suggested in Ref. \citenum{4C-DSRG-MRPT2-2024}.
As can be seen from Table \ref{augBasis}, the SOSs of \ce{B} to \ce{F} are indeed improved discernibly with such augmented basis sets. Still, however,
even larger basis sets are required to obtain converged results, e.g., the uncontracted cc-pV5Z basis sets\cite{Dunning1989} further reduce
the MAPE by more than 2\%.
As an additional indicator for the request on new basis sets for spinor-based relativistic calculations, we mention that
the ordering of the $^3\text{P}_2$, $^3\text{P}_0$, and $^3\text{P}_1$ states of \ce{Te} could not be reproduced correctly
with all the considered basis sets, except for the fortuitous case of valance-only correlation with
the scalar relativistically contracted ANO-RCC basis set\cite{ANO2} (see Table S4).

%The situation worsens for fourth-row elements, where the absolute error is typically 50 to 100 cm$^{-1}$.
%As shown in the next section, increasing the basis set to uncontracted \textcolor[rgb]{1.00,0.00,0.00}{cc-pV5Z-DK} cannot improve the results significantly.

\begin{table}[htbp]
	\tiny
	\centering
	\caption{Comparison between different methods for the spin-orbit splittings (in cm$^{-1}$) of selected $p$-block atoms
		with the DCB Hamiltonian and uncontracted cc-pVTZ basis sets\cite{Dunning1989,Dunning1993, Dunning1999, Dixon2001}.
		All electrons were correlated for second-row atoms.
		Only valence electrons were correlated for third- and fourth-row atoms}
	\begin{threeparttable}
		\begin{tabular}{clrrrrrrr}
			\toprule
			atom
			&  splitting
			& \multicolumn{1}{c}{Expt.} & \multicolumn{1}{l}{CASSCF} & \multicolumn{1}{l}{CASPT2} & \multicolumn{1}{c}{MRCISD+Q} & \multicolumn{1}{c}{DSRG-MRPT2} & \multicolumn{1}{c}{DSRG-MRPT3} & \multicolumn{1}{c}{4C-iCIPT2} \\
			\midrule
			B
			& $^2\text{P}_{1/2}\rightarrow$$^2\text{P}_{3/2}$
			& 15.29  & 13.25  & 13.99  & 13.91  & 13.99  & 14.32  & 13.97  \\
			\multirow{2}[0]{*}{C}
			& $^3\text{P}_{0}\rightarrow$$^3\text{P}_{1}$
			& 16.42  & 14.94  & 14.93  & 15.40  & 14.93  & 17.20  & 15.61  \\
			& $^3\text{P}_{1}\rightarrow$$^3\text{P}_{2}$
			& 27.00  & 23.82  & 23.95  & 24.88  & 28.48  & 28.72  & 24.61  \\
			\multirow{2}[0]{*}{N}
			& $^4\text{S}_{3/2}\rightarrow$$^2\text{D}_{5/2}$
			& 19224.46  & 22910.28  & 20414.63  & 20095.77\tnote{a}  & 20231.34  & 19952.54  & 20183.94\tnote{a}  \\
			& $^2\text{D}_{5/2}\rightarrow$$^2\text{D}_{3/2}$
			& 8.71  & 11.09  & 9.47  & 9.40\tnote{a}  & 9.41  & 7.89  & 9.54\tnote{a}  \\
			\multirow{2}[0]{*}{O}
			& $^3\text{P}_{2}\rightarrow$$^3\text{P}_{1}$
			& 158.27  & 153.24  & 145.35  & 152.52  & 145.35  & 138.92  & 152.08  \\
			& $^3\text{P}_{1}\rightarrow$$^3\text{P}_{0}$
			& 68.71  & 66.99  & 62.55  & 65.85  & 54.49  & 58.32  & 66.12  \\
			F
			& $^2\text{P}_{3/2}\rightarrow$$^2\text{P}_{1/2}$
			& 404.14  & 382.59  & 384.70  & 388.38  & 384.70  & 391.68  & 387.66  \\
			\midrule
			Al
			& $^2\text{P}_{1/2}\rightarrow$$^2\text{P}_{3/2}$
			& 112.06  & 96.81  & 106.60  & 106.96  & 106.70  & 108.35  & 106.43  \\
			\multirow{2}[0]{*}{Si}
			& $^3\text{P}_{0}\rightarrow$$^3\text{P}_{1}$
			& 77.12  & 72.89  & 78.90  & 73.76  & 69.94  & 81.52  & 73.80  \\
			& $^3\text{P}_{1}\rightarrow$$^3\text{P}_{2}$
			& 146.04  & 138.28  & 133.25  & 140.43  & 149.20  & 154.03  & 140.35  \\
			\multirow{2}[0]{*}{P}
			& $^4\text{S}_{3/2}\rightarrow$$^2\text{D}_{3/2}$
			& 11361.02  & 15305.18  & 12561.03  & \multicolumn{1}{c}{N/A} & 13302.41  & 12666.93  & 12308.90  \\
			& $^2\text{D}_{3/2}\rightarrow$$^2\text{D}_{5/2}$
			& 15.61  & 15.34  & 14.01  & \multicolumn{1}{c}{N/A} & 11.08  & 13.38  & 13.62  \\
			\multirow{2}[0]{*}{S}
			& $^3\text{P}_{2}\rightarrow$$^3\text{P}_{1}$
			& 396.06  & 398.64  & 386.02  & 383.94  & 355.94  & 400.61  & 386.11  \\
			& $^3\text{P}_{1}\rightarrow$$^3\text{P}_{0}$
			& 177.59  & 180.62  & 161.92  & 173.46  & 175.36  & 181.66  & 174.36  \\
			Cl
			& $^2\text{P}_{3/2}\rightarrow$$^2\text{P}_{1/2}$
			& 882.35  & 886.86  & 894.86  & 861.80  & 867.69  & 888.44  & 864.17  \\
			\midrule
			Ga
			& $^2\text{P}_{1/2}\rightarrow$$^2\text{P}_{3/2}$
			& 826.19  & 685.92  & 776.51  & 745.97  & 743.28  & 791.62  & 739.16  \\
			\multirow{2}[0]{*}{Ge}
			& $^3\text{P}_{0}\rightarrow$$^3\text{P}_{1}$
			& 557.13  & 512.35  & 553.07  & 502.94\tnote{b}  & 485.56  & 570.25  & 504.81  \\
			& $^3\text{P}_{1}\rightarrow$$^3\text{P}_{2}$
			& 852.83  & 809.65  & 711.39  & 814.20\tnote{b}  & 857.09  & 878.13  & 795.11  \\
			\multirow{2}[0]{*}{As}
			& $^4\text{S}_{3/2}\rightarrow$$^2\text{D}_{3/2}$
			& 10592.50  & 14289.54  & 11799.71  & \multicolumn{1}{c}{N/A} & 12567.66  & 12034.45  & 11444.25  \\
			& $^2\text{D}_{3/2}\rightarrow$$^2\text{D}_{5/2}$
			& 322.10  & 354.53  & 324.93  & \multicolumn{1}{c}{N/A} & 227.88  & 327.81  & 285.05  \\
			\multirow{2}[0]{*}{Se}
			& $^3\text{P}_{2}\rightarrow$$^3\text{P}_{1}$
			& 1989.50  & 1949.63  & 1917.35  & 1900.34\tnote{b}  & 1745.74  & 1991.36  & 1886.09  \\
			& $^3\text{P}_{1}\rightarrow$$^3\text{P}_{0}$
			& 544.86  & 572.88  & 532.84  & 566.74\tnote{b}  & 573.81  & 590.89  & 548.63  \\
			Br
			& $^2\text{P}_{3/2}\rightarrow$$^2\text{P}_{1/2}$
			& 3685.24  & 3683.62  & 3704.50  & 3540.14  & 3546.46  & 3683.90  & 3548.41  \\ \midrule
			& MAE\tnote{c}   & - & 19.29 & 19.35 &26.82\tnote{f} &37.77 & 9.88 &26.51 \\
			& MAPE\tnote{d}  & - & 7.09  &  6.15 & 5.24\tnote{f} & 9.38 & 5.24 & 5.90 \\\midrule
			& MAPE\tnote{e}  & - & 5.36  &  4.39 & 0.76\tnote{f} & 6.09 & 7.87 & -\\
			\bottomrule
		\end{tabular}%
		\begin{tablenotes}
			\item [a] Two core spinors ($1s_{1/2}$) were frozen for nitrogen. Otherwise, the symmetry breaking, small though, renders the
state assignment difficult.
			\item [b] Uncontracted cc-pVDZ basis set.
			\item [c] Mean absolute error compared to experimental values\cite{NIST1}
			($^4\text{S}_{3/2}\rightarrow$$^2\text{D}_{5/2}$/$^2\text{D}_{3/2}$ of \ce{N}, \ce{P}, and \ce{As} were excluded).
			\item [d] Mean absolute percentage error compared to experimental values\cite{NIST1}
			($^4\text{S}_{3/2}\rightarrow$$^2\text{D}_{5/2}$/$^2\text{D}_{3/2}$ of \ce{N}, \ce{P}, and \ce{As} were excluded).
			\item [e] Mean absolute percentage error compared to 4C-iCIPT2 results
			($^4\text{S}_{3/2}\rightarrow$$^2\text{D}_{5/2}$/$^2\text{D}_{3/2}$ of \ce{N}, \ce{P}, and \ce{As} were excluded).
			\item [f] Unavailable data were omitted from the averaging.
		\end{tablenotes}
	\end{threeparttable}
	\label{comparison}
\end{table}%

\begin{table}[htbp]
	\tiny
	\centering
	\caption{Spin-orbit splittings (in cm$^{-1}$) and percentage error for the second-row elements with the original (orig.) uncontracted cc-pVTZ basis
		and the augmented (aug.) sets with a tight $p$-function (68.8243, 106.9535, 152.2273, 196.9866, and 250.83491 for \ce{B}, \ce{C}, \ce{N}, \ce{O}, and \ce{F}, respectively.
		).
	}
	\begin{threeparttable}
		\begin{tabular}{clrrrrrrrr} \toprule
			\multicolumn{1}{c}{\multirow{2}[0]{*}{Atom}}
			& \multicolumn{1}{c}{\multirow{2}[0]{*}{Transition}}
			& \multicolumn{2}{c}{DSRG-MRPT2} & \multicolumn{2}{c}{DSRG-MRPT3} & \multicolumn{3}{c}{4C-iCIPT2}
			& \multicolumn{1}{c}{\multirow{2}[0]{*}{Exp.}}\\\cline{3-9}
			&       & \multicolumn{1}{c}{orig.} & \multicolumn{1}{c}{aug.} & \multicolumn{1}{c}{orig.} & \multicolumn{1}{c}{aug.} & \multicolumn{1}{c}{orig.} & \multicolumn{1}{c}{aug.} & \multicolumn{1}{c}{5Z orig.\tnote{a}} \\ \toprule
			B
			& $^2\text{P}_{1/2}\rightarrow$$^2\text{P}_{3/2}$
			& 13.99 (-8.5) & 14.81 (-3.1) & 14.32 (-6.3) & 14.88 (-2.7) & 13.97 (-8.6)  & 14.43 (-5.6) & 14.96 (-2.1) &15.29  \\
			\multirow{2}[0]{*}{C}
			& $^3\text{P}_{0}\rightarrow$$^3\text{P}_{1}$
			& 14.93 (-9.1) & 17.51 (6.6) & 17.20 (4.8)  & 17.75 (8.1) & 15.61 (-4.9)  & 16.68 (1.6) & 16.12 (-1.8)  &16.42  \\
			& $^3\text{P}_{1}\rightarrow$$^3\text{P}_{2}$
			& 28.48 (5.5) & 29.56 (9.5) & 28.72 (6.4) & 29.81 (10.4) & 24.61 (-8.9)  & 24.67 (-8.6) & 26.58 (-1.6) &27.00  \\
			\multirow{2}[0]{*}{N\tnote{b}}
			& $^4\text{S}_{3/2}\rightarrow$$^2\text{D}_{5/2}$
			& 20231.34 (5.2) & 20235.14 (5.3) & 19952.54 (3.8) & 19956.74 (3.8) & 20183.94 (5.0)  & 20187.97 (5.0) & 19356.79 (0.7) & 19224.46 \\
			& $^2\text{D}_{5/2}\rightarrow$$^2\text{D}_{3/2}$
			& 9.41 (8.0)  & 8.02 (-7.9)  & 7.89 (-9.4) & 7.73 (-11.3)  & 9.54 (9.5)  & 9.44 (8.4) & 8.83 (1.4) & 8.71 \\
			\multirow{2}[0]{*}{O}
			& $^3\text{P}_{2}\rightarrow$$^3\text{P}_{1}$
			& 145.35 (-8.2) & 133.37 (-15.7) & 138.92 (-12.2) & 142.57 (-9.9) & 152.08 (-3.9)  & 156.00 (-1.4)  & 153.69 (2.9) &158.27 \\
			& $^3\text{P}_{1}\rightarrow$$^3\text{P}_{0}$
			& 54.49 (-20.7) & 56.12 (-18.3) & 58.32 (-15.1) & 60.07 (-12.6) & 66.12 (-3.8)  & 68.25 (-0.7) & 68.70 (0.0) &68.71   \\
			F
			& $^2\text{P}_{3/2}\rightarrow$$^2\text{P}_{1/2}$
			& 384.70 (-4.8) & 391.15 (-3.2) & 391.68 (-3.1) & 403.10 (-0.4s) & 387.66 (-4.1)  & 398.81 (-1.3) & 399.73 (-1.1) &404.14   \\\midrule
			MAX\tnote{c} && 20.7 & 18.3&15.1 &12.6 & 9.5 & 8.6& 2.9 & \\
			MAPE\tnote{d}&& 9.3 & 9.2 & 8.2 & 7.9 & 6.2 & 4.0 & 1.6 &\\
			\bottomrule
		\end{tabular}%
		\begin{tablenotes}
\item [a] Original uncontracted cc-pV5Z (see also Table S1).
\item [b] Two core spinors ($1s_{1/2}$) were frozen in 4C-iCIPT2 calculations of nitrogen. Otherwise, the symmetry breaking, small though, renders the
state assignment difficult.
\item [c] Maximal percentage error compared to experimental values\cite{NIST1} ($^4\text{S}_{3/2}\rightarrow$$^2\text{D}_{5/2}$ of \ce{N} was excluded).
\item [d] Mean absolute percentage error
		compared to experimental values\cite{NIST1} ($^4\text{S}_{3/2}\rightarrow$$^2\text{D}_{5/2}$ of \ce{N} was excluded).
		\end{tablenotes}
	\end{threeparttable}\label{augBasis}
\end{table}%

\section{CONCLUSIONS and OUTLOOK}\label{Conclusion}
It has been shown that relativistic (spin-dependent) wave function methods can straightforwardly be implemented on top of the templates used for their
nonrelativistic (spin-free) counterparts by leveraging the power of meta-programming furnished by C++,
provided that the same diagrammatical representation is employed for the spin-dependent and spin-free Hamiltonians.
For the case of 4C-iCIPT2 considered here, once the code for constructing the Hamiltonian matrix is made ready,
all the rest can be generated automatically from the existing templates used for the spin-free iCIPT2. Moreover,
both time reversal and binary double point group symmetries can readily be incorporated, so as to facilitate
not only the computation but also the state assignment. Although the $p$-block atoms taken here as showcases are
not sufficient to reveal the real efficacy of the near-exact 4C-iCIPT2, it is undoubted that 4C-iCIPT2 is applicable to
more complex systems containing heavy atoms. Noticeably, the transformation of 4C two-electron integrals from the atomic to
molecular orbital representation is awfully expensive and memory intensive.
This can be surmounted by going to the picture change-free
Q4C Hamiltonian\cite{Q4C,Q4CX2C}, where the integral transformation can be performed as a whole
instead of component-wise as in the 4C case\cite{UnifiedH}.
Just like that iCISCF\cite{iCISCF} results from the use of iCIPT2
as the CI solver of CASSCF and can handle ca. 60 active orbitals, 4C-iCISCF for large active spaces
can readily be formulated by taking 4C-iCIPT2 as the CI solver of 4C-CASSCF. Guess orbials can here be provided
by the recently proposed Kramers restricted open-shell DHF\cite{UnifiedH}.
 Work along these directions is being carried out at our laboratory.

\section*{Acknowledgments}
This work was supported by the National Natural Science Foundation of China (Grant No. 22373057) and
Mount Tai Scholar Climbing Project of Shandong Province.

\section*{Data Availability Statement}
The data that support the findings of this study are available within the article.

\section*{Conflicts of interest}
There are no conflicts to declare.

\iffalse
\subsubsection{$p\neq q\neq r\neq s, p<r, q<s$}
\begin{equation}
	\begin{split}
		& (pq|rs) a_p^{\dagger} a_r^{\dagger} a_s a_q +
		(\bar{p}\bar{q}|\bar{r}\bar{s}) a_{\bar{p}}^{\dagger} a_{\bar{r}}^{\dagger} a_{\bar{s}} a_{\bar{q}} + \\
		& (\bar{p}q|rs) a_{\bar{p}}^{\dagger} a_r^{\dagger} a_s a_q +
		(p\bar{q}|rs) a_p^{\dagger}  a_r^{\dagger} a_s a_{\bar{q}} +
		(pq|\bar{r}s) a_p^{\dagger}  a_{\bar{r}}^{\dagger} a_s a_q +
		(pq|r\bar{s}) a_p^{\dagger} a_r^{\dagger} a_{\bar{s}} a_q + \\
		& (\bar{p}\bar{q}|rs) a_{\bar{p}}^{\dagger} a_r^{\dagger} a_s a_{\bar{q}} +
		(\bar{p}q|\bar{r}s) a_{\bar{p}}^{\dagger} a_{\bar{r}}^{\dagger} a_s a_q +
		(\bar{p}q|r\bar{s}) a_{\bar{p}}^{\dagger} a_r^{\dagger} a_{\bar{s}} a_q +
		(p\bar{q}|\bar{r}s) a_p^{\dagger} a_{\bar{r}}^{\dagger} a_s a_{\bar{q}} + \\
		& (p\bar{q}|r\bar{s}) a_p^{\dagger} a_{r}^{\dagger} a_{\bar{s}} a_{\bar{q}} +
		(pq|\bar{r}\bar{s}) a_p^{\dagger} a_{\bar{r}}^{\dagger} a_{\bar{s}} a_{q} + \\
		& (p\bar{q}|\bar{r}\bar{s}) a_p^{\dagger} a_{\bar{r}}^{\dagger} a_{\bar{s}} a_{\bar{q}} +
		(\bar{p}q|\bar{r}\bar{s}) a_{\bar{p}}^{\dagger} a_{\bar{r}}^{\dagger} a_{\bar{s}} a_{q}  +
		(\bar{p}\bar{q}|r\bar{s}) a_{\bar{p}}^{\dagger} a_r^{\dagger} a_{\bar{s}} a_{\bar{q}} +
		(\bar{p}\bar{q}|\bar{r}s) a_{\bar{p}}^{\dagger} a_{\bar{r}}^{\dagger} a_s a_{\bar{q}} \\	
	\end{split}
\end{equation}
\fi

\newpage

\appendix

\section{Symmetry in Molecular Integrals}\label{SecSymmERI}
\subsection{Time reversal and permutation}\label{StorageIntPermutation}
Apart from the four-fold permutation symmetries
\begin{align}
	(pq|rs)&=(rs|pq)=(qp|sr)^*=(sr|qp)^*,\quad p,q,r,s\in A,B,\label{permC}\\
	(p\balpha q|r\balpha s)&=(r\balpha s|p\balpha q)=(q\balpha p|s\balpha r)^*=(s\balpha r|q\balpha p)^*,\quad p,q,r,s\in A,B, \label{permB}
\end{align}
the Coulomb and Gaunt/Breit types of ERIs also have the following time reversal-based relations
\begin{align}
	(pq|rs)& = (pq|\bar{s}\bar{r})=(\bar{q}\bar{p}|rs) =(\bar{q}\bar{p}|\bar{s}\bar{r}),\quad p,q,r,s\in A; \bar{p},\bar{q},\bar{r}, \bar{s}\in B,\nonumber\\
	(pq|r\bar{s})& =-(pq|s\bar{r})=(\bar{q}\bar{p}|r\bar{s})= -(\bar{q}\bar{p}|s\bar{r}),\nonumber\\
	(p\bar{q}|r\bar{s})& =-(p\bar{q}|s\bar{r})= -(q\bar{p}|r\bar{s})= (q\bar{p}|s\bar{r}),\nonumber\\
	(\bar{p}q|r\bar{s})& =-(\bar{p}q|s\bar{r})= -(\bar{q}p|r\bar{s})= (\bar{q}p|s\bar{r}),\label{TR-C-ERI}\\
	(p \boldsymbol{\alpha} q   | r \boldsymbol{\alpha} s)
	&=-(p \boldsymbol{\alpha}q | \bar{s} \boldsymbol{\alpha}\bar{r})
	=-(\bar{q}\boldsymbol{\alpha} \bar{p} | r\boldsymbol{\alpha} s)
	=(\bar{q}\boldsymbol{\alpha} \bar{p}  | \bar{s}\boldsymbol{\alpha} \bar{r}),\nonumber\\
	(p \boldsymbol{\alpha}q   | r\boldsymbol{\alpha} \bar{s})
	&=(p\boldsymbol{\alpha} q | s\boldsymbol{\alpha} \bar{r})
	=-(\bar{q}\boldsymbol{\alpha} \bar{p}  | r \boldsymbol{\alpha} \bar{s})
	= -(\bar{q}\boldsymbol{\alpha} \bar{p} | s\boldsymbol{\alpha} \bar{r}),\nonumber\\
	(p\boldsymbol{\alpha} \bar{q}   | r\boldsymbol{\alpha} \bar{s})
	&=(p\boldsymbol{\alpha} \bar{q} | s\boldsymbol{\alpha} \bar{r})
	=(q\boldsymbol{\alpha} \bar{p} | r\boldsymbol{\alpha} \bar{s})
	=(q\boldsymbol{\alpha} \bar{p} | s\boldsymbol{\alpha} \bar{r}),\nonumber\\
	(\bar{p}\boldsymbol{\alpha} q   | r\boldsymbol{\alpha} \bar{s})
	&=(\bar{p}\boldsymbol{\alpha} q | s\boldsymbol{\alpha} \bar{r})
	=(\bar{q}\boldsymbol{\alpha} p | r\boldsymbol{\alpha} \bar{s})
	=(\bar{q}\boldsymbol{\alpha} p | s\boldsymbol{\alpha} \bar{r}),\label{TR-B-ERI}
\end{align}	
thanks to the following identities
\begin{align}
	\psi_p^\dag\psi_q&=\psi_{\bar{q}}^\dag\psi_{\bar{p}},\quad \psi_{\bar{p}}^\dag\psi_{q}=-\psi_{\bar{q}}^\dag\psi_{p},\quad \psi_{p}^\dag\psi_{\bar{q}}=-\psi_{q}^\dag\psi_{\bar{p}},\\
	\psi_p^\dag\balpha\psi_q&=-\psi_{\bar{q}}^\dag\balpha\psi_{\bar{p}},\quad \psi_{\bar{p}}^\dag\balpha\psi_{q}=\psi_{\bar{q}}^\dag\balpha\psi_{p},\quad \psi_{p}^\dag\balpha\psi_{\bar{q}}=\psi_{q}^\dag\balpha\psi_{\bar{p}}.
\end{align}
Eq. \eqref{TR-C-ERI}/\eqref{TR-B-ERI} allows one to classify the 16-fold ERIs $(pq|rs)_{p,q,r,s\in A, B}$/$(p\balpha q|r\balpha s)_{p,q,r,s\in A, B}$
into four unique groups,
\begin{enumerate}[(1)]
\item $G_{AAAA}=\{(AA|AA), (AA|BB), (BB|AA), (BB|BB)\}$,
\item $G_{ABBA}=\{(AB|BA), (BA|AB)\}$,
\item $G_{AAAB}=\{(AA|AB), (BA|AA), (AB|AA), (AA|BA), (AB|BB), (BA|BB), (AB|BB), (BB|BA)\}$,
\item $G_{ABAB}=\{(AB|AB), (BA|BA)\}$.
\end{enumerate}
Within each group, the other types of ERIs can all be transformed to the first one .
Therefore, only the four generic $(AA|AA)$, $(AB|BA)$, $(AA|AB)$, and $(AB|AB)$ types of ERIs need to be stored.
This already gives rise to a four-times reduction of the store, as compared to the storage of the whole 16-fold ERIs.
For the $(AA|AA)$ type of ERIs $(pq|rs)$, the four-fold permutation
 (i.e., $p\geq r$ and $p\geq q$), which covers all the four cases in $G_{AAAA}$, gives rise to a further four-times reduction and hence an overall 16-times
reduction of the storage.
For the $(AB|BA)$ type of ERIs $(p\bar{q}|\bar{r}s)$ [$=(\bar{r}s|p\bar{q})]$, which
have a two-fold permutation (i.e., $p\ge r$) and are anti-symmetric with respect to both the $p,q$ and $r,s$ pairs,
an eight-times reduction of the storage can be achieved. Further considering the fact that
the group $G_{ABBA}$ has a two-in-one feature, an overall sixteen-times reduction of the storage can be achieved.
%For the $(AB|AB)$ type of ERIs $(p\bar{q}|r\bar{s})$ [$=(r\bar{s}|p\bar{q})]$, which
%has a two-fold permutation (i.e., $p\ge r$) and are anti-symmetric with respect to both the $p,q$ and $r,s$ pairs,
%an eight-times reduction of the storage can be achieved. Further considering the fact that
%the group $G_{ABAB}$ has a two-in-one feature, an overall sixteen-times reduction of the storage can be achieved.
The type $(AB|AB)$ of ERIs behaves the same as $(AB|BA)$.
As for the $(AA|AB)$ type of ERIs $(pq|r\bar{s})$, which have no permutation symmetries but are anti-symmetric
with respect to the $r,s$ pair, a two-times reduction of the storage can be achieved. Further considering the fact that
the group $G_{AAAB}$ has an eight-in-one feature, again an overall sixteen-times reduction of the storage can be achieved.
Therefore, an overall sixteen-times reduction of the storage of the Coulomb/Breit ERIs
can be achieved by making use only of time reversal and permutation symmetries.

As noted before\cite{RQC_BOOK_Dyall}, both the $(AA|AA)$ and $(AB|BA)$ types of ERIs appear in the main diagonal blocks,
whereas the $(AA|AB)$ and $(AB|AB)$ types of ERIs enter the first and second off-diagonal blocks, respectively,
of the block pentadiagonal Hamiltonian matrix over SDs that are ordered
in a descent order of their $M_K$ values. It will be shown in Sec. \ref{doublegroup1}
that the $(AA|AB)$ type of ERIs vanishes for binary double point groups other than $C_1^\ast$ and $C_i^\ast$,
and that all the ERIs are real-valued for the $C_{2v}^\ast$, $D_2^\ast$, and $D_{2h}^\ast$ groups.

\subsection{Double Group Symmetry}\label{doublegroup1}
Suppose a spinor $\psi$ and its time-reversed partner $\bar{\psi}$ transform as the basis of an irreducible representation (irrep)
of a binary double point group ($D_{2h}^{\ast}$ and its subgroups), which can be ascribed to one of
three Frobenius–Schur classes: (a) $C_1^\ast$ and $C_i^\ast$, where $\psi$ and $\bar{\psi}$ belong to the same, non-degenerate fermion irrep;
(b) $C_2^\ast$, $C_s^\ast$, and $C_{2h}^\ast$, where $\psi$ and $\bar{\psi}$ belong to different, non-degenerate fermion irreps (which together form a doubly
degenerate fermion representation); (c) $C_{2v}^\ast$, $D_2^\ast$, and $D_{2h}^\ast$, where $\psi$ and $\bar{\psi}$ belong to the first and second columns of a doubly
degenerate irrep. Such Kramers paired symmetry functions can readily be constructed (even for any double point group)\cite{Symm2009}, so as to
render the matrices of Hermitian operators well structured (quaternion, complex, or real for the three Frobenius–Schur classes,
respectively). However, the point of focus here is to uncover the distribution of boson irreps of binary double point groups among
the individual components of the already symmetrized spinor $\psi$,
\begin{equation}
	\psi=\left(\begin{array}{l}
		\psi^{L \alpha} \\
		\psi^{L \beta} \\
		\psi^{S \alpha} \\
		\psi^{S \beta}
	\end{array}\right)
	=\left(\begin{array}{l}
		\operatorname{Re}\left(\psi^{L \alpha}\right)+\ii \operatorname{Im}\left(\psi^{L \alpha}\right) \\
		\operatorname{Re}\left(\psi^{L \beta}\right) +\ii \operatorname{Im}\left(\psi^{L \beta}\right) \\
		\operatorname{Re}\left(\psi^{S \alpha}\right)+\ii \operatorname{Im}\left(\psi^{S \alpha}\right) \\
		\operatorname{Re}\left(\psi^{S \beta}\right) +\ii \operatorname{Im}\left(\psi^{S \beta}\right) \\
	\end{array}\right).\label{8-scalar}
\end{equation}
To this end, we rewrite the one-electron Dirac equation as
\begin{align}
	\psi^L&=\frac{c}{E-V}
	(\boldsymbol{\sigma} \cdot \mathbf{p}) B(E) (\boldsymbol{\sigma} \cdot \mathbf{p}) \psi^L, \label{LS_Separationa}  \\
	\psi^S & =B(E) (\boldsymbol{\sigma} \cdot \mathbf{p}) \psi^L, \label{LS_Separationb}\\
	B(E)&=\frac{c}{2c^2+E-V},
\end{align}
where the potential $V$ and hence the operator $B(E)$ are totally symmetric. The
$(\boldsymbol{\sigma} \cdot \mathbf{p}) B(E) (\boldsymbol{\sigma} \cdot \mathbf{p})$ operator can further be written in block form
\begin{align}
	(\boldsymbol{\sigma} \cdot \mathbf{p}) B(E) (\boldsymbol{\sigma} \cdot \mathbf{p})&=\left(\begin{array}{cc}
		\mathbf{p}B(E)\cdot\mathbf{p}+\ii L_z & L_y +\ii L_x\\
		-L_y+\ii L_x & \mathbf{p}B(E)\cdot\mathbf{p} - \ii L_z
	\end{array}\right),\label{pBepOp}\\
	L_i&=(\mathbf{p}B(E)\times \mathbf{p})_i,\quad i=x, y, z,
\end{align}
where $\mathbf{p}B(E)\cdot\mathbf{p}$ transforms as $r^2$ and hence spans the totally symmetric irrep $\Gamma_0$,
whereas $L_i$ is the $i$-th component of the angular momentum and transforms as the rotation $R_i$, thereby leading to
\begin{equation}
	\Gamma\left((\boldsymbol{\sigma} \cdot \mathbf{p}) B(E) (\boldsymbol{\sigma} \cdot \mathbf{p})\right)=\left(\begin{array}{cc}
		\Gamma_0+\ii \Gamma\left(R_z\right) & \Gamma\left(R_y\right)+\ii \Gamma\left(R_x\right) \\
		\Gamma\left(R_y\right)+\ii \Gamma\left(R_x\right) & \Gamma_0+\ii \Gamma\left(R_z\right)
	\end{array}\right).\label{GammaBe}
\end{equation}
Note that the negative signs in Eq. \eqref{pBepOp} are immaterial for the symmetry issue.
To reveal the symmetry of Eq. \eqref{LS_Separationa},
we start with a spinor \eqref{8-scalar} with only $\operatorname{Re}\left(\psi^{L \alpha}\right)$ nonzero,
so as to obtain
\begin{align}
	\Gamma\left(\psi^L\right)
	&=\left(\begin{array}{c}
		\Gamma\left(\operatorname{Re}\psi^{L\alpha}\right)+\ii \Gamma\left(\operatorname{Im}\psi^{L\alpha}\right) \\
		\Gamma\left(\operatorname{Re}\psi^{L\beta}\right)+\ii \Gamma\left(\operatorname{Im}\psi^{L\beta}\right)
	\end{array}\right)\nonumber \\
	&=\left(\begin{array}{c}
		\Gamma_0+\ii \Gamma\left(R_z\right) \\
		\Gamma\left(R_y\right)+\ii \Gamma\left(R_x\right)
	\end{array}\right)\otimes\Gamma\left(\operatorname{Re}\psi^{L\alpha}\right).\label{Bosons-L}
\end{align}
Note that the explicit appearance of imaginary $\ii$ emphasizes the symmetry of the imaginary part of
$\psi^{L}$. By virtue of the following identities (cf. Table \ref{table_LS_connect})
\begin{align}
	&\Gamma\left(R_i\right)\otimes\Gamma\left(R_j\right)=\Gamma\left(R_k\right),\quad (i, j, k) = (x, y, z),\nonumber\\
	&\Gamma\left(R_i\right)\otimes\Gamma\left(R_i\right)=\Gamma_0,\quad i=x, y, z,\nonumber\\
	&\sum_{i=x,y,z} \Gamma(i) \otimes \Gamma\left(R_i\right) = \Gamma(\mathbf{r}\cdot\mathbf{R}),\label{SymmIdentity}
\end{align}
it can readily be checked that the repeated application of the operator \eqref{GammaBe} on $\Gamma\left(\psi^L\right)$ \eqref{Bosons-L}
does not change the result. Therefore, Eq. \eqref{Bosons-L} does reveal
the full internal symmetry structure of the large component $\psi^{L}$.

To reveal the internal symmetry structure
of the small component $\psi^{S}$, we invoke the symmetry content of the $\boldsymbol{\sigma} \cdot \mathbf{p}$ operator, viz.,
\begin{equation}
	\Gamma(\boldsymbol{\sigma} \cdot \mathbf{p})=\left(\begin{array}{cc}
		\ii \Gamma(z) & \Gamma(y)+\ii \Gamma(x) \\
		\Gamma(y)+\ii \Gamma(x) & \ii \Gamma(z)
	\end{array}\right).
\end{equation}
Multiplying $\Gamma(\boldsymbol{\sigma} \cdot \mathbf{p})$ from the left of Eq. \eqref{Bosons-L} gives rise to
\begin{align}
	\Gamma\left(\psi^S \right)&=\left(\begin{array}{cc}
		\ii \Gamma(z) & \Gamma(y)+\ii \Gamma(x) \\
		\Gamma(y)+\ii \Gamma(x) & \ii \Gamma(z)
	\end{array}\right) \left(\begin{array}{c}
		\Gamma_0+\ii \Gamma\left(R_z\right) \\
		\Gamma\left(R_y\right)+\ii \Gamma\left(R_x\right)
	\end{array}\right)\otimes\Gamma\left(\operatorname{Re}\psi^{L\alpha}\right)\nonumber\\
	&=\left(\begin{array}{cc} \Gamma(\mathbf{r}\cdot\mathbf{R})\left[\Gamma_0 + \ii \Gamma\left(R_z\right)\right] \\
		\Gamma(\mathbf{r}\cdot\mathbf{R})\left[\Gamma\left(R_y\right)+\ii \Gamma\left(R_x\right)\right]
	\end{array}\right)\otimes\Gamma\left(\operatorname{Re}\psi^{L\alpha}\right)\\
	&=\Gamma(\mathbf{r}\cdot\mathbf{R})\otimes\Gamma\left(\psi^L\right)\label{Bosons-S}
\end{align}
in view of Eq. \eqref{LS_Separationb} and
the identities in Eq. \eqref{SymmIdentity}.
It follows from Eqs. \eqref{Bosons-L} and \eqref{Bosons-S} that the boson irreps of the eight scalar functions in $\psi$ (cf. Eq. \eqref{8-scalar})
are not independent, but are determined\cite{Quaternion1999,RQC_BOOK_Dyall} solely by $\Gamma\left(\operatorname{Re}\psi^{L\alpha}\right)$, the irrep of
the real part of the $L\alpha$ component. In particular,
the small component $\psi^S$ differs from the large component $\psi^L$ only in parity (because $\mathbf{r}\cdot\mathbf{R}$
changes sign under spatial inversion).

Given the internal symmetry structure of $\psi$, that of the time-reversed partner $\bar{\psi}$ can be obtained
simply by interchanging the boson irreps of the $\alpha$ and $\beta$ components of $\psi^X$ ($X=L, S$), viz.,
\begin{align}
	\Gamma\left(\bar{\psi}^L\right)
	&=\left(\begin{array}{c}
		\Gamma\left(\operatorname{Re}\psi^{L\beta}\right)+\ii \Gamma\left(\operatorname{Im}\psi^{L\beta}\right)\\
		\Gamma\left(\operatorname{Re}\psi^{L\alpha}\right)+\ii \Gamma\left(\operatorname{Im}\psi^{L\alpha}\right)
	\end{array}\right)\nonumber\\
	&=\left(\begin{array}{c}
		\Gamma\left(R_y\right)+\ii \Gamma\left(R_x\right)\\
		\Gamma_0+\ii \Gamma\left(R_z\right) \\
	\end{array}\right)\otimes\Gamma\left(\operatorname{Re}\psi^{L\alpha}\right),\label{Bosons-Lbar}\\
	\Gamma\left(\bar{\psi}^S\right)
	&=\Gamma(\mathbf{r}\cdot\mathbf{R})\otimes \Gamma\left(\bar{\psi}^L\right).\label{Bosons-Sbar}
\end{align}

With the above relations, the binary boson irreps for the components of spinors $\psi$ and $\bar{\psi}$
(of a given Frobenius–Schur class) can readily be generated
when $\Gamma\left(\operatorname{Re}\psi^{L\alpha}\right)$ is chosen to be the totally symmetric irrep $\Gamma_0$,
see Tables \ref{spinor_sym_a}-\ref{spinor_sym_c}. However, the spatial inversion
has not yet been taken into account. As such,
the procedure should be repeated for $C_i^\ast$, $C_{2h}^\ast$, and $D_{2h}^\ast$
by setting $\Gamma\left(\operatorname{Re}\psi^{L\alpha}\right)$ to
the totally anti-symmetric irrep $A_u$, so as to complete the lists for the three groups.
As a matter of fact, $\Gamma\left(\operatorname{Re}\psi^{L\alpha}\right)$
can be set to any boson irrep $\Gamma_b$ (and its opposite-parity counterpart, if any), with the corresponding distributions of boson irreps
being those obtained by multiplying $\Gamma_b$ to the ones documented in Tables \ref{spinor_sym_a}-\ref{spinor_sym_c}.
Such information can be employed reversely for the construction of Kramers paired symmetry functions of binary double point groups\cite{Quaternion1999}.

\begin{table}[htbp!]
	\centering
	\caption{Symmetry of operators connecting the small and large components.}
	\begin{tabular}{cccc}\toprule
		double group &  $(\Gamma(x), \Gamma(y), \Gamma(z))$     &   $(\Gamma(R_x), \Gamma(R_y), \Gamma(R_z)) $    &  $\Gamma(\mathbf{r}\cdot\mathbf{R})$ \\\toprule
		$C_1^\ast$    & $(A,A,A)$         & $(A,A,A)$        & $A$   \\
		$C_i^\ast$    & $(A_u,A_u,A_u)$   & $(A_g,A_g,A_g)$  & $A_u$ \\
		$C_2^\ast$    & $(B,B,A)$         & $(B,B,A)$        & $A$   \\
		$C_s^\ast$    & $(A',A',A'')$     & $(A'',A'',A')$   & $A''$ \\
		$C_{2h}^\ast$ & $(B_u,B_u,A_u)$   & $(B_g,B_g,A_g)$  & $A_u$ \\
		$C_{2v}^\ast$ & $(B_1,B_2,A_1)$   & $(B_2,B_1,A_2)$  & $A_2$ \\
		$D_2^\ast$    & $(B_3,B_2,B_1)$   & $(B_3,B_2,B_1)$  & $A$ \\
		$D_{2h}^\ast$ & $(B_{3u},B_{2u},B_{1u})$ & $(B_{3g},B_{2g},B_{1g})$ & $A_u$  \\\bottomrule
	\end{tabular}%
	\label{table_LS_connect}
\end{table}%

\begin{table}[htbp]
	\centering
	\small
	\caption{Symmetry of each component of spinors $\psi$ and $\bar{\psi}$ for $C_1^\ast$ and $C_i^\ast$ (Frobenius-Schur class (a))}
	\begin{tabular}{cc|ccc}\toprule
		& $C_1^\ast$ & \multicolumn{2}{c}{$C_i^\ast$} \\\cline{2-4}
		& $\psi/\bar{\psi}\in A_{1/2}$ & $\psi_g/\bar{\psi}_g\in A_{1/2,g}$   & $\psi_u/\bar{\psi}_u\in A_{1/2,u}$            \\\toprule
		$\psi^{L\alpha}$	& $A+iA$            & $A_g+iA_g$ & $A_u+iA_u$ \\
		$\psi^{L\beta}$	& $A+iA$                & $A_g+iA_g$ & $A_u+iA_u$ \\
		$\psi^{S\alpha}$	& $A+iA$            & $A_u+iA_u$ & $A_g+iA_g$ \\
		$\psi^{S\beta}$	& $A+iA$                & $A_u+iA_u$ & $A_g+iA_g$ \\\bottomrule
	\end{tabular}%
	\label{spinor_sym_a}
\end{table}%

\begin{table}[htbp]
	\centering
	\small
	\caption{Symmetry of each component of spinors $\psi$ and $\bar{\psi}$ for $C_2^\ast$, $C_s^\ast$ and $C_{2h}^\ast$ (Frobenius-Schur class (b))}
	\begin{tabular}{ccc|cc|cccc}\toprule
		& \multicolumn{2}{c|}{$C_2^\ast$} & \multicolumn{2}{c|}{$C_s^\ast$} & \multicolumn{4}{c}{$C_{2h}^\ast$} \\\cline{2-9}
		&  $\psi\in$ $^{1}E_{1/2}$  & $\bar{\psi}\in$ $^2E_{1/2}$  &  $\psi\in$ $^1E_{1/2}$  & $\bar{\psi}\in$ $^2E_{1/2}$
		& $\psi_g\in$ $^1E_{1/2,g}$ &  $\bar{\psi}_g\in$ $^2E_{1/2,g}$     & $\psi_u\in$ $^1E_{1/2,u}$  & $\bar{\psi}_u\in$ $^2E_{1/2,u}$  \\\toprule
		$\psi^{L\alpha}$	&  $A+iA$     &  $B+iB$     &  $A'+iA'$     & $A''+iA''$   & $A_g+iA_g$  & $B_g+iB_g$  & $A_u+iA_u$     & $B_u+iB_u$ \\
		$\psi^{L\beta}$	    &  $B+iB$     &  $A+iA$     &  $A''+iA''$   & $A'+iA'$     & $B_g+iB_g$  & $A_g+iA_g$  & $B_u+iB_u$     & $A_u+iA_u$ \\
		$\psi^{S\alpha}$	&  $A+iA$     &  $B+iB$     &  $A''+iA''$   & $A'+iA'$     & $A_u+iA_u$  & $B_u+iB_u$  & $A_g+iA_g$     & $B_g+iB_g$ \\
		$\psi^{S\beta}$	    &  $B+iB$     &  $A+iA$     &  $A'+iA'$     & $A''+iA''$   & $B_u+iB_u$  & $A_u+iA_u$  & $B_g+iB_g$     & $A_g+iA_g$ \\\bottomrule
	\end{tabular}%
	\label{spinor_sym_b}
\end{table}%

\begin{table}[htbp]
	\centering
	\small
	\caption{Symmetry of each component of spinors $\psi$ and $\bar{\psi}$ for $C_{2v}^\ast$, $D_2^\ast$ and $D_{2h}^\ast$ (Frobenius-Schur class (c))}
	\begin{tabular}{cc|c|cccccc}\toprule
		& \multicolumn{1}{c|}{$C_{2v}^\ast$} & \multicolumn{1}{c|}{$D_2^\ast$}  & \multicolumn{2}{c}{$D_{2h}^\ast$} \\\cline{2-5}
		&  $(\psi,\bar{\psi})\in E_{1/2}$    &  $(\psi, \bar{\psi})\in E_{1/2}$ & $(\psi_g, \bar{\psi}_g)\in E_{1/2,g}$
		& $(\psi_u, \bar{\psi}_u)\in E_{1/2,u}$  \\\toprule
		$\psi^{L\alpha}$	&  $(A_1+iA_2, B_1+iB_2)$ &  $(A+iB_1,  B_2+iB_3)$  & $(A_g+iB_{1g}, B_{2g}+iB_{3g})$  & $(A_u+iB_{1u}, B_{2u}+iB_{3u})$ \\
		$\psi^{L\beta}$	    &  $(B_1+iB_2, A_1+iA_2)$ &  $(B_2+iB_3, A+iB_1) $  & $(B_{2g}+iB_{3g}, A_g+iB_{1g})$  & $(B_{2u}+iB_{3u}, A_u+iB_{1u})$ \\
		$\psi^{S\alpha}$	&  $(A_2+iA_1, B_2+iB_1)$ &  $(A+iB_1, B_2+iB_3)$   & $(A_u+iB_{1u}, B_{2u}+iB_{3u})$  & $(A_g+iB_{1g}, B_{2g}+iB_{3g})$ \\
		$\psi^{S\beta}$	    &  $(B_2+iB_1, A_2+iA_1)$ &  $(B_2+iB_3, A+iB_1)$   & $(B_{2u}+iB_{3u}, A_u+iB_{1u})$  & $(B_{2g}+iB_{3g}, A_g+iB_{1g})$ \\\bottomrule
	\end{tabular}%
	\label{spinor_sym_c}
\end{table}%

By virtue of the relations \eqref{Bosons-L}, \eqref{Bosons-S}, \eqref{Bosons-Lbar}, \eqref{Bosons-Sbar}, and \eqref{SymmIdentity},
the symmetries of the (scalar) overlap charge density and current density components
can readily be calculated as
\begin{align}
	\Gamma\left(\psi_p^{\dagger} \psi_q|_{p,q\in A, B}\right)&=
	\begin{cases}
		\left[\Gamma_0+\ii \Gamma\left(R_z\right)\right]\otimes\Gamma_{2LA}\quad  (\mbox{even A})\\
		\left[\Gamma\left(R_y\right)+\ii \Gamma\left(R_x\right)\right]\otimes\Gamma_{2LA}\quad  (\mbox{odd A})
	\end{cases},\label{pqpair}\\
	\Gamma\left(\psi_p^{\dagger}\alpha_x \psi_q|_{p,q\in A, B}\right)&=
	\begin{cases}
		\Gamma\left(\mathbf{r}\cdot\mathbf{R}\right)\otimes\left[\Gamma\left(R_y\right)+\ii \Gamma\left(R_x\right)\right]\otimes\Gamma_{2LA}\quad  (\mbox{even A})\\
		\Gamma\left(\mathbf{r}\cdot\mathbf{R}\right)\otimes\left[\Gamma_0+\ii \Gamma\left(R_z\right)\right]\otimes\Gamma_{2LA}\quad  (\mbox{odd A})
	\end{cases}, \\
	\Gamma\left(\psi_p^{\dagger}\alpha_y \psi_q|_{p,q\in A, B}\right) &=
	\begin{cases}
		\Gamma\left(\mathbf{r}\cdot\mathbf{R}\right)\otimes\left[\Gamma\left(R_x\right)+\ii \Gamma\left(R_y\right)\right]\otimes\Gamma_{2LA}\quad  (\mbox{even A})\\
		\Gamma\left(\mathbf{r}\cdot\mathbf{R}\right)\otimes\left[\Gamma\left(R_z\right)+\ii \Gamma_0\right]\otimes\Gamma_{2LA}\quad  (\mbox{odd A})
	\end{cases},\\
	\Gamma\left(\psi_p^{\dagger}\alpha_z \psi_q|_{p,q\in A, B}\right)&=
	\begin{cases}
		\Gamma\left(\mathbf{r}\cdot\mathbf{R}\right)\otimes\left[\Gamma_0+\ii \Gamma\left(R_z\right)\right]\otimes\Gamma_{2LA}\quad  (\mbox{even A})\\
		\Gamma\left(\mathbf{r}\cdot\mathbf{R}\right)\otimes\left[\Gamma\left(R_y\right)+\ii \Gamma\left(R_x\right)\right]\otimes\Gamma_{2LA}\quad  (\mbox{odd A})
	\end{cases},
\end{align}
where
\begin{align}
	\Gamma_{2LA}&=
	\Gamma\left(\operatorname{Re}\psi^{L\alpha}_{p}\right) \Gamma\left(\operatorname{Re}\psi^{L\alpha}_{q}\right)|_{p,q\in A},\label{Gamma2LA}
\end{align}
and `even/odd A' is a shorthand notation for `even/odd number of A spinors'. We then have
\begin{align}
	\Gamma\left(\psi_p^{\dagger} \psi_q\psi_r^{\dagger} \psi_s|_{p,q,r,s\in A, B}\right)
	&= \Gamma\left(\psi_p^{\dagger} \alpha_i \psi_q\psi_r^{\dagger} \alpha_i \psi_s|_{p,q,r,s\in A, B}\right),\quad i\in x, y, z\nonumber\\
	&=\begin{cases}
		\left[\Gamma_0+\ii \Gamma\left(R_z\right)\right]\otimes \Gamma_{4LA}\quad  (\mbox{even A})\\
		\left[\Gamma\left(R_y\right)+\ii \Gamma\left(R_x\right)\right]\otimes \Gamma_{4LA}\quad (\mbox{odd A})
	\end{cases},\label{SymmERI}
\end{align}
where
\begin{align}
	\Gamma_{4LA}&=\Gamma\left(\operatorname{Re}\psi^{L\alpha}_{p}\right) \Gamma\left(\operatorname{Re}\psi^{L\alpha}_{q}\right)
	\Gamma\left(\operatorname{Re}\psi^{L\alpha}_{r}\right) \Gamma\left(\operatorname{Re}\psi^{L\alpha}_{s}\right)|_{p,q,r,s\in A}.\label{Gamma4LA}
\end{align}
It follows that the Coulomb/Gaunt/Breit two-electron repulsion integrals (ERI) have exactly the same point group symmetries and are nonzero only when
the real or imaginary part of the quartet $\psi_p^{\dagger} \psi_q\psi_r^{\dagger} \psi_s|_{p,q,r,s\in A, B}$ contains
the totally symmetric irrep $\Gamma_0$.
As pointed out previously, all $\{\operatorname{Re}\psi^{L\alpha}_{\cdot}\}$ must be assigned to
the same boson irrep (or its opposite-parity counterpart, if any) for internal consistency.
Therefore, $\Gamma_{4LA}$ \eqref{Gamma4LA} is always equal to $\Gamma_0$ [NB: in the presence
of spatial inversion as in $C_i^\ast$, $C_{2h}^\ast$, and $D_{2h}^\ast$, the number of ungerade
$\{\operatorname{Re}\psi^{L\alpha}_{p}; p\in A\}$
(equivalently, the number of ungerade A- or B-spinors, for $\psi_p$ and $\psi_{\bar p}$ have the same parity as $\operatorname{Re}\psi^{L\alpha}_{p}$) in the quartet
must be even for nonzero ERIs]. As such, Eq. \eqref{SymmERI} can actually be simplified to
\begin{align}
	\Gamma\left(\psi_p^{\dagger} \psi_q\psi_r^{\dagger} \psi_s|_{p,q,r,s\in A, B}\right)
	&= \Gamma\left(\psi_p^{\dagger} \alpha_i \psi_q\psi_r^{\dagger} \alpha_i \psi_s|_{p,q,r,s\in A, B}\right),\quad i\in x, y, z\nonumber\\
	&=\begin{cases}
		\left[\Gamma_0+\ii \Gamma\left(R_z\right)\right]\quad  (\mbox{even A})\\
		\left[\Gamma\left(R_y\right)+\ii \Gamma\left(R_x\right)\right]\quad (\mbox{odd A})
	\end{cases}.\label{SymmERI-2}
\end{align}
It is then clear that
only $C_1^\ast$ and $C_i^\ast$ permit both even and odd numbers of A-spinors, for both $\Gamma\left(R_z\right)$ and $\Gamma\left(R_x\right)$
are equal to $\Gamma_0$ in this case. In contrast, $\Gamma\left(R_x\right)$ is different from $\Gamma_0$ for the other
binary double point groups, such that an odd number of A-spinors (or equivalently B-spinors) is not permitted therein.
Since $\Gamma\left(R_z\right)$ is also equal to $\Gamma_0$ for $C_2^\ast$, $C_s^\ast$, and $C_{2h}^\ast$,
the ERIs are complex-valued, just like those of $C_1^\ast$ and $C_i^\ast$.
On the other hand,
$\Gamma\left(R_z\right)$ is different from $\Gamma_0$ for $C_{2v}^\ast$, $D_2^\ast$, and $D_{2h}^\ast$,
thereby leading to real-valued ERIs.

In short, we have the following selection rules for the ERIs solely due to symmetry reasons:
\begin{itemize}
	\item The Coulomb and Gaunt/Breit ERIs have exactly the same point group symmetries;
	\item Except for $C_1^\ast$ and $C_i^\ast$, the quartet $\psi_p^\dag \psi_q \psi_r^\dag \psi_s|_{p,q,r,s\in A, B}$ must contain an even number of A/B-spinors;
	\item For $C_i^\ast$, $C_{2h}^\ast$, and $D_{2h}^\ast$, the quartet $\psi_p^\dag \psi_q \psi_r^\dag \psi_s|_{p,q,r,s\in A, B}$
	must contain an even number of gerade/ungerade spinors;
	\item The ERIs are real-valued for $C_{2v}^\ast$, $D_2^\ast$, and $D_{2h}^\ast$, but are complex-valued for the other binary double point groups.
\end{itemize}
Such rules will be employed to construct symmetry-adapted many-electron basis functons in Sec. \ref{doublegroup2}. Note in passing that
the same analysis can also be performed for the one-electron case: just like $\Gamma_{4LA}$ \eqref{Gamma4LA}, $\Gamma_{2LA}$ \eqref{Gamma2LA}
in Eq. \eqref{pqpair} must also be the totally symmetric irrep $\Gamma_0$ in order for
the integrals of totally symmetric one-body operators to be nonzero, thereby leading to the following selection rules:
\begin{itemize}
	\item Except for $C_1^\ast$ and $C_i^\ast$, the doublet $\psi_p^\dag \psi_q|_{p,q\in A, B}$ must contain an even number of A/B-spinors;
	\item For $C_i^\ast$, $C_{2h}^\ast$, and $D_{2h}^\ast$, the doublet $\psi_p^\dag \psi_q|_{p,q\in A, B}$
	must contain an even number of gerade/ungerade spinors;
	\item The one-electron integrals are real-valued for $C_{2v}^\ast$, $D_2^\ast$, and $D_{2h}^\ast$, but are complex-valued for the other binary
	double point groups.
\end{itemize}

Finally, it deserves to be mentioned that all the one- and two-electron integrals discussed so far can actually be made real-valued (for
arbitrary double point groups)\cite{Symm2009}. However, this must be companied by proper recombinations of the excitation operators, which complicate
the implementation of correlated wave function methods. Nevertheless, this option deserves to be considered in future.

\section{Symmetry in Many-Electron Functions}\label{doublegroup2}
After having discussed extensively the time reversal and double point group symmetries inherent in the relativistic ERIs (see Appendix \ref{SecSymmERI}),
we are ready to discuss the symmetries of MBFs.
%First of all,
%the MBFs belonging to the same irrep must have the same parity under spatial inversion, whether
%the number of electrons is odd or even. Apart from this, the cases with odd and even numbers of electrons
%have to be discussed separately, again confined to binary double point groups.

\subsection{Odd Number of Electrons}
For systems of an odd number of electrons, the MBFs are associated to a fermion irrep of the given double point group.
When calculating the eigenstates, only one partner of a degenerate Kramers pair needs to be solved,
which results in a two-times reduction of the computational cost. This holds irrespective of point group symmetries.
As shown in Sec. \ref{doublegroup1}, for the $C_2^\ast$, $C_s^\ast$, $C_{2h}^\ast$, $C_{2v}^\ast$, $D_2^\ast$, and $D_{2h}^\ast$ groups,
the molecular ERIs can be nonzero only when the quartets $\psi_p^\dag\psi_q\psi_r^\dag\psi_s|_{p,q,r,s\in A, B}$ contain an even number of
B-spinors.
For an odd number of electrons, we have $\tilde{M}_K= 1$ or $3$ for every MBF.
The MBFs of different $\tilde{M}_K$ values belong to different fermion irreps in the case of $C_2^\ast$, $C_s^\ast$, and $C_{2h}^\ast$,
and to different columns of the same fermion irrep in the case of $C_{2v}^\ast$, $D_2^\ast$, and $D_{2h}^\ast$  (cf. Table. \ref{even_sym}).
Therefore, the Hamiltonian matrix elements in between are absolutely zero, which results in an additional two-times
reduction of the computational cost for these groups. For $C_{2v}^\ast$, $D_2^\ast$, and $D_{2h}^\ast$,
a further two-times reduction can be gained due to the fact
that the Hamiltonian matrix elements are all real-valued. Finally, the spatial inversion in $C_i$, $C_{2h}^\ast$, and $D_{2h}^\ast$
also gives a two-times reduction. Therefore, for systems of an odd number of electrons,
the combined use of time reversal and point group symmetries
reduces the computational cost by factors of
2,          4,       4,          4,        8,              8,            8,            and 16 for
the $C_1^\ast$, $C_i$, $C_2^\ast$, $C_s^\ast$, $C_{2h}^\ast$, $C_{2v}^\ast$, $D_2^\ast$, and $D_{2h}^\ast$ point groups, respectively,
which are precisely their group orders.

\begin{threeparttable}
	\centering
\small
	\caption{Binary double point group symmetries of $N$-electron basis functions built up with $N_A$ A-spinors and $N_B$ B-spinors. }
\begin{tabular}{ccc|cccc}\toprule
        &\multicolumn{2}{c|}{odd N} &\multicolumn{4}{c}{even N}\\
        \cline{2-7}
		& 1\tnote{a}  & 3\tnote{a} & $(0,1)$\tnote{b} & $(0,-1)$\tnote{b}  & $(2,1)$\tnote{b}  & $(2,-1)$\tnote{b}   \\\toprule
		$C_1^\ast$    & $A_{1/2}$  & $A_{1/2}$ & $A$       & $A$       & $A$       & $A$         \\
		$C_i^\ast$\tnote{c} & $A_{1/2,g}/A_{1/2,u}$ & $A_{1/2,g}/A_{1/2,u}$ & $A_g/A_u$ & $A_g/A_u$ & $A_g/A_u$ & $A_g/A_u$   \\
		$C_2^\ast$    & $^1E_{1/2}$ & $^2E_{1/2}$ & $A$       & $A$       & $B$       & $B$        \\
		$C_s^\ast$    & $^1E_{1/2}$ & $^2E_{1/2}$  & $A'$      & $A'$      & $A''$     & $A''$      \\
		$C_{2h}^\ast$\tnote{c}  & $^1E_{1/2,g}/^1E_{1/2,u}$ & $^2E_{1/2,g}/^2E_{1/2,u}$ & $A_g/A_u$ & $A_g/A_u$ & $B_g/B_u$ & $B_g/B_u$   \\
		$C_{2v}^\ast$ & $E_{1/2,1/2}$ & $E_{1/2,-1/2}$ & $A_1$     & $A_2$     & $B_1$     & $B_2$       \\
		$D_2^\ast$    & $E_{1/2,1/2}$ & $E_{1/2,-1/2}$ & $A$       & $B_1$     & $B_2$     & $B_3$       \\
		$D_{2h}^\ast$\tnote{c}  & $E_{1/2,1/2,g}/E_{1/2,1/2,u}$ & $E_{1/2,-1/2,g}/E_{1/2,-1/2,u}$ & $A_g/A_u$ & $B_{1g}/B_{1u}$ & $B_{2g}/B_{2u}$ & $B_{3g}/B_{3u}$  \\ \bottomrule
\end{tabular}
\begin{tablenotes}
\item [a]$\tilde{M}_K = (N_A-N_B) \mod 4$. The assignment of $\tilde{M}_K=1$ and 3 to the first and second columns
of the irrep $E_{1/2}$ of $C_{2v}^\ast$, $D_2^\ast$, and $D_2^\ast$ is not unambiguous, but which
does not affect the computation nor the state assignment.
\item [b] $(\tilde{M}_K, s)$ with $s=\pm 1$ in Eq. \eqref{even_basis}.
\item [c]The gerade (g) and ungerade (u) parities are determined automatically by those of the occupied spinors.
\end{tablenotes}
\label{even_sym}
\end{threeparttable}

\subsection{Even Number of Electrons}
For systems of an even number of electrons, the MBFs must belong to a boson irrep of the given double binary point group.
Since the time reversal-based transformation \eqref{even_basis} renders
the Hamiltonian matrix elements \eqref{even_hmat} real-valued, a two-times reduction of the computational cost can always be achieved,
irrespective of point group symmetries.
%The spatial inversion gives rise to an additional two-times reduction for the $C_i$, $C_{2h}$, and $D_{2h}$ groups.
Like the case of an odd number of electrons,
the quartet $\psi_p^\dag\psi_q\psi_r^\dag\psi_s|_{p,q,r,s\in A, B}$ must contain an even number of B-spinors
for the $C_2^\ast$, $C_s^\ast$, $C_{2h}^\ast$, $C_{2v}^\ast$, $D_2^\ast$, and $D_{2h}^\ast$ groups.
The MBFs of $\tilde{M}_K=0$ and 2 are associated to the totally symmetric and the other boson irreps, respectively,
in the case of $C_2^\ast$ and $C_s^\ast$. This gives rise to an additional two-times reduction, leading to an overall
four-times reduction for $C_2^\ast$ and $C_s^\ast$. The situation is similar for $C_{2h}^\ast$, except that
the parity has to be distinguished here (which gives rise to an additional two-times reduction, thereby leading to
an overall eight-times reduction). For the $C_{2v}^\ast$, $D_2^\ast$, and $D_{2h}^\ast$ groups, which have
more than two boson irreps, the two $\tilde{M}_K$ values ($0$ or $2$) are themselves not enough to distinguish the MBFs. Instead,
the $s$ values ($=\pm 1$) in Eq. \eqref{even_basis} should further be employed (cf. Table. \ref{even_sym}). It turns out
that the boson irrep for each $(\tilde{M}_K,s)$ pair can readily be determined for the case of two electrons, by first calculating
the actions of symmetry operations $\hat{R}$ on the fermion basis $(\psi,\bar{\psi})$, viz.,
\begin{equation}
\hat{R}(\psi,\bar{\psi})=(\psi^\prime,\bar{\psi}^\prime)=(\psi,\bar{\psi})\mathbf{D}^{(E_{1/2})}(R),\nonumber
\end{equation}
where $\{\mathbf{D}^{(E_{1/2})}(R), \forall R\in E_{1/2}\}$ are the matrix representations of the irrep $E_{1/2}$ of $C_{2v}^\ast$, $D_2^\ast$, and $D_{2h}^\ast$, and then calculating
\begin{align}
\hat{R}M^{\pm}&=c_{\pm}(R) M^{\pm},\nonumber\\
M^{\pm}&\sim \psi_p\psi_q\pm \psi_{\bar p}\psi_{\bar q} \quad (\tilde{M}_K=2),\nonumber\\
M^{\pm}&\sim \psi_p\psi_{\bar q}\mp \psi_{\bar p}\psi_q \quad (\tilde{M}_K=0),\nonumber
\end{align}
where the coefficients $c_{\pm}(R)$ are just the matrix elements (characters) of a (one-dimensional) boson irrep.
It can then be proven by induction
that the results are the same for $2m$ ($m>1$) electrons. Together with time reversal, the four $(\tilde{M}_K,s)$ pairs
give rise to an overall eight-times reduction for $C_{2v}^\ast$, $D_2^\ast$, and an overall sixteen-times reduction for $D_{2h}^\ast$ (due to parity).
In short, for systems of an even number of electrons, the combined use of time reversal and point group symmetries also
reduces the computational cost by the orders of the binary double point groups.

\section{Special treat of Hamiltonian matrix elements}\label{decomposition}

\subsection{Matrix elements $\langle I\mu|H_1^0 + H_2^0|I\mu\rangle$}\label{SecC1}
The diagonal matrix elements $\langle I\mu|H_1^0 + H_2^0|I\mu\rangle$ are composed of two terms
\begin{align}
	\mathcal{A}^0_1&=\sum_{p} h_{p p} \tilde{n}_p
	+ \sum_{p} [(\bar{p}\bar{p}|pp)- (\bar{p}p|p\bar{p})] n_pn_{\bar{p}}
	+ \sum_{p<q}  (pp|qq) \tilde{n}_p \tilde{n}_q, \label{Dg_Term1} \\
	\mathcal{A}^0_2&= -\sum_{p<q} \left[
	 (pq|qp) (n_pn_q + n_{\bar{p}}n_{\bar{q}})
	+ (\bar{p}q|q\bar{p}) (n_{\bar{p}}n_{q} + n_{p}n_{\bar{q}}) \right], \label{Dg_Term2}
\end{align}
where the summations run over Kramers pairs.
A close inspection reveals that $\mathcal{A}^0_1$ depends only
on the PON $\tilde{n}_p$, since $n_pn_{\bar{p}}$ is non-vanishing if and only if $\tilde{n}_p=2$, where $n_p=n_{\bar{p}}=1$.
On the other hand, if the PON of Kramers pair $\mathcal{P}_p$ or $\mathcal{P}_q$ is 2, e.g., $\tilde{n}_p=2$
and hence $n_p=n_{\bar{p}}=1$, $\mathcal{A}^0_2$ \eqref{Dg_Term2} would read
\begin{align*}
	&- (pq|qp) (n_pn_q + n_{\bar{p}}n_{\bar{q}})
	- (\bar{p}q|q\bar{p}) (n_{\bar{p}}n_{q} + n_{p}n_{\bar{q}})\\
	&= - (pq|qp) \tilde{n}_q - (\bar{p}q|q\bar{p}) \tilde{n}_q\\
	&= - [(pq|qp)+(\bar{p}q|q\bar{p})]\tilde{n}_q,
\end{align*}
which depends only on the PON $\tilde{n}_q$. Therefore, $\mathcal{A}^0_2$ \eqref{Dg_Term2} is to be evaluated case by case only when $\tilde{n}_p=\tilde{n}_q=1$.
However, the expressions $\mathcal{A}^0_1$  \eqref{Dg_Term1} and $\mathcal{A}^0_2$ \eqref{Dg_Term2}
are not yet optimal, as they are both quadratic with respect to the number of occupied Kramers pairs.
The problem can be mitigated\cite{iCIPT2} by introducing a closed-shell reference oCFG $|\omega\rangle$
with $\tilde{\omega}_p=\omega_p+\omega_{\bar{p}}=2$ or 0 (NB: the total number of electrons
is chosen to be $N-1$ for odd $N$). Then, any oCFG $|I\rangle$ can be characterized
by the differential occupations $\Delta_p = n_p^I - \omega_p$,  $\Delta_{\bar{p}} = n_{\bar{p}}^I - \omega_{\bar{p}}$, and
$\tilde{\Delta}_p = \Delta_p + \Delta_{\bar{p}}$.
The three terms in $\mathcal{A}_1^0$ \eqref{Dg_Term1} can therefore be rewritten, respectively, as
\begin{align}
	\sum_{p} h_{p p} (
	\omega_p + \omega_{\bar{p}} + \Delta_p + \Delta_{\bar{p}}) & =
	\sum_{p} h_{p p} \tilde{\omega}_p + h_{pp} (\Delta_p + \Delta_{\bar{p}})\nonumber\\
	&= \sum_{p} h_{p p} \tilde{\omega}_p +  h_{pp} \tilde{\Delta}_p,  \label{A10-a}
\end{align}
\begin{align}
	&\sum_{p} [(\bar{p}\bar{p}|pp) - (\bar{p}p|p\bar{p})] n_pn_{\bar{p}}  =
	\sum_{p} [(\bar{p}\bar{p}|pp) - (\bar{p}p|p\bar{p})] (\omega_p + \Delta_p)(\omega_{\bar{p}} +\Delta_{\bar{p}}) \nonumber \\
	& =
	\sum_{p} [(\bar{p}\bar{p}|pp) - (\bar{p}p|p\bar{p})]\omega_p\omega_{\bar{p}}
	+ \sum_{p} [(\bar{p}\bar{p}|pp) - (\bar{p}p|p\bar{p})](\Delta_p\omega_{\bar{p}}+
	\Delta_{\bar{p}}\omega_{p}) \nonumber \\
	& + \sum_{p} [(\bar{p}\bar{p}|pp) - (\bar{p}p|p\bar{p})]\Delta_p \Delta_{\bar{p}}  \nonumber \\
	& = \sum_{p} [(\bar{p}\bar{p}|pp) - (\bar{p}p|p\bar{p})]\delta_{\tilde{\omega}_p,2}
	+ \sum_{p} [(\bar{p}\bar{p}|pp) - (\bar{p}p|p\bar{p})]\tilde{\Delta}_p\delta_{\tilde{\omega}_p,2} \nonumber\\
	& + \sum_{p} [(\bar{p}\bar{p}|pp) - (\bar{p}p|p\bar{p})][\delta_{\tilde{\Delta}_p,2} + \delta_{\tilde{\Delta}_p,-2}],\label{A10-b}
\end{align}
\begin{align}
	\sum_{p<q} (pp|qq) \tilde{n}_p\tilde{n}_q & =
	\sum_{p<q} (pp|qq) (\tilde{\omega}_p+\tilde{\Delta}_p)(\tilde{\omega}_q+\tilde{\Delta}_q) \nonumber \\
	& = \sum_{p<q} (pp|qq) (\tilde{\omega}_p\tilde{\omega}_q + \tilde{\omega}_p\tilde{\Delta}_q
	+\tilde{\omega}_q\tilde{\Delta}_p
	+\tilde{\Delta}_p\tilde{\Delta}_q) \nonumber \\
	& = \sum_{p<q} 4(pp|qq)  \delta_{\tilde{\omega}_p,2}\delta_{\tilde{\omega}_q,2} + \sum_{q} \left\{ \sum_{p\neq q} 2(pp|qq)\delta_{\tilde{\omega}_p,2} \right\} \tilde{\Delta}_q \nonumber\\
	&+ \sum_{p<q} (pp|qq) \tilde{\Delta}_p\tilde{\Delta}_q. \label{A10-c}
\end{align}
Likewise, $\mathcal{A}^0_2$ \eqref{Dg_Term2} can be written as
\begin{align}
	\small
	\mathcal{A}^0_2= & -\sum_{p<q}  \left\{(pq|qp)
	\left[ (\omega_p+\Delta_p)(\omega_q+\Delta_q) +
	(\omega_{\bar{p}}+\Delta_{\bar{p}})(\omega_{\bar{q}}+\Delta_{\bar{q}}) \right]\right. \nonumber \\
	& + \left.(\bar{p}q|q\bar{p})
	\left[(\omega_{\bar{p}}+\Delta_{\bar{p}})(\omega_q+\Delta_q) + (\omega_p+\Delta_p)(\omega_{\bar{q}}+\Delta_{\bar{q}})\right]\right\}
	\nonumber \\	
	= & - \sum_{p<q} \left\{(pq|qp)\left[\omega_p\omega_q+\omega_{\bar{p}}\omega_{\bar{q}}\right]
	+ (\bar{p}q|q\bar{p})\left[\omega_{\bar{p}}\omega_q + \omega_{p}\omega_{\bar{q}}\right]\right\} \nonumber \\
	- & \sum_{p<q} (pq|qp)\left[\omega_p\Delta_q+\omega_q\Delta_p
	+\omega_{\bar{p}}\Delta_{\bar{q}}
	+\omega_{\bar{q}}\Delta_{\bar{p}}\right]  \nonumber \\
	- & \sum_{p<q} (\bar{p}q|q\bar{p})\left[
	\omega_p\Delta_{\bar{q}}
	+\omega_{\bar{p}}\Delta_{q}
	+\omega_{\bar{q}}\Delta_p
	+\omega_{q}\Delta_{\bar{p}}\right] \nonumber \\
	- & \sum_{p<q} \left\{(pq|qp)\left[\Delta_p\Delta_q+\Delta_{\bar{p}}\Delta_{\bar{q}}\right]
	+ (\bar{p}q|q\bar{p})\left[\Delta_{\bar{p}}\Delta_q + \Delta_{p}\Delta_{\bar{q}}\right]\right\} \nonumber \\
	= &- 2 \sum_{p<q}\left\{ (pq|qp)\delta_{\tilde{\omega}_p,2}\delta_{\tilde{\omega}_q,2}
	+(\bar{p}q|q\bar{p})\delta_{\tilde{\omega}_p,2}\delta_{\tilde{\omega}_q,2} \right\}\nonumber  \\
	- & \sum_{p<q} [(pq|qp)+(\bar{p}q|q\bar{p})] \left[ \delta_{\tilde{\omega}_p,2} \tilde{\Delta}_q
	+ \delta_{\tilde{\omega}_q,2} \tilde{\Delta}_p \right] \nonumber \\
	- & \sum_{p<q} \left\{ (pq|qp)\left[\Delta_p\Delta_q+\Delta_{\bar{p}}\Delta_{\bar{q}}\right]
	+ (\bar{p}q|q\bar{p})\left[\Delta_{\bar{p}}\Delta_q + \Delta_{p}\Delta_{\bar{q}}\right]\right\} \nonumber \\
	= & - 2\sum_{p<q} \left[(pq|qp)
	+ (\bar{p}q|q\bar{p})\right] \delta_{\tilde{\omega}_p,2}\delta_{\tilde{\omega}_q,2}
	%\nonumber  \\
	-
	% &
	% \sum_p
	\sum_{q\neq p} \left[(pq|qp)+(\bar{p}q|q\bar{p})\right] \delta_{\tilde{\omega}_q,2}\tilde{\Delta}_p\nonumber \\
	- & \sum_{p<q} \left\{(pq|qp)\left[\Delta_p\Delta_q+\Delta_{\bar{p}}\Delta_{\bar{q}}\right]
	+ (\bar{p}q|q\bar{p})\left[\Delta_{\bar{p}}\Delta_q + \Delta_{p}\Delta_{\bar{q}}\right]\right\}. \label{Dg_Term3_3}
\end{align}
Here, it is to be noted that if one of the Kramers pairs, say $\mathcal{P}_q$,
satisfies $\tilde{\Delta}_q=\pm2$ (i.e., $\Delta_q=\Delta_{\bar{q}}=\pm 1$), then both $\Delta_p\Delta_q+\Delta_{\bar{p}}\Delta_{\bar{q}}$ and $\Delta_{\bar{p}}\Delta_q + \Delta_{p}\Delta_{\bar{q}}$ are equal to $\pm\tilde{\Delta}_p$. Therefore, Eq. \eqref{Dg_Term3_3}
can be reorganized to
\begin{align}
	\small
	\mathcal{A}^0_2= & - 2\sum_{p<q} \left[(pq|qp)
	+ (\bar{p}q|q\bar{p})\right] \delta_{\tilde{\omega}_p,2}\delta_{\tilde{\omega}_q,2}
	-
	% \sum_p
	\sum_{q\neq p} \left[(pq|qp)+(\bar{p}q|q\bar{p})\right] \delta_{\tilde{\omega}_q,2}\tilde{\Delta}_p \nonumber \\
	& -\sum_{p\neq q} [(pq|qp)+(\bar{p}q|q\bar{p})] \delta_{\tilde{\Delta}_p,2}\tilde{\Delta}_q
	  +\sum_{p\neq q} [(pq|qp)+(\bar{p}q|q\bar{p})] \delta_{\tilde{\Delta}_p,-2}\tilde{\Delta}_q \nonumber \\
	& - \sum_{p<q}^\prime \left\{(pq|qp)\left[\Delta_p\Delta_q+\Delta_{\bar{p}}\Delta_{\bar{q}}\right]
	+ (\bar{p}q|q\bar{p})\left[\Delta_{\bar{p}}\Delta_q + \Delta_{p}\Delta_{\bar{q}}\right]\right\},\label{A02final}
\end{align}
where $\sum^\prime$ indicates that both Kramers pairs $\mathcal{P}_p$ and $\mathcal{P}_q$ are singly occupied.
$\mathcal{A}^0_2$ \eqref{A02final} can be combined with $\mathcal{A}_1^0$ \eqref{A10-a}--\eqref{A10-c} to yield the final expression for $H^0_{\text{D}}=\mathcal{A}^0_1+\mathcal{A}^0_2$ in Eq. \eqref{Dg_3}.
In most cases, only a few Kramers pairs possess non-zero  $\Delta_p$ or $\Delta_{\bar{p}}$, thereby simplifying
the evaluation of the diagonal matrix elements significantly.

\iffalse
\textcolor[rgb]{1.00,0.00,0.00}{The following was not used for $H^0_{\mathrm{O}}$ \eqref{H0off} ??}
\begin{align}
\mathcal{A}^0_3&=  \sum_{i<j} \left\{
		[(\bar{i}i|\bar{j}j) - (\bar{i}j|\bar{j}i)]
		a_{\bar{i}}^{\dagger} a_i a_{\bar{j}}^{\dagger} a_j  +
		[(\bar{i}i|j\bar{j}) - (\bar{i}\bar{j}|ji)]a_{\bar{i}}^{\dagger} a_i a_j^{\dagger} a_{\bar{j}} \right. \nonumber \\
		& + \left.
		[(i\bar{i}|\bar{j}j)-(ij|\bar{j}\bar{i})]
		a_i^{\dagger} a_{\bar{i}} a_{\bar{j}}^{\dagger} a_j  +
		[(i\bar{i}|j\bar{j}) - (i\bar{j}|j\bar{i})]
		a_i^{\dagger} a_{\bar{i}} a_{j}^{\dagger} a_{\bar{j}} \right\} \label{Dg_Term3} \\
\mathcal{A}^0_4&=  \sum_{i<j} \left\{
		- (\bar{i}j|ji) a_{\bar{i}}^{\dagger} a_i (n_j-n_{\bar{j}})
		- (i\bar{j}|ji) a_j^{\dagger} a_{\bar{j}} (n_i-n_{\bar{i}}) \right. \nonumber \\
		& \left.
		- (ij|\bar{j}i) a_{\bar{j}}^{\dagger} a_j (n_i-n_{\bar{i}})
		- (ij|j\bar{i}) a_i^{\dagger} a_{\bar{i}} (n_j-n_{\bar{j}})  \right\}, \label{Dg_Term4}
\end{align}
\fi

\subsection{Matrix Elements $\langle I\mu|H_1^1+H_2^1|J\nu\rangle$}\label{SecC2}
%\begin{align}
%\sum_{p \ne q\in A,B}(pq|pp)a^{pp}_{pq}&=\sum_{p\ne q\in A}[(pq|pp)a^{pp}_{pq}+(\bar{p}q|\bar{p}\bar{p})a^{\bar{p}\bar{p}}_{\bar{p}q}+(p\bar{q}|pp)a^{pp}_{p\bar{q}}
%+(\bar{p}\bar{q}|\bar{p}\bar{p})a^{\bar{p}\bar{p}}_{\bar{p}\bar{q}}]\\
%&=\sum_{p\ne q\in A}\left[(pq|pp)(n_p-1)a^p_q+(\bar{p}q|\bar{p}\bar{p})(n_{\bar{p}}-1)a^{\bar{p}}_q\right.\nonumber\\
%&\left. +(p\bar{q}|pp)(n_p-1) a^{p}_{\bar{q}}
%+(\bar{p}\bar{q}|\bar{p}\bar{p})(n_{\bar{p}}-1)a^{\bar{p}}_{\bar{q}}\right] \label{junk1}\\
%&=\sum_{p \ne q\in A,B}(pq|pp)(n_p-1)a^{p}_{q}\label{junk2}\\
%&=\sum_{p_\sigma \ne q_\tau \in A}(p_\sigma q_\tau |p_\sigma p_\sigma)(n_{p_\sigma}-1)a^{p_\sigma}_{q_\tau}.\label{junk3}
%\end{align}	
%Eq. \eqref{junk1} can be obtained from Eq. \eqref{junk2} by inserting bars as done in Eq. \eqref{junk3}.

When an oCFG $|I\rangle$ can be obtained from oCFG $|J\rangle$ by exciting a single electron from spinor $q_\tau$ to $p_\sigma$ ($\ne q_\tau$),
the matrix elements of the first term of $H_2^1$ \eqref{H2-1a} (denoted as $H_{2,\mathrm{c}}^1$) read
\begin{align}
\langle I\mu|H_{2,\mathrm{c}}^1|J\nu\rangle& =\left[
(p_\sigma q_\tau|p_{\bar{\sigma}}p_{\bar{\sigma}})n_{p_{\bar{\sigma}}}^{J\nu}
+ (p_\sigma q_\tau|q_{\bar{\tau}}q_{\bar{\tau}})n^{J\nu}_{q_{\bar{\tau}}}\right]\gamma^{I\mu,J\nu}_{p_\sigma q_\tau}.\label{H21-C}
\end{align}	
The matrix elements of the first, direct term in the second term of $H_2^1$ \eqref{H2-1a} (denoted as $H_{2,\mathrm{d}}^1$),
along with $H_1^1$ \eqref{H1-1}, take the following form
\begin{align}
\langle I\mu|H_1^1+H_{2,\mathrm{d}}^1|J\nu\rangle&=[h_{p_\sigma q_\tau}
+  \sum_{\lambda}\sum_{r}^\prime(p_\sigma q_\tau|r_\lambda r_\lambda) n^{J\nu}_{r_\lambda}]\gamma^{I\mu,J\nu}_{p_\sigma q_\tau}\nonumber\\
&= [h_{p_\sigma q_\tau}+\sum_{r}^\prime (p_\sigma q_\tau|rr) \tilde{n}^J_r]\gamma^{I\mu,J\nu}_{p_\sigma q_\tau},\label{H11-H21-D}
\end{align}
where use of the relation $(p_\sigma q_\tau|rr)=(p_\sigma q_\tau|\bar{r}\bar{r})$ has been made to arrive at the second equality, and
the summation $\sum^\prime_r $ over Kramers pair excludes $r= p$ or $q$. Taking the common closed-shell oCFG as the reference
(i.e., $\tilde{n}_r^J=\tilde{\omega}_r+\tilde{\Delta}_r^J$),
Eq. \eqref{H11-H21-D} can further be written as
\begin{align}
\langle I\mu|H_1^1+H_{2,\mathrm{d}}^1|J\nu\rangle&=
 [f^{\mathrm{c}}_{p_\sigma q_\tau} + \sum_{r}^\prime  (p_\sigma q_\tau|rr) \tilde{\Delta}_r^J]\gamma^{I\mu,J\nu}_{p_\sigma q_\tau},\label{Temp1}\\
f^{\mathrm{c}}_{p_\sigma q_\tau} &= h_{p_\sigma q_\tau}+ 2 \sum_{r}^\prime (p_\sigma q_\tau|rr)\delta_{\tilde{\omega}_r,2},\label{fcdef}
\end{align}
where $f^{\mathrm{c}}_{p_\sigma q_\tau}$ depends only on the closed-shell reference oCFG.

The matrix elements of the second, exchange term in the second term of $H_2^1$ \eqref{H2-1a} (denoted as $H_{2,\mathrm{e}}^1$) read
\begin{align}
\langle I\mu|H_{2,\mathrm{e}}^1|J\nu\rangle& =	 \sum_{r}^\prime
	  (p_\sigma r|r q_\tau)\Gamma_{p_\sigma rr q_\tau}^{I\mu J\nu}
	+ (p_\sigma \bar{r}|\bar{r} q_\tau) \Gamma_{p_\sigma \bar{r}\bar{r} q_\tau}^{I\mu J\nu}\nonumber \\
	& + (p_\sigma r|\bar{r} q_\tau) \Gamma_{p_\sigma r\bar{r}q_\tau}^{I\mu J\nu}
	  + (p_\sigma \bar{r}|r q_\tau) \Gamma_{p_\sigma \bar{r}r q_\tau}^{I\mu J\nu}\nonumber \\
	= & - \sum_{\lambda} \sum_{r}^\prime
	(p_\sigma r_\lambda|r_\lambda q_\tau) n_{r_\lambda}^{J\nu} \gamma_{p_\sigma q_\tau}^{I\mu J\nu}
	+ \sum_{\lambda} \sum_{r}^\prime
	(p_\sigma r_\lambda|r_{\bar{\lambda}} q_\tau) \Gamma_{p_\sigma r\bar{r} q_\tau}^{I\mu J\nu}\nonumber \\
	= & -\sum_{\lambda}\sum_{r}^\prime
	(p_\sigma r_\lambda|r_\lambda q_\tau) \delta_{\tilde{n}_r,2} \gamma_{p_\sigma q_\tau}^{I\mu J\nu}
	-\sum_{\lambda}\sum_{r}^\prime
	(p_\sigma r_\lambda|r_\lambda q_\tau)\delta_{\tilde{n}_r,1}n_{r_\lambda}\gamma_{p_\sigma q_\tau}^{I\mu J\nu}\nonumber \\
	& + \sum_{\lambda} \sum_{r}^\prime
	(p_\sigma r_\lambda|r_{\bar{\lambda}} q_\tau) \Gamma_{p_\sigma r\bar{r} q_\tau}^{I\mu J\nu},\label{three_part_5}
\end{align}
which can be combined with Eqs. \eqref{H21-C} and \eqref{H11-H21-D} to yield the final expression \eqref{SingleME}.

\subsection{Matrix elements of the Gaunt/Breit interaction}\label{GauntME}
Since the Gaunt/Breit ERIs differ formally from the Coulomb ERIs only in time reversal symmetry (cf. Eqs. \eqref{TR-B-ERI} and \eqref{TR-C-ERI}),
the same procedures as adopted in Appendices \ref{SecC1} and \ref{SecC2} can be adopted to derive the corresponding matrix elements.

The diagonal term for the matrix elements between SDs of the same oCFG $|I\rangle$ reads
\begin{align}
	H^{0,\mathrm{G}}_{\text{D}}	
	& =E^{\mathrm{G}}_{\mathrm{ref}} +
	\left\{\sum_{p} [(\bar{p}\balpha\bar{p}|p\balpha p) - (\bar{p}\balpha p|p\balpha\bar{p})]
	\tilde{\Delta}_p\delta_{\tilde{\omega}_p,2}\right.\nonumber\\
    &\left.+\sum_{p} [(\bar{p}\balpha\bar{p}|p\balpha p) - (\bar{p}\balpha p|p\balpha \bar{p})]
	[\delta_{\tilde{\Delta}_p,2} +  \delta_{\tilde{\Delta}_p,-2}]\right\}   \nonumber \\
	& - \sum_q \tilde{\Delta}_q
	\sum_{p\neq q} [(p\balpha q|q\balpha p)+(\bar{p}\balpha q|q\balpha\bar{p})] \delta_{\tilde{\omega}_p,2}
	  \nonumber \\
	& -   \sum_{p\neq q} [(p\balpha q|q\balpha  p) + (\bar{p}\balpha  q|q\balpha \bar{p})]\tilde{\Delta}_p\delta_{\tilde{\Delta}_q,2} \nonumber\\
	&+   \sum_{p\neq q} [(p\balpha q|q\balpha p) + (\bar{p}\balpha q|q\balpha\bar{p})] \tilde{\Delta}_p\delta_{\tilde{\Delta}_q,-2} \nonumber\\
	& -   \sum_{p<q}^\prime \left\{
	(p\balpha q|q\balpha p)
	\left[\Delta_p\Delta_q+\Delta_{\bar{p}}\Delta_{\bar{q}}\right]
	+
	(\bar{p}\balpha q|q\balpha\bar{p})
	\left[\Delta_{\bar{p}}\Delta_q + \Delta_{p}\Delta_{\bar{q}}\right]\right\}\nonumber\\
	& + \sum_{p<q}^\prime (p\balpha p|\balpha q)(n_p-n_{\bar{p}})(n_q-n_{\bar{q}}),
\end{align}
where the summations run over Kramers pairs and $\sum^\prime$ means that both Kramers pairs $\mathcal{P}_p$ and $\mathcal{P}_q$ are singly occupied.
The reference energy $E_{\mathrm{ref}}^{\mathrm{G}}$ reads
\begin{align}
	E_{\mathrm{ref}}^{\mathrm{G}}&=\sum_p [(p\balpha p|\bar{p}\balpha\bar{p}) - (p\balpha\bar{p}|\bar{p}\balpha p)]\delta_{\tilde{\omega}_p,2}  \nonumber\\
	&-\sum_{p<q} 2[(p\balpha q|q\balpha p) + (\bar{p}\balpha q|q\balpha \bar{p})] \delta_{\tilde{\omega}_p,2}\delta_{\tilde{\omega}_q,2}.
\end{align}
The off-diagonal matrix elements take the following form
\begin{align}
	\langle I\mu|H_{\mathrm{O}}^{0,\mathrm{G}}|I\nu\rangle
	& = \sum_{p<q}
	[(\bar{p}\balpha p|\bar{q}\balpha q) - (\bar{p}\balpha q|\bar{q}\balpha p)]\Gamma^{I\mu I\nu}_{\bar{p} p, \bar{q}q}\nonumber\\
&      +\sum_{p<q}
	[(\bar{p}\balpha p|q\balpha \bar{q}) - (\bar{p}\balpha\bar{q}|q\balpha p)]\Gamma^{I\mu I\nu}_{\bar{p} p, q\bar{q}} \nonumber  \\
&      +\sum_{p<q}
	[(p\balpha\bar{p}|\bar{q}\balpha q) - (p\balpha q|\bar{q}\balpha\bar{p})]\Gamma^{I\mu I\nu}_{p\bar{p}, \bar{q}q}\nonumber  \\
&      +\sum_{p<q}
	[(p\balpha \bar{p}|q\balpha \bar{q}) - (p\balpha \bar{q}|q\balpha \bar{p})]\Gamma^{I\mu I\nu}_{p\bar{p},q\bar{q}} \nonumber \\
	& - \sum_{p<q}
	[(\bar{p}\balpha q|q\balpha p) - (\bar{p}\balpha  p|q\balpha q)] (n_q-n_{\bar{q}})\gamma^{I\mu I\nu}_{\bar{p}p}   \nonumber \\
	& - \sum_{p<q}
    [(p\balpha \bar{q}|q\balpha p) - (p\balpha p|q\balpha \bar{q})] (n_p-n_{\bar{p}})\gamma^{I\mu I\nu}_{q\bar{q}}   \nonumber \\
	& - \sum_{p<q}
	[(p\balpha q|\bar{q}\balpha p)-(p\balpha p|\bar{q}\balpha q)](n_p-n_{\bar{p}})\gamma^{I\mu I\nu}_{\bar{q}q}  \nonumber \\
	& - \sum_{p<q}
	[(p\balpha q|q\balpha \bar{p})-(p\balpha \bar{p}|q\balpha q)](n_q-n_{\bar{q}})\gamma^{I\mu I\nu}_{p\bar{p}}.
\end{align}

When an SD $|I\mu\rangle$ can be obtained from $|J\nu\rangle$ by exciting a single electron from spinor $q_\tau$ to $p_\sigma$ ($\ne q_\tau$), we have
\begin{align}
	\langle I\mu|H_1^{1,\mathrm{G}}+H_2^{1,\mathrm{G}}|J\nu\rangle
	&=\left\{
	 (p_\sigma\balpha  q_\tau|p_{\bar{\sigma}}\balpha  p_{\bar{\sigma}})
	 n_{p_{\bar{\sigma}}}^{J\nu}
	+
	(p_\sigma\balpha  q_\tau|q_{\bar{\tau}}\balpha q_{\bar{\tau}})
	n_{q_{\bar{\tau}}}^{J\nu} \right.\nonumber\\
   &\left.
	-\sum_\lambda\sum_{r}^\prime (p_\sigma\balpha  r_\lambda|r_{\lambda}\balpha q_\tau) n_{r_\lambda}^{J\nu}\right\} \gamma^{I\mu J\nu}_{p_\sigma q_\tau} \nonumber\\
	&+\sum_\lambda\sum_{r}^\prime
	[(p_\sigma\balpha  r_\lambda|r_{\bar{\lambda}}\balpha q_\tau )- (p_\sigma s\balpha q_\tau |r_{\bar{\lambda}}\balpha r_\lambda)]
	\Gamma_{p_\sigma r_\lambda, r_{\bar{\lambda}}q_\tau}^{I\mu J\nu} \nonumber \\
	& + \sum_{r}^\prime (p_\sigma\balpha q_\tau|r\balpha r) (n_r^{J\nu}-n_{\bar{r}}^{J\nu})\gamma^{I\mu J\nu}_{p_\sigma q_\tau} ,
\end{align}
where the summation $\sum_r^\prime$ over Kramers pairs excludes $r= p$ or $q$.
All the terms need to be calculated only for Kramers pairs with different PONs or
those that are commonly singly occupied.

Finally, for the case of double excitations, we have
\begin{align}
	\langle I\mu|H_2^{2,\mathrm{G}}|J\nu\rangle& =
	\left[ 2^{-\delta_{p_\sigma r_\lambda}\delta_{q_\tau s_\delta}}
	(p_\sigma\balpha q_\tau|r_\lambda\balpha s_\delta)\right.\nonumber\\
	 &\left. - (1-\delta_{p_\sigma r_\lambda})(1-\delta_{q_\tau s_\delta})(p_\sigma\balpha s_\delta|r_{\lambda}\balpha q_{\tau})\right]
	\Gamma_{p_\sigma q_\tau, r_\lambda s_\delta}^{I\mu J\nu}.
\end{align}

\newpage

\newpage

\bibliography{4C}

\end{document}

%% file: Diagram.tex
%\section{Diagrammatic representation of the spin--free Hamiltonian}
%\include{H-diagrams}

\begin{figure*}[!htp]
	\centering
	\begin{tabular}{cccccc}
		\begin{tikzpicture}[scale = 0.40,thick]
			\draw (0,0)--(0,5);
			\draw (3,0)--(3,5);
			%\draw[fill] (3,2.5) circle [radius=0.1];
			\draw[fill] (3,2.5) node[diamond, fill, inner sep=1.3pt] {};
			%\draw[fill] (0,2.5) circle [radius=0.1];
			\draw[fill] (0,2.5) node[diamond, fill, inner sep=1.3pt] {};
			\draw (3,2.5) -- (0,2.5);
			\node at (-0.5,2.5) {\Large $p$};
			\node at (3.5,2.5) {\Large $p$};
		\end{tikzpicture}
		&
		\begin{tikzpicture}[scale = 0.40,thick]
			\draw (0,0)--(0,5);
			\draw (3,0)--(3,5);
			%\draw[fill] (3,2.5) circle [radius=0.1];
			%\draw[fill] (0,2.5) circle [radius=0.1];
			\draw[fill] (3,2.5) node[diamond, fill, inner sep=1.3pt] {};
			\draw[fill] (0,2.5) node[diamond, fill, inner sep=1.3pt] {};
			\draw (3,2.5) arc [start angle = 0,end angle = 180,radius = 1.5];
			\draw (3,2.5) arc [start angle = 360,end angle = 180,radius = 1.5];
			\node at (-0.5,2.5) {\Large $p$};
			\node at (3.5,2.5) {\Large $p$};
		\end{tikzpicture}
		&
		\begin{tikzpicture}[scale = 0.40,thick]
			\draw (0,0)--(0,5);
			\draw (3,0)--(3,5);
			%\draw[fill] (3,1) circle [radius=0.1];
			%\draw[fill] (3,4) circle [radius=0.1];
			%\draw[fill] (0,1) circle [radius=0.1];
			%\draw[fill] (0,4) circle [radius=0.1];
			\draw[fill] (3,1) node[diamond, fill, inner sep=1.3pt] {};
			\draw[fill] (3,4) node[diamond, fill, inner sep=1.3pt] {};
			\draw[fill] (0,1) node[diamond, fill, inner sep=1.3pt] {};
			\draw[fill] (0,4) node[diamond, fill, inner sep=1.3pt] {};
			\draw (3,1) -- (0,1);
			\draw (3,4) -- (0,4);
			\node at (-0.5,1) {\Large $p$};
			\node at (3.5,1) {\Large $p$};
			\node at (-0.5,4) {\Large $q$};
			\node at (3.5,4) {\Large $q$};
		\end{tikzpicture}
		&
		\begin{tikzpicture}[scale = 0.40,thick]
			\draw (0,0)--(0,5);
			\draw (3,0)--(3,5);
			%\draw[fill] (3,1) circle [radius=0.1];
			%\draw[fill] (3,4) circle [radius=0.1];
			%\draw[fill] (0,1) circle [radius=0.1];
			%\draw[fill] (0,4) circle [radius=0.1];
			\draw[fill] (3,1) node[diamond, fill, inner sep=1.3pt] {};
			\draw[fill] (3,4) node[diamond, fill, inner sep=1.3pt] {};
			\draw[fill] (0,1) node[diamond, fill, inner sep=1.3pt] {};
			\draw[fill] (0,4) node[diamond, fill, inner sep=1.3pt] {};
			\draw (3,1) -- (0,4);
			\draw (3,4) -- (0,1);
			\node at (-0.5,1) {\Large $p$};
			\node at (3.5,1) {\Large $p$};
			\node at (-0.5,4) {\Large $q$};
			\node at (3.5,4) {\Large $q$};
		\end{tikzpicture}
		\\
		(a) w1& (b) w2 & (c) w3 & (d) b7=c7\\
	\end{tabular}
	\caption{Diagrammatic representation of $H_1^0$ (a) and $H_2^0$ (b-d)}
	\label{Diagrams-0}
\end{figure*}

\begin{figure*}[!htp]
	\centering
	\begin{tabular}{cccccc}
		\begin{tikzpicture}[scale = 0.40,thick]
			\draw (0,0)--(0,5);
			\draw (3,0)--(3,5);
			%\draw[fill] (3,4) circle [radius=0.1];
			%\draw[fill] (0,1) circle [radius=0.1];
			\draw[fill] (3,4) node[diamond, fill, inner sep=1.3pt] {};
			\draw[fill] (0,1) node[diamond, fill, inner sep=1.3pt] {};
			\draw (3,4) -- (0,1);
			\node at (-0.5,1) {\Large $p$};
			\node at (3.5,4) {\Large $q$};
		\end{tikzpicture}
		&
		\begin{tikzpicture}[scale = 0.40,thick]
			\draw (0,0)--(0,5);
			\draw (3,0)--(3,5);
			%\draw[fill] (3,1) circle [radius=0.1];
			%\draw[fill] (0,4) circle [radius=0.1];
			\draw[fill] (3,1) node[diamond, fill, inner sep=1.3pt] {};
			\draw[fill] (0,4) node[diamond, fill, inner sep=1.3pt] {};
			\draw (3,1) -- (0,4);
			\node at (3.5,1) {\Large $q$};
			\node at (-0.5,4) {\Large $p$};
		\end{tikzpicture}
		&
		\begin{tikzpicture}[scale = 0.40,thick]
			\draw (0,0)--(0,5);
			\draw (3,0)--(3,5);
			%\draw[fill] (3,4) circle [radius=0.1];
			%\draw[fill] (3,1) circle [radius=0.1];
			%\draw[fill] (0,1) circle [radius=0.1];
			\draw[fill] (3,4) node[diamond, fill, inner sep=1.3pt] {};
			\draw[fill] (3,1) node[diamond, fill, inner sep=1.3pt] {};
			\draw[fill] (0,1) node[diamond, fill, inner sep=1.3pt] {};
			\draw (3,4) -- (0,1);
			\draw (3,1) -- (0,1);
			\node at (-0.5,1) {\Large $p$};
			\node at (3.5,1) {\Large $p$};
			\node at (3.5,4) {\Large $q$};
		\end{tikzpicture}
		&
		\begin{tikzpicture}[scale = 0.40,thick]
			\draw (0,0)--(0,5);
			\draw (3,0)--(3,5);
			%\draw[fill] (3,1) circle [radius=0.1];
			%\draw[fill] (0,4) circle [radius=0.1];
			%\draw[fill] (3,4) circle [radius=0.1];
			\draw[fill] (3,1) node[diamond, fill, inner sep=1.3pt] {};
			\draw[fill] (0,4) node[diamond, fill, inner sep=1.3pt] {};
			\draw[fill] (3,4) node[diamond, fill, inner sep=1.3pt] {};
			\draw (3,1) -- (0,4);
			\draw (3,4) -- (0,4);
			\node at (3.5,1) {\Large $q$};
			\node at (-0.5,4) {\Large $p$};
			\node at (3.5,4) {\Large $p$};
		\end{tikzpicture}
		&
		\begin{tikzpicture}[scale = 0.40,thick]
			\draw (0,0)--(0,5);
			\draw (3,0)--(3,5);
			%\draw[fill] (3,4) circle [radius=0.1];
			%\draw[fill] (0,1) circle [radius=0.1];
			%\draw[fill] (0,4) circle [radius=0.1];
			\draw[fill] (3,4) node[diamond, fill, inner sep=1.3pt] {};
			\draw[fill] (0,1) node[diamond, fill, inner sep=1.3pt] {};
			\draw[fill] (0,4) node[diamond, fill, inner sep=1.3pt] {};
			\draw (3,4) -- (0,1);
			\draw (3,4) -- (0,4);
			\node at (-0.5,1) {\Large $p$};
			\node at (3.5,4) {\Large $q$};
			\node at (-0.5,4) {\Large $q$};
		\end{tikzpicture}
		&
		\begin{tikzpicture}[scale = 0.40,thick]
			\draw (0,0)--(0,5);
			\draw (3,0)--(3,5);
			%\draw[fill] (3,1) circle [radius=0.1];
			%\draw[fill] (0,4) circle [radius=0.1];
			%\draw[fill] (0,1) circle [radius=0.1];
			\draw[fill] (3,1) node[diamond, fill, inner sep=1.3pt] {};
			\draw[fill] (0,4) node[diamond, fill, inner sep=1.3pt] {};
			\draw[fill] (0,1) node[diamond, fill, inner sep=1.3pt] {};
			\draw (3,1) -- (0,4);
			\draw (3,1) -- (0,1);
			\node at (3.5,1) {\Large $q$};
			\node at (-0.5,4) {\Large $p$};
			\node at (-0.5,1) {\Large $q$};
		\end{tikzpicture}
		\\
		(a) s1& (b) s2& (c) s3 & (d) s4 &(e) s5 & (f) s6\\ \\
		\begin{tikzpicture}[scale = 0.40,thick]
			\draw (0,0)--(0,5);
			\draw (3,0)--(3,5);
			%\draw[fill] (3,4) circle [radius=0.1];
			%\draw[fill] (0,4) circle [radius=0.1];
			%\draw[fill] (3,2.5) circle [radius=0.1];
			%\draw[fill] (0,1) circle [radius=0.1];
			\draw[fill] (3,4) node[diamond, fill, inner sep=1.3pt] {};
			\draw[fill] (0,4) node[diamond, fill, inner sep=1.3pt] {};
			\draw[fill] (3,2.5) node[diamond, fill, inner sep=1.3pt] {};
			\draw[fill] (0,1) node[diamond, fill, inner sep=1.3pt] {};
			\draw (3,2.5) -- (0,1);
			\draw (3,4) -- (0,4);
			\node at (-0.5,4) {\Large $r$};
			\node at (3.5,2.5) {\Large $q$};
			\node at (-0.5,1) {\Large $p$};
			\node at (3.5,4) {\Large $r$};
		\end{tikzpicture}
		&
		\begin{tikzpicture}[scale = 0.40,thick]
			\draw (0,0)--(0,5);
			\draw (3,0)--(3,5);
			%\draw[fill] (3,4) circle [radius=0.1];
			%\draw[fill] (0,4) circle [radius=0.1];
			%\draw[fill] (3,1) circle [radius=0.1];
			%\draw[fill] (0,2.5) circle [radius=0.1];
			\draw[fill] (3,4) node[diamond, fill, inner sep=1.3pt] {};
			\draw[fill] (0,4) node[diamond, fill, inner sep=1.3pt] {};
			\draw[fill] (3,1) node[diamond, fill, inner sep=1.3pt] {};
			\draw[fill] (0,2.5) node[diamond, fill, inner sep=1.3pt] {};
			\draw (3,1) -- (0,2.5);
			\draw (3,4) -- (0,4);
			\node at (-0.5,4) {\Large $r$};
			\node at (3.5,1) {\Large $q$};
			\node at (-0.5,2.5) {\Large $p$};
			\node at (3.5,4) {\Large $r$};
		\end{tikzpicture}
		&
		\begin{tikzpicture}[scale = 0.40,thick]
			\draw (0,0)--(0,5);
			\draw (3,0)--(3,5);
			%\draw[fill] (3,1) circle [radius=0.1];
			%\draw[fill] (3,4) circle [radius=0.1];
			%\draw[fill] (0,1) circle [radius=0.1];
			%\draw[fill] (0,2.5) circle [radius=0.1];
			\draw[fill] (3,1) node[diamond, fill, inner sep=1.3pt] {};
			\draw[fill] (3,4) node[diamond, fill, inner sep=1.3pt] {};
			\draw[fill] (0,1) node[diamond, fill, inner sep=1.3pt] {};
			\draw[fill] (0,2.5) node[diamond, fill, inner sep=1.3pt] {};
			\draw (3,1) -- (0,1);
			\draw (3,4) -- (0,2.5);
			\node at (-0.5,2.5) {\Large $p$};
			\node at (3.5,1) {\Large $r$};
			\node at (-0.5,1) {\Large $r$};
			\node at (3.5,4) {\Large $q$};
		\end{tikzpicture}
		&
		\begin{tikzpicture}[scale = 0.40,thick]
			\draw (0,0)--(0,5);
			\draw (3,0)--(3,5);
			%\draw[fill] (3,1) circle [radius=0.1];
			%\draw[fill] (3,2.5) circle [radius=0.1];
			%\draw[fill] (0,1) circle [radius=0.1];
			%\draw[fill] (0,1) circle [radius=0.1];
			\draw[fill] (3,1) node[diamond, fill, inner sep=1.3pt] {};
			\draw[fill] (3,2.5) node[diamond, fill, inner sep=1.3pt] {};
			\draw[fill] (0,1) node[diamond, fill, inner sep=1.3pt] {};
			\draw[fill] (0,4) node[diamond, fill, inner sep=1.3pt] {};
			\draw (3,1) -- (0,1);
			\draw (3,2.5) -- (0,4);
			\node at (-0.5,4) {\Large $p$};
			\node at (3.5,1) {\Large $r$};
			\node at (-0.5,1) {\Large $r$};
			\node at (3.5,2.5) {\Large $q$};
		\end{tikzpicture}
		&
		\begin{tikzpicture}[scale = 0.40,thick]
			\draw (0,0)--(0,5);
			\draw (3,0)--(3,5);
			%\draw[fill] (3,2.5) circle [radius=0.1];
			%\draw[fill] (3,4) circle [radius=0.1];
			%\draw[fill] (0,1) circle [radius=0.1];
			%\draw[fill] (0,2.5) circle [radius=0.1];
			\draw[fill] (3,2.5) node[diamond, fill, inner sep=1.3pt] {};
			\draw[fill] (3,4) node[diamond, fill, inner sep=1.3pt] {};
			\draw[fill] (0,1) node[diamond, fill, inner sep=1.3pt] {};
			\draw[fill] (0,2.5) node[diamond, fill, inner sep=1.3pt] {};
			\draw (3,2.5) -- (0,2.5);
			\draw (3,4) -- (0,1);
			\node at (-0.5,1) {\Large $p$};
			\node at (3.5,2.5) {\Large $r$};
			\node at (-0.5,2.5) {\Large $r$};
			\node at (3.5,4) {\Large $q$};
		\end{tikzpicture}
		&
		\begin{tikzpicture}[scale = 0.40,thick]
			\draw (0,0)--(0,5);
			\draw (3,0)--(3,5);
			%\draw[fill] (3,2.5) circle [radius=0.1];
			%\draw[fill] (3,1) circle [radius=0.1];
			%\draw[fill] (0,4) circle [radius=0.1];
			%\draw[fill] (0,2.5) circle [radius=0.1];
			\draw[fill] (3,2.5) node[diamond, fill, inner sep=1.3pt] {};
			\draw[fill] (3,1) node[diamond, fill, inner sep=1.3pt] {};
			\draw[fill] (0,4) node[diamond, fill, inner sep=1.3pt] {};
			\draw[fill] (0,2.5) node[diamond, fill, inner sep=1.3pt] {};
			\draw (3,2.5) -- (0,2.5);
			\draw (3,1) -- (0,4);
			\node at (-0.5,4) {\Large $p$};
			\node at (3.5,2.5) {\Large $r$};
			\node at (-0.5,2.5) {\Large $r$};
			\node at (3.5,1) {\Large $q$};
		\end{tikzpicture}\\
		(g) s7& (h) s8& (i) s9& (j) s10& (k) s11& (l) s12\\ \\
		\begin{tikzpicture}[scale = 0.4,thick]
			\draw (0,0)--(0,5);
			\draw (3,0)--(3,5);
			%\draw[fill] (3,4) circle [radius=0.1];
			%\draw[fill] (3,2.5) circle [radius=0.1];
			%\draw[fill] (0,1) circle [radius=0.1];
			%\draw[fill] (0,4) circle [radius=0.1];
			\draw[fill] (3,4) node[diamond, fill, inner sep=1.3pt] {};
			\draw[fill] (3,2.5) node[diamond, fill, inner sep=1.3pt] {};
			\draw[fill] (0,1) node[diamond, fill, inner sep=1.3pt] {};
			\draw[fill] (0,4) node[diamond, fill, inner sep=1.3pt] {};
			\draw (3,2.5) -- (0,4);
			\draw (3,4) -- (0,1);
			\node at (-0.5,4) {\Large $r$};
			\node at (3.5,4) {\Large $r$};
			\node at (-0.5,1) {\Large $p$};
			\node at (3.5,2.5) {\Large $q$};
		\end{tikzpicture}
		&
		\begin{tikzpicture}[scale = 0.4,thick]
			\draw (0,0)--(0,5);
			\draw (3,0)--(3,5);
			%\draw[fill] (3,4) circle [radius=0.1];
			%\draw[fill] (3,1) circle [radius=0.1];
			%\draw[fill] (0,2.5) circle [radius=0.1];
			%\draw[fill] (0,4) circle [radius=0.1];
			\draw[fill] (3,4) node[diamond, fill, inner sep=1.3pt] {};
			\draw[fill] (3,1) node[diamond, fill, inner sep=1.3pt] {};
			\draw[fill] (0,2.5) node[diamond, fill, inner sep=1.3pt] {};
			\draw[fill] (0,4) node[diamond, fill, inner sep=1.3pt] {};
			\draw (3,1) -- (0,4);
			\draw (3,4) -- (0,2.5);
			\node at (-0.5,4) {\Large $r$};
			\node at (3.5,4) {\Large $r$};
			\node at (-0.5,2.5) {\Large $p$};
			\node at (3.5,1) {\Large $q$};
		\end{tikzpicture}
		&
		\begin{tikzpicture}[scale = 0.4,thick]
			\draw (0,0)--(0,5);
			\draw (3,0)--(3,5);
			%\draw[fill] (3,1) circle [radius=0.1];
			%\draw[fill] (3,4) circle [radius=0.1];
			%\draw[fill] (0,1) circle [radius=0.1];
			%\draw[fill] (0,3) circle [radius=0.1];
			\draw[fill] (3,1) node[diamond, fill, inner sep=1.3pt] {};
			\draw[fill] (3,4) node[diamond, fill, inner sep=1.3pt] {};
			\draw[fill] (0,1) node[diamond, fill, inner sep=1.3pt] {};
			\draw[fill] (0,3) node[diamond, fill, inner sep=1.3pt] {};
			\draw (3,1) -- (0,3);
			\draw (3,4) -- (0,1);
			\node at (-0.5,1) {\Large $r$};
			\node at (3.5,1) {\Large $r$};
			\node at (-0.5,3) {\Large $p$};
			\node at (3.5,4) {\Large $q$};
		\end{tikzpicture}
		&
		\begin{tikzpicture}[scale = 0.4,thick]
			\draw (0,0)--(0,5);
			\draw (3,0)--(3,5);
			%\draw[fill] (3,1) circle [radius=0.1];
			%\draw[fill] (3,3) circle [radius=0.1];
			%\draw[fill] (0,1) circle [radius=0.1];
			%\draw[fill] (0,4) circle [radius=0.1];
			\draw[fill] (3,1) node[diamond, fill, inner sep=1.3pt] {};
			\draw[fill] (3,3) node[diamond, fill, inner sep=1.3pt] {};
			\draw[fill] (0,1) node[diamond, fill, inner sep=1.3pt] {};
			\draw[fill] (0,4) node[diamond, fill, inner sep=1.3pt] {};
			\draw (3,1) -- (0,4);
			\draw (3,3) -- (0,1);
			\node at (-0.5,1) {\Large $r$};
			\node at (3.5,1) {\Large $r$};
			\node at (-0.5,4) {\Large $p$};
			\node at (3.5,3) {\Large $q$};
		\end{tikzpicture}
		&
		\begin{tikzpicture}[scale = 0.4,thick]
			\draw (0,0)--(0,5);
			\draw (3,0)--(3,5);
			%\draw[fill] (3,4) circle [radius=0.1];
			%\draw[fill] (3,2.5) circle [radius=0.1];
			%\draw[fill] (0,2.5) circle [radius=0.1];
			%\draw[fill] (0,1) circle [radius=0.1];
			\draw[fill] (3,4) node[diamond, fill, inner sep=1.3pt] {};
			\draw[fill] (3,2.5) node[diamond, fill, inner sep=1.3pt] {};
			\draw[fill] (0,2.5) node[diamond, fill, inner sep=1.3pt] {};
			\draw[fill] (0,1) node[diamond, fill, inner sep=1.3pt] {};
			\draw (3,2.5) -- (0,1);
			\draw (3,4) -- (0,2.5);
			\node at (-0.5,2.5) {\Large $r$};
			\node at (3.5,2.5) {\Large $r$};
			\node at (-0.5,1) {\Large $p$};
			\node at (3.5,4) {\Large $q$};
		\end{tikzpicture}
		&
		\begin{tikzpicture}[scale = 0.4,thick]
			\draw (0,0)--(0,5);
			\draw (3,0)--(3,5);
			%\draw[fill] (3,1) circle [radius=0.1];
			%\draw[fill] (3,2.5) circle [radius=0.1];
			%\draw[fill] (0,2.5) circle [radius=0.1];
			%\draw[fill] (0,4) circle [radius=0.1];
			\draw[fill] (3,1) node[diamond, fill, inner sep=1.3pt] {};
			\draw[fill] (3,2.5) node[diamond, fill, inner sep=1.3pt] {};
			\draw[fill] (0,2.5) node[diamond, fill, inner sep=1.3pt] {};
			\draw[fill] (0,4) node[diamond, fill, inner sep=1.3pt] {};
			\draw (3,2.5) -- (0,4);
			\draw (3,1) -- (0,2.5);
			\node at (-0.5,2.5) {\Large $r$};
			\node at (3.5,2.5) {\Large $r$};
			\node at (-0.5,4) {\Large $p$};
			\node at (3.5,1) {\Large $q$};
		\end{tikzpicture}
		\\
		(m) b6& (n) c6& (o) b4& (p) c4& (q) a2& (r) d2\\
	\end{tabular}
	\caption{Diagrammatic representation of $H_1^1$ (a-b) and $H_2^1$ (c-r)}
	\label{Diagrams-1}
\end{figure*}

\clearpage
\newpage

\begin{figure*}[!htp]
	\centering
	\begin{tabular}{cccccc}
		\begin{tikzpicture}[scale = 0.4,thick]
			\draw (0,0)--(0,5);
			\draw (3,0)--(3,5);
			%\draw[fill] (3,4) circle [radius=0.1];
			%\draw[fill] (3,2) circle [radius=0.1];
			%\draw[fill] (0,3) circle [radius=0.1];
			%\draw[fill] (0,1) circle [radius=0.1];
			\draw[fill] (3,4) node[diamond, fill, inner sep=1.3pt] {};
			\draw[fill] (3,2) node[diamond, fill, inner sep=1.3pt] {};
			\draw[fill] (0,3) node[diamond, fill, inner sep=1.3pt] {};
			\draw[fill] (0,1) node[diamond, fill, inner sep=1.3pt] {};
			\draw (3,2) -- (0,1);
			\draw (3,4) -- (0,3);
			\node at (-0.5,1) {\Large $p$};
			\node at (-0.5,3) {\Large $r$};
			\node at (3.5,2) {\Large $q$};
			\node at (3.5,4) {\Large $s$};
		\end{tikzpicture}
		&
		\begin{tikzpicture}[scale = 0.4,thick]
			\draw (0,0)--(0,5);
			\draw (3,0)--(3,5);
			%\draw[fill] (3,4) circle [radius=0.1];
			%\draw[fill] (3,3) circle [radius=0.1];
			%\draw[fill] (0,2) circle [radius=0.1];
			%\draw[fill] (0,1) circle [radius=0.1];
			\draw[fill] (3,4) node[diamond, fill, inner sep=1.3pt] {};
			\draw[fill] (3,3) node[diamond, fill, inner sep=1.3pt] {};
			\draw[fill] (0,2) node[diamond, fill, inner sep=1.3pt] {};
			\draw[fill] (0,1) node[diamond, fill, inner sep=1.3pt] {};
			\draw (3,3) -- (0,1);
			\draw (3,4) -- (0,2);
			\node at (-0.5,1) {\Large $p$};
			\node at (-0.5,2) {\Large $r$};
			\node at (3.5,3) {\Large $q$};
			\node at (3.5,4) {\Large $s$};
		\end{tikzpicture}
		&
		\begin{tikzpicture}[scale = 0.4,thick]
			\draw (0,0)--(0,5);
			\draw (3,0)--(3,5);
			%\draw[fill] (0,1) circle [radius=0.1];
			%\draw[fill] (3,4) circle [radius=0.1];
			%\draw[fill] (3,2.5) circle [radius=0.1];
			\draw[fill] (0,1) node[diamond, fill, inner sep=1.3pt] {};
			\draw[fill] (3,4) node[diamond, fill, inner sep=1.3pt] {};
			\draw[fill] (3,2.5) node[diamond, fill, inner sep=1.3pt] {};
			\draw (0,1) -- (3,4);
			\draw (0,1) -- (3,2.5);
			\node at (3.5,2.5) {\Large $q$};
			\node at (3.5,4) {\Large $s$};
			\node at (-0.5,1) {\Large $p$};
		\end{tikzpicture}
		&
		\begin{tikzpicture}[scale = 0.4,thick]
			\draw (6,0)--(6,5);
			\draw (9,0)--(9,5);
			%\draw[fill] (9,4) circle [radius=0.1];
			%\draw[fill] (9,3) circle [radius=0.1];
			%\draw[fill] (6,2) circle [radius=0.1];
			%\draw[fill] (6,1) circle [radius=0.1];
			\draw[fill] (9,4) node[diamond, fill, inner sep=1.3pt] {};
			\draw[fill] (9,3) node[diamond, fill, inner sep=1.3pt] {};
			\draw[fill] (6,2) node[diamond, fill, inner sep=1.3pt] {};
			\draw[fill] (6,1) node[diamond, fill, inner sep=1.3pt] {};
			\draw (9,3) -- (6,2);
			\draw (9,4) -- (6,1);
			\node at (5.5,1) {\Large $p$};
			\node at (5.5,2) {\Large $r$};
			\node at (9.5,3) {\Large $q$};
			\node at (9.5,4) {\Large $s$};
		\end{tikzpicture}
		&
		\begin{tikzpicture}[scale = 0.4,thick]
			\draw (0,0)--(0,5);
			\draw (3,0)--(3,5);
			%\draw[fill] (3,4) circle [radius=0.1];
			%\draw[fill] (0,2.5) circle [radius=0.1];
			%\draw[fill] (0,1) circle [radius=0.1];
			\draw[fill] (3,4) node[diamond, fill, inner sep=1.3pt] {};
			\draw[fill] (0,2.5) node[diamond, fill, inner sep=1.3pt] {};
			\draw[fill] (0,1) node[diamond, fill, inner sep=1.3pt] {};
			\draw (3,4) -- (0,2.5);
			\draw (3,4) -- (0,1);
			\node at (-0.5,1) {\Large $p$};
			\node at (3.5,4) {\Large $q$};
			\node at (-0.5,2.5) {\Large $r$};
		\end{tikzpicture}
		&
		\begin{tikzpicture}[scale = 0.4,thick]
			\draw (0,0)--(0,5);
			\draw (3,0)--(3,5);
			%\draw[fill] (3,4) circle [radius=0.1];
			%\draw[fill] (0,1) circle [radius=0.1];
			\draw[fill] (3,4) node[diamond, fill, inner sep=1.3pt] {};
			\draw[fill] (0,1) node[diamond, fill, inner sep=1.3pt] {};
			\draw (0,1) arc [start angle = 270,end angle = 360,radius = 3];
			\draw (3,4) arc [start angle = 90,end angle = 180,radius = 3];
			\node at (3.5,4) {\Large $q$};
			\node at (-0.5,1) {\Large $p$};
		\end{tikzpicture}
		\\
		(a) a1 & (b) a3 & (c) a4 & (d) a5 & (e) a6 & (f) a7 \\
		\begin{tikzpicture}[scale = 0.4,thick]
			\draw (0,0)--(0,5);
			\draw (3,0)--(3,5);
			%\draw[fill] (3,4) circle [radius=0.1];
			%\draw[fill] (3,1) circle [radius=0.1];
			%\draw[fill] (0,3) circle [radius=0.1];
			%\draw[fill] (0,2) circle [radius=0.1];
			\draw[fill] (3,4) node[diamond, fill, inner sep=1.3pt] {};
			\draw[fill] (3,1) node[diamond, fill, inner sep=1.3pt] {};
			\draw[fill] (0,3) node[diamond, fill, inner sep=1.3pt] {};
			\draw[fill] (0,2) node[diamond, fill, inner sep=1.3pt] {};
			\draw (3,1) -- (0,2);
			\draw (3,4) -- (0,3);
			\node at (-0.5,2) {\Large $p$};
			\node at (-0.5,3) {\Large $r$};
			\node at (3.5,1) {\Large $q$};
			\node at (3.5,4) {\Large $s$};
		\end{tikzpicture}
		&
		\begin{tikzpicture}[scale = 0.4, thick]
			\draw (0,0)--(0,5);
			\draw (3,0)--(3,5);
			%\draw[fill] (0,2.5) circle [radius=0.1];
			%\draw[fill] (3,1) circle [radius=0.1];
			%\draw[fill] (3,4) circle [radius=0.1];
			\draw[fill] (0,2.5) node[diamond, fill, inner sep=1.3pt] {};
			\draw[fill] (3,1) node[diamond, fill, inner sep=1.3pt] {};
			\draw[fill] (3,4) node[diamond, fill, inner sep=1.3pt] {};
			\draw (0,2.5) -- (3,4);
			\draw (0,2.5) -- (3,1);
			\node at (3.5,1) {\Large $q$};
			\node at (3.5,4) {\Large $s$};
			\node at (-0.5,2.5) {\Large $p$};
		\end{tikzpicture}
		&
		\begin{tikzpicture}[scale = 0.4,thick]
			\draw (6,0)--(6,5);
			\draw (9,0)--(9,5);
			%\draw[fill] (9,4) circle [radius=0.1];
			%\draw[fill] (9,1) circle [radius=0.1];
			%\draw[fill] (6,3) circle [radius=0.1];
			%\draw[fill] (6,2) circle [radius=0.1];
			\draw[fill] (9,4) node[diamond, fill, inner sep=1.3pt] {};
			\draw[fill] (9,1) node[diamond, fill, inner sep=1.3pt] {};
			\draw[fill] (6,3) node[diamond, fill, inner sep=1.3pt] {};
			\draw[fill] (6,2) node[diamond, fill, inner sep=1.3pt] {};
			\draw (9,1) -- (6,3);
			\draw (9,4) -- (6,2);
			\node at (5.5,2) {\Large $p$};
			\node at (5.5,3) {\Large $r$};
			\node at (9.5,1) {\Large $q$};
			\node at (9.5,4) {\Large $s$};
		\end{tikzpicture}
		&
		\begin{tikzpicture}[scale = 0.4,thick]
			\draw (6,0)--(6,5);
			\draw (9,0)--(9,5);
			%\draw[fill] (9,4) circle [radius=0.1];
			%\draw[fill] (9,2) circle [radius=0.1];
			%\draw[fill] (6,3) circle [radius=0.1];
			%\draw[fill] (6,1) circle [radius=0.1];
			\draw[fill] (9,4) node[diamond, fill, inner sep=1.3pt] {};
			\draw[fill] (9,2) node[diamond, fill, inner sep=1.3pt] {};
			\draw[fill] (6,3) node[diamond, fill, inner sep=1.3pt] {};
			\draw[fill] (6,1) node[diamond, fill, inner sep=1.3pt] {};
			\draw (9,2) -- (6,3);
			\draw (9,4) -- (6,1);
			\node at (5.5,1) {\Large $p$};
			\node at (5.5,3) {\Large $r$};
			\node at (9.5,2) {\Large $q$};
			\node at (9.5,4) {\Large $s$};
		\end{tikzpicture}
		&

		&
		
		\\
		(g) b1& (h) b2 & (i) b3 & (j) b5 &\\
		\begin{tikzpicture}[scale = 0.4,thick]
			\draw (0,0)--(0,5);
			\draw (3,0)--(3,5);
			%\draw[fill] (3,3) circle [radius=0.1];
			%\draw[fill] (3,2) circle [radius=0.1];
			%\draw[fill] (0,4) circle [radius=0.1];
			%\draw[fill] (0,1) circle [radius=0.1];
			\draw[fill] (3,3) node[diamond, fill, inner sep=1.3pt] {};
			\draw[fill] (3,2) node[diamond, fill, inner sep=1.3pt] {};
			\draw[fill] (0,4) node[diamond, fill, inner sep=1.3pt] {};
			\draw[fill] (0,1) node[diamond, fill, inner sep=1.3pt] {};
			\draw (3,2) -- (0,1);
			\draw (3,3) -- (0,4);
			\node at (-0.5,1) {\Large $p$};
			\node at (-0.5,4) {\Large $r$};
			\node at (3.5,2) {\Large $q$};
			\node at (3.5,3) {\Large $s$};
		\end{tikzpicture}
		&
		\begin{tikzpicture}[scale = 0.4,thick]
			\draw (0,0)--(0,5);
			\draw (3,0)--(3,5);
			%\draw[fill] (3,2.5) circle [radius=0.1];
			%\draw[fill] (0,4) circle [radius=0.1];
			%\draw[fill] (0,1) circle [radius=0.1];
			\draw[fill] (3,2.5) node[diamond, fill, inner sep=1.3pt] {};
			\draw[fill] (0,4) node[diamond, fill, inner sep=1.3pt] {};
			\draw[fill] (0,1) node[diamond, fill, inner sep=1.3pt] {};
			\draw (3,2.5) -- (0,4);
			\draw (3,2.5) -- (0,1);
			\node at (-0.5,1) {\Large $p$};
			\node at (-0.5,4) {\Large $r$};
			\node at (3.5,2.5) {\Large $q$};
		\end{tikzpicture}
		&
		\begin{tikzpicture}[scale = 0.4,thick]
			\draw (6,0)--(6,5);
			\draw (9,0)--(9,5);
			%\draw[fill] (9,3) circle [radius=0.1];
			%\draw[fill] (9,2) circle [radius=0.1];
			%\draw[fill] (6,4) circle [radius=0.1];
			%\draw[fill] (6,1) circle [radius=0.1];
			\draw[fill] (9,3) node[diamond, fill, inner sep=1.3pt] {};
			\draw[fill] (9,2) node[diamond, fill, inner sep=1.3pt] {};
			\draw[fill] (6,4) node[diamond, fill, inner sep=1.3pt] {};
			\draw[fill] (6,1) node[diamond, fill, inner sep=1.3pt] {};
			\draw (9,2) -- (6,4);
			\draw (9,3) -- (6,1);
			\node at (5.5,1) {\Large $p$};
			\node at (5.5,4) {\Large $r$};
			\node at (9.5,2) {\Large $q$};
			\node at (9.5,3) {\Large $s$};
		\end{tikzpicture}
		&
		\begin{tikzpicture}[scale = 0.4,thick]
			\draw (6,0)--(6,5);
			\draw (9,0)--(9,5);
			%\draw[fill] (9,3) circle [radius=0.1];
			%\draw[fill] (9,1) circle [radius=0.1];
			%\draw[fill] (6,4) circle [radius=0.1];
			%\draw[fill] (6,2) circle [radius=0.1];
			\draw[fill] (9,3) node[diamond, fill, inner sep=1.3pt] {};
			\draw[fill] (9,1) node[diamond, fill, inner sep=1.3pt] {};
			\draw[fill] (6,4) node[diamond, fill, inner sep=1.3pt] {};
			\draw[fill] (6,2) node[diamond, fill, inner sep=1.3pt] {};
			\draw (9,1) -- (6,4);
			\draw (9,3) -- (6,2);
			\node at (5.5,2) {\Large $p$};
			\node at (5.5,4) {\Large $r$};
			\node at (9.5,1) {\Large $q$};
			\node at (9.5,3) {\Large $s$};
		\end{tikzpicture}
		&
		
		&

		\\
		(k) c1& (l) c2 & (m) c3 & (n) c5 & \\
		\begin{tikzpicture}[scale = 0.4,thick]
			\draw (0,0)--(0,5);
			\draw (3,0)--(3,5);
			%\draw[fill] (3,3) circle [radius=0.1];
			%\draw[fill] (3,1) circle [radius=0.1];
			%\draw[fill] (0,4) circle [radius=0.1];
			%\draw[fill] (0,2) circle [radius=0.1];
			\draw[fill] (3,3) node[diamond, fill, inner sep=1.3pt] {};
			\draw[fill] (3,1) node[diamond, fill, inner sep=1.3pt] {};
			\draw[fill] (0,4) node[diamond, fill, inner sep=1.3pt] {};
			\draw[fill] (0,2) node[diamond, fill, inner sep=1.3pt] {};
			\draw (3,1) -- (0,2);
			\draw (3,3) -- (0,4);
			\node at (-0.5,2) {\Large $p$};
			\node at (-0.5,4) {\Large $r$};
			\node at (3.5,1) {\Large $q$};
			\node at (3.5,3) {\Large $s$};
		\end{tikzpicture}
		&
		\begin{tikzpicture}[scale = 0.4,thick]
			\draw (0,0)--(0,5);
			\draw (3,0)--(3,5);
			%\draw[fill] (3,1) circle [radius=0.1];
			%\draw[fill] (3,2) circle [radius=0.1];
			%\draw[fill] (0,3) circle [radius=0.1];
			%\draw[fill] (0,4) circle [radius=0.1];
			\draw[fill] (3,1) node[diamond, fill, inner sep=1.3pt] {};
			\draw[fill] (3,2) node[diamond, fill, inner sep=1.3pt] {};
			\draw[fill] (0,3) node[diamond, fill, inner sep=1.3pt] {};
			\draw[fill] (0,4) node[diamond, fill, inner sep=1.3pt] {};
			\draw (3,1) -- (0,3);
			\draw (3,2) -- (0,4);
			\node at (-0.5,3) {\Large $p$};
			\node at (-0.5,4) {\Large $r$};
			\node at (3.5,1) {\Large $q$};
			\node at (3.5,2) {\Large $s$};
		\end{tikzpicture}
		&
		\begin{tikzpicture}[scale = 0.4,thick]
			\draw (0,0)--(0,5);
			\draw (3,0)--(3,5);
			%\draw[fill] (3,1) circle [radius=0.1];
			%\draw[fill] (0,4) circle [radius=0.1];
			%\draw[fill] (0,2.5) circle [radius=0.1];
			\draw[fill] (3,1) node[diamond, fill, inner sep=1.3pt] {};
			\draw[fill] (0,4) node[diamond, fill, inner sep=1.3pt] {};
			\draw[fill] (0,2.5) node[diamond, fill, inner sep=1.3pt] {};
			\draw (3,1) -- (0,4);
			\draw (3,1) -- (0,2.5);
			\node at (-0.5,2.5) {\Large $p$};
			\node at (-0.5,4) {\Large $r$};
			\node at (3.5,1) {\Large $q$};
		\end{tikzpicture}
		&
		\begin{tikzpicture}[scale = 0.4,thick]
			\draw (6,0)--(6,5);
			\draw (9,0)--(9,5);
			%\draw[fill] (9,2) circle [radius=0.1];
			%\draw[fill] (9,1) circle [radius=0.1];
			%\draw[fill] (6,4) circle [radius=0.1];
			%\draw[fill] (6,3) circle [radius=0.1];
			\draw[fill] (9,2) node[diamond, fill, inner sep=1.3pt] {};
			\draw[fill] (9,1) node[diamond, fill, inner sep=1.3pt] {};
			\draw[fill] (6,4) node[diamond, fill, inner sep=1.3pt] {};
			\draw[fill] (6,3) node[diamond, fill, inner sep=1.3pt] {};
			\draw (9,1) -- (6,4);
			\draw (9,2) -- (6,3);
			\node at (5.5,3) {\Large $p$};
			\node at (5.5,4) {\Large $r$};
			\node at (9.5,1) {\Large $q$};
			\node at (9.5,2) {\Large $s$};
		\end{tikzpicture}
		&
		\begin{tikzpicture}[scale = 0.4,thick]
			\draw (0,0)--(0,5);
			\draw (3,0)--(3,5);
			%\draw[fill] (0,4) circle [radius=0.1];
			%\draw[fill] (3,1) circle [radius=0.1];
			%\draw[fill] (3,2.5) circle [radius=0.1];
			\draw[fill] (0,4) node[diamond, fill, inner sep=1.3pt] {};
			\draw[fill] (3,1) node[diamond, fill, inner sep=1.3pt] {};
			\draw[fill] (3,2.5) node[diamond, fill, inner sep=1.3pt] {};
			\draw (0,4) -- (3,2.5);
			\draw (0,4) -- (3,1);
			\node at (3.5,1) {\Large $q$};
			\node at (3.5,2.5) {\Large $s$};
			\node at (-0.5,4) {\Large $p$};
		\end{tikzpicture}
		&
		\begin{tikzpicture}[scale = 0.4,thick]
			\draw (0,0)--(0,5);
			\draw (3,0)--(3,5);
			%\draw[fill] (3,1) circle [radius=0.1];
			%\draw[fill] (0,4) circle [radius=0.1];
			\draw[fill] (3,1) node[diamond, fill, inner sep=1.3pt] {};
			\draw[fill] (0,4) node[diamond, fill, inner sep=1.3pt] {};
			\draw (3,1) arc [start angle = 0,end angle = 90,radius = 3];
			\draw (0,4) arc [start angle = 180,end angle = 270,radius = 3];
			\node at (3.5,1) {\Large $q$};
			\node at (-0.5,4) {\Large $p$};
		\end{tikzpicture}
		\\
		(o) d1 & (p) d3 & (q) d4 & (r) d5 & (s) d6 & (t) d7 \\

	\end{tabular}
	\caption{Diagrammatic representation of $H_2^2$ }
	\label{Diagrams-2}
\end{figure*}

%% file: 4C.bbl
\providecommand{\latin}[1]{#1}
\makeatletter
\providecommand{\doi}
  {\begingroup\let\do\@makeother\dospecials
  \catcode`\{=1 \catcode`\}=2 \doi@aux}
\providecommand{\doi@aux}[1]{\endgroup\texttt{#1}}
\makeatother
\providecommand*\mcitethebibliography{\thebibliography}
\csname @ifundefined\endcsname{endmcitethebibliography}
  {\let\endmcitethebibliography\endthebibliography}{}
\begin{mcitethebibliography}{149}
\providecommand*\natexlab[1]{#1}
\providecommand*\mciteSetBstSublistMode[1]{}
\providecommand*\mciteSetBstMaxWidthForm[2]{}
\providecommand*\mciteBstWouldAddEndPuncttrue
  {\def\EndOfBibitem{\unskip.}}
\providecommand*\mciteBstWouldAddEndPunctfalse
  {\let\EndOfBibitem\relax}
\providecommand*\mciteSetBstMidEndSepPunct[3]{}
\providecommand*\mciteSetBstSublistLabelBeginEnd[3]{}
\providecommand*\EndOfBibitem{}
\mciteSetBstSublistMode{f}
\mciteSetBstMaxWidthForm{subitem}{(\alph{mcitesubitemcount})}
\mciteSetBstSublistLabelBeginEnd
  {\mcitemaxwidthsubitemform\space}
  {\relax}
  {\relax}

\bibitem[Liu(2023)]{LiuWIRES2023}
Liu,~W. Perspective: Simultaneous treatment of relativity, correlation, and
  QED. \emph{WIRES Comput. Mol. Sci.} \textbf{2023}, \emph{13}, e1652\relax
\mciteBstWouldAddEndPuncttrue
\mciteSetBstMidEndSepPunct{\mcitedefaultmidpunct}
{\mcitedefaultendpunct}{\mcitedefaultseppunct}\relax
\EndOfBibitem
\bibitem[Niskanen \latin{et~al.}(2017)Niskanen, J{\"a}nk{\"a}l{\"a}, Huttula,
  and F{\"o}hlisch]{IP-QED2017}
Niskanen,~J.; J{\"a}nk{\"a}l{\"a},~K.; Huttula,~M.; F{\"o}hlisch,~A. QED
  effects in 1s and 2s single and double ionization potentials of the noble
  gases. \emph{J. Chem. Phys.} \textbf{2017}, \emph{146}, 144312\relax
\mciteBstWouldAddEndPuncttrue
\mciteSetBstMidEndSepPunct{\mcitedefaultmidpunct}
{\mcitedefaultendpunct}{\mcitedefaultseppunct}\relax
\EndOfBibitem
\bibitem[Pa{\v{s}}teka \latin{et~al.}(2017)Pa{\v{s}}teka, Eliav, Borschevsky,
  Kaldor, and Schwerdtfeger]{SchwerdtfegerAuPRL2017}
Pa{\v{s}}teka,~L.~F.; Eliav,~E.; Borschevsky,~A.; Kaldor,~U.; Schwerdtfeger,~P.
  Relativistic coupled cluster calculations with variational quantum
  electrodynamics resolve the discrepancy between experiment and theory
  concerning the electron affinity and ionization potential of gold.
  \emph{Phys. Rev. Lett.} \textbf{2017}, \emph{118}, 023002\relax
\mciteBstWouldAddEndPuncttrue
\mciteSetBstMidEndSepPunct{\mcitedefaultmidpunct}
{\mcitedefaultendpunct}{\mcitedefaultseppunct}\relax
\EndOfBibitem
\bibitem[Sunaga \latin{et~al.}(2022)Sunaga, Salman, and Saue]{SaueeQED2022}
Sunaga,~A.; Salman,~M.; Saue,~T. 4-component relativistic Hamiltonian with
  effective QED potentials for molecular calculations. \emph{J. Chem. Phys.}
  \textbf{2022}, \emph{157}, 164101\relax
\mciteBstWouldAddEndPuncttrue
\mciteSetBstMidEndSepPunct{\mcitedefaultmidpunct}
{\mcitedefaultendpunct}{\mcitedefaultseppunct}\relax
\EndOfBibitem
\bibitem[Liu and Lindgren(2013)Liu, and Lindgren]{eQED}
Liu,~W.; Lindgren,~I. Going beyond ``no-pair relativistic quantum chemistry''.
  \emph{J. Chem. Phys.} \textbf{2013}, \emph{139}, 014108, (E)\textbf{144},
  049901 (2016).\relax
\mciteBstWouldAddEndPunctfalse
\mciteSetBstMidEndSepPunct{\mcitedefaultmidpunct}
{}{\mcitedefaultseppunct}\relax
\EndOfBibitem
\bibitem[Liu(2014)]{LiuPhysRep}
Liu,~W. Advances in relativistic molecular quantum mechanics. \emph{Phys. Rep.}
  \textbf{2014}, \emph{537}, 59--89\relax
\mciteBstWouldAddEndPuncttrue
\mciteSetBstMidEndSepPunct{\mcitedefaultmidpunct}
{\mcitedefaultendpunct}{\mcitedefaultseppunct}\relax
\EndOfBibitem
\bibitem[Liu(2014)]{IJQCrelH}
Liu,~W. Perspective: relativistic hamiltonians. \emph{Int. J. Quantum Chem.}
  \textbf{2014}, \emph{114}, 983--986\relax
\mciteBstWouldAddEndPuncttrue
\mciteSetBstMidEndSepPunct{\mcitedefaultmidpunct}
{\mcitedefaultendpunct}{\mcitedefaultseppunct}\relax
\EndOfBibitem
\bibitem[Liu(2015)]{IJQCeQED}
Liu,~W. Effective quantum electrodynamics Hamiltonians: a tutorial review.
  \emph{Int. J. Quantum Chem.} \textbf{2015}, \emph{115}, 631--640,
  (E)\textbf{116}, 971 (2016).\relax
\mciteBstWouldAddEndPunctfalse
\mciteSetBstMidEndSepPunct{\mcitedefaultmidpunct}
{}{\mcitedefaultseppunct}\relax
\EndOfBibitem
\bibitem[Liu(2016)]{X2C2016}
Liu,~W. Big picture of relativistic molecular quantum mechanics. \emph{Natl.
  Sci. Rev.} \textbf{2016}, \emph{3}, 204--221\relax
\mciteBstWouldAddEndPuncttrue
\mciteSetBstMidEndSepPunct{\mcitedefaultmidpunct}
{\mcitedefaultendpunct}{\mcitedefaultseppunct}\relax
\EndOfBibitem
\bibitem[Liu(2020)]{LiuPerspective2020}
Liu,~W. Essentials of relativistic quantum chemistry. \emph{J. Chem. Phys.}
  \textbf{2020}, \emph{152}, 180901\relax
\mciteBstWouldAddEndPuncttrue
\mciteSetBstMidEndSepPunct{\mcitedefaultmidpunct}
{\mcitedefaultendpunct}{\mcitedefaultseppunct}\relax
\EndOfBibitem
\bibitem[Liu(2020)]{LiuSciChina2020}
Liu,~W. Relativistic quantum chemistry: today and tomorrow. \emph{Sci. Sin.
  Chim.} \textbf{2020}, \emph{50}, 1672--1696\relax
\mciteBstWouldAddEndPuncttrue
\mciteSetBstMidEndSepPunct{\mcitedefaultmidpunct}
{\mcitedefaultendpunct}{\mcitedefaultseppunct}\relax
\EndOfBibitem
\bibitem[Kutzelnigg(2012)]{Kutzelnigg2012}
Kutzelnigg,~W. Solved and unsolved problems in relativistic quantum chemistry.
  \emph{Chem. Phys.} \textbf{2012}, \emph{395}, 16--34\relax
\mciteBstWouldAddEndPuncttrue
\mciteSetBstMidEndSepPunct{\mcitedefaultmidpunct}
{\mcitedefaultendpunct}{\mcitedefaultseppunct}\relax
\EndOfBibitem
\bibitem[Liu(2012)]{PCCPNES}
Liu,~W. Perspectives of relativistic quantum chemistry: the negative energy cat
  smiles. \emph{Phys. Chem. Chem. Phys.} \textbf{2012}, \emph{14}, 35--48\relax
\mciteBstWouldAddEndPuncttrue
\mciteSetBstMidEndSepPunct{\mcitedefaultmidpunct}
{\mcitedefaultendpunct}{\mcitedefaultseppunct}\relax
\EndOfBibitem
\bibitem[Li \latin{et~al.}(2012)Li, Shao, and Liu]{RelR12}
Li,~Z.; Shao,~S.; Liu,~W. Relativistic explicit correlation: Coalescence
  conditions and practical suggestions. \emph{J. Chem. Phys.} \textbf{2012},
  \emph{136}, 144117\relax
\mciteBstWouldAddEndPuncttrue
\mciteSetBstMidEndSepPunct{\mcitedefaultmidpunct}
{\mcitedefaultendpunct}{\mcitedefaultseppunct}\relax
\EndOfBibitem
\bibitem[Jeszenszki \latin{et~al.}(2021)Jeszenszki, Ferenc, and
  M{\'a}tyus]{cutting-projector}
Jeszenszki,~P.; Ferenc,~D.; M{\'a}tyus,~E. All-order explicitly correlated
  relativistic computations for atoms and molecules. \emph{J. Chem. Phys.}
  \textbf{2021}, \emph{154}, 224110\relax
\mciteBstWouldAddEndPuncttrue
\mciteSetBstMidEndSepPunct{\mcitedefaultmidpunct}
{\mcitedefaultendpunct}{\mcitedefaultseppunct}\relax
\EndOfBibitem
\bibitem[Jeszenszki \latin{et~al.}(2022)Jeszenszki, Ferenc, and
  M{\'a}tyus]{punching-projector}
Jeszenszki,~P.; Ferenc,~D.; M{\'a}tyus,~E. Variational Dirac--Coulomb
  explicitly correlated computations for atoms and molecules. \emph{J. Chem.
  Phys.} \textbf{2022}, \emph{156}, 084111\relax
\mciteBstWouldAddEndPuncttrue
\mciteSetBstMidEndSepPunct{\mcitedefaultmidpunct}
{\mcitedefaultendpunct}{\mcitedefaultseppunct}\relax
\EndOfBibitem
\bibitem[Holl{\'o}sy \latin{et~al.}(2024)Holl{\'o}sy, Jeszenszki, and
  M{\'a}tyus]{hollosy2024one}
Holl{\'o}sy,~P.; Jeszenszki,~P.; M{\'a}tyus,~E. One-particle operator
  representation over two-particle basis sets for relativistic QED
  computations. \emph{J. Chem. Theory Comput.} \textbf{2024}, \emph{20},
  5122--5132\relax
\mciteBstWouldAddEndPuncttrue
\mciteSetBstMidEndSepPunct{\mcitedefaultmidpunct}
{\mcitedefaultendpunct}{\mcitedefaultseppunct}\relax
\EndOfBibitem
\bibitem[J{\o}rgen Aa.~Jensen \latin{et~al.}(1996)J{\o}rgen Aa.~Jensen, Dyall,
  Saue, and F{\ae}gri~Jr]{4C-MCSCF1996}
J{\o}rgen Aa.~Jensen,~H.; Dyall,~K.~G.; Saue,~T.; F{\ae}gri~Jr,~K. Relativistic
  four-component multiconfigurational self-consistent-field theory for
  molecules: Formalism. \emph{J. Chem. Phys.} \textbf{1996}, \emph{104},
  4083--4097\relax
\mciteBstWouldAddEndPuncttrue
\mciteSetBstMidEndSepPunct{\mcitedefaultmidpunct}
{\mcitedefaultendpunct}{\mcitedefaultseppunct}\relax
\EndOfBibitem
\bibitem[Fleig \latin{et~al.}(1997)Fleig, Marian, and Olsen]{2C-CASSCF1996}
Fleig,~T.; Marian,~C.~M.; Olsen,~J. Spinor optimization for a relativistic
  spin-dependent CASSCF program. \emph{Theor. Chem. Acc.} \textbf{1997},
  \emph{97}, 125--135\relax
\mciteBstWouldAddEndPuncttrue
\mciteSetBstMidEndSepPunct{\mcitedefaultmidpunct}
{\mcitedefaultendpunct}{\mcitedefaultseppunct}\relax
\EndOfBibitem
\bibitem[Kim and Lee(2003)Kim, and Lee]{2C-CASSCF2003}
Kim,~Y.~S.; Lee,~Y.~S. The Kramers’ restricted complete active space
  self-consistent-field method for two-component molecular spinors and
  relativistic effective core potentials including spin--orbit interactions.
  \emph{J. Chem. Phys.} \textbf{2003}, \emph{119}, 12169--12178\relax
\mciteBstWouldAddEndPuncttrue
\mciteSetBstMidEndSepPunct{\mcitedefaultmidpunct}
{\mcitedefaultendpunct}{\mcitedefaultseppunct}\relax
\EndOfBibitem
\bibitem[Kim and Lee(2013)Kim, and Lee]{2C-CASSCF2013}
Kim,~I.; Lee,~Y.~S. Two-component Kramers restricted complete active space
  self-consistent field method with relativistic effective core potential
  revisited: Theory, implementation, and applications to spin-orbit splitting
  of lower p-block atoms. \emph{J. Chem. Phys.} \textbf{2013}, \emph{139},
  134115\relax
\mciteBstWouldAddEndPuncttrue
\mciteSetBstMidEndSepPunct{\mcitedefaultmidpunct}
{\mcitedefaultendpunct}{\mcitedefaultseppunct}\relax
\EndOfBibitem
\bibitem[Thyssen \latin{et~al.}(2008)Thyssen, Fleig, and Jensen]{4C-MCSCF2008}
Thyssen,~J.; Fleig,~T.; Jensen,~H. J.~A. A direct relativistic four-component
  multiconfiguration self-consistent-field method for molecules. \emph{J. Chem.
  Phys.} \textbf{2008}, \emph{129}, 034109\relax
\mciteBstWouldAddEndPuncttrue
\mciteSetBstMidEndSepPunct{\mcitedefaultmidpunct}
{\mcitedefaultendpunct}{\mcitedefaultseppunct}\relax
\EndOfBibitem
\bibitem[Bates and Shiozaki(2015)Bates, and Shiozaki]{4C-CASSCF2015}
Bates,~J.~E.; Shiozaki,~T. Fully relativistic complete active space
  self-consistent field for large molecules: Quasi-second-order minimax
  optimization. \emph{J. Chem. Phys.} \textbf{2015}, \emph{142}, 044112\relax
\mciteBstWouldAddEndPuncttrue
\mciteSetBstMidEndSepPunct{\mcitedefaultmidpunct}
{\mcitedefaultendpunct}{\mcitedefaultseppunct}\relax
\EndOfBibitem
\bibitem[Reynolds \latin{et~al.}(2018)Reynolds, Yanai, and
  Shiozaki]{4C-CASSCF2018}
Reynolds,~R.~D.; Yanai,~T.; Shiozaki,~T. Large-scale relativistic complete
  active space self-consistent field with robust convergence. \emph{J. Chem.
  Phys.} \textbf{2018}, \emph{149}, 014106\relax
\mciteBstWouldAddEndPuncttrue
\mciteSetBstMidEndSepPunct{\mcitedefaultmidpunct}
{\mcitedefaultendpunct}{\mcitedefaultseppunct}\relax
\EndOfBibitem
\bibitem[Jenkins \latin{et~al.}(2019)Jenkins, Liu, Kasper, Frisch, and
  Li]{LixiaosongX2CCASSCF}
Jenkins,~A.~J.; Liu,~H.; Kasper,~J.~M.; Frisch,~M.~J.; Li,~X. Variational
  Relativistic Two-Component Complete-Active-Space Self-Consistent Field
  Method. \emph{J. Chem. Theory Comput.} \textbf{2019}, \emph{15},
  2974--2982\relax
\mciteBstWouldAddEndPuncttrue
\mciteSetBstMidEndSepPunct{\mcitedefaultmidpunct}
{\mcitedefaultendpunct}{\mcitedefaultseppunct}\relax
\EndOfBibitem
\bibitem[Dyall(1994)]{4C-MP21994}
Dyall,~K.~G. Second-order M{\o}ller-Plesset perturbation theory for molecular
  Dirac-Hartree-Fock wavefunctions. Theory for up to two open-shell electrons.
  \emph{Chem. Phys. Lett.} \textbf{1994}, \emph{224}, 186--194\relax
\mciteBstWouldAddEndPuncttrue
\mciteSetBstMidEndSepPunct{\mcitedefaultmidpunct}
{\mcitedefaultendpunct}{\mcitedefaultseppunct}\relax
\EndOfBibitem
\bibitem[Vilkas \latin{et~al.}(1999)Vilkas, Ishikawa, and Koc]{4C-MRPT21999}
Vilkas,~M.~J.; Ishikawa,~Y.; Koc,~K. Relativistic multireference many-body
  perturbation theory for quasidegenerate systems: Energy levels of ions of the
  oxygen isoelectronic sequence. \emph{Phys. Rev. A} \textbf{1999}, \emph{60},
  2808\relax
\mciteBstWouldAddEndPuncttrue
\mciteSetBstMidEndSepPunct{\mcitedefaultmidpunct}
{\mcitedefaultendpunct}{\mcitedefaultseppunct}\relax
\EndOfBibitem
\bibitem[Abe \latin{et~al.}(2006)Abe, Nakajima, and Hirao]{4C-CASPT22006}
Abe,~M.; Nakajima,~T.; Hirao,~K. The relativistic complete active-space
  second-order perturbation theory with the four-component Dirac Hamiltonian.
  \emph{J. Chem. Phys.} \textbf{2006}, \emph{125}, 234110\relax
\mciteBstWouldAddEndPuncttrue
\mciteSetBstMidEndSepPunct{\mcitedefaultmidpunct}
{\mcitedefaultendpunct}{\mcitedefaultseppunct}\relax
\EndOfBibitem
\bibitem[Kim and Lee(2014)Kim, and Lee]{2C-MRPT22014}
Kim,~I.; Lee,~Y.~S. Two-component multi-configurational second-order
  perturbation theory with Kramers restricted complete active space
  self-consistent field reference function and spin-orbit relativistic
  effective core potential. \emph{J. Chem. Phys.} \textbf{2014}, \emph{141},
  164104\relax
\mciteBstWouldAddEndPuncttrue
\mciteSetBstMidEndSepPunct{\mcitedefaultmidpunct}
{\mcitedefaultendpunct}{\mcitedefaultseppunct}\relax
\EndOfBibitem
\bibitem[Shiozaki and Mizukami(2015)Shiozaki, and
  Mizukami]{4C-icMRCI-CASPT2-NEVPT22015}
Shiozaki,~T.; Mizukami,~W. Relativistic internally contracted multireference
  electron correlation methods. \emph{J. Chem. Theory Comput.} \textbf{2015},
  \emph{11}, 4733--4739\relax
\mciteBstWouldAddEndPuncttrue
\mciteSetBstMidEndSepPunct{\mcitedefaultmidpunct}
{\mcitedefaultendpunct}{\mcitedefaultseppunct}\relax
\EndOfBibitem
\bibitem[Lu \latin{et~al.}(2022)Lu, Hu, Jenkins, and Li]{X2C-MRPT20222}
Lu,~L.; Hu,~H.; Jenkins,~A.~J.; Li,~X. Exact-two-component relativistic
  multireference second-order perturbation theory. \emph{J. Chem. Theory
  Comput.} \textbf{2022}, \emph{18}, 2983--2992\relax
\mciteBstWouldAddEndPuncttrue
\mciteSetBstMidEndSepPunct{\mcitedefaultmidpunct}
{\mcitedefaultendpunct}{\mcitedefaultseppunct}\relax
\EndOfBibitem
\bibitem[Zhao and Evangelista(2024)Zhao, and Evangelista]{4C-DSRG-MRPT2-2024}
Zhao,~Z.; Evangelista,~F.~A. Toward Accurate Spin–Orbit Splittings from
  Relativistic Multireference Electronic Structure Theory. \emph{J. Phys. Chem.
  Lett.} \textbf{2024}, \emph{15}, 7103--7110\relax
\mciteBstWouldAddEndPuncttrue
\mciteSetBstMidEndSepPunct{\mcitedefaultmidpunct}
{\mcitedefaultendpunct}{\mcitedefaultseppunct}\relax
\EndOfBibitem
\bibitem[Visscher \latin{et~al.}(1995)Visscher, Dyall, and Lee]{4C-CCSD1995}
Visscher,~L.; Dyall,~K.~G.; Lee,~T.~J. Kramers-restricted closed-shell ccsd
  theory. \emph{Int. J. Quantum Chem.} \textbf{1995}, \emph{56}, 411--419\relax
\mciteBstWouldAddEndPuncttrue
\mciteSetBstMidEndSepPunct{\mcitedefaultmidpunct}
{\mcitedefaultendpunct}{\mcitedefaultseppunct}\relax
\EndOfBibitem
\bibitem[Visscher \latin{et~al.}(1996)Visscher, Lee, and Dyall]{4C-CCSD1996}
Visscher,~L.; Lee,~T.~J.; Dyall,~K.~G. Formulation and implementation of a
  relativistic unrestricted coupled-cluster method including noniterative
  connected triples. \emph{J. Chem. Phys.} \textbf{1996}, \emph{105},
  8769--8776\relax
\mciteBstWouldAddEndPuncttrue
\mciteSetBstMidEndSepPunct{\mcitedefaultmidpunct}
{\mcitedefaultendpunct}{\mcitedefaultseppunct}\relax
\EndOfBibitem
\bibitem[Ilia{\v{s}} \latin{et~al.}(2001)Ilia{\v{s}}, Kell{\"o}, Visscher, and
  Schimmelpfennig]{2C-CC2001}
Ilia{\v{s}},~M.; Kell{\"o},~V.; Visscher,~L.; Schimmelpfennig,~B. Inclusion of
  mean-field spin--orbit effects based on all-electron two-component spinors:
  Pilot calculations on atomic and molecular properties. \emph{J. Chem. Phys.}
  \textbf{2001}, \emph{115}, 9667--9674\relax
\mciteBstWouldAddEndPuncttrue
\mciteSetBstMidEndSepPunct{\mcitedefaultmidpunct}
{\mcitedefaultendpunct}{\mcitedefaultseppunct}\relax
\EndOfBibitem
\bibitem[Lee \latin{et~al.}(2005)Lee, Cho, Choi, and Lee]{2C-CC2005}
Lee,~H.~S.; Cho,~W.~K.; Choi,~Y.~J.; Lee,~Y.~S. Spin--orbit effects for the
  diatomic molecules containing halogen elements studied with relativistic
  effective core potentials: HX, X2 (X= Cl, Br and I) and IZ (Z= F, Cl and Br)
  molecules. \emph{Chem. Phys.} \textbf{2005}, \emph{311}, 121--127\relax
\mciteBstWouldAddEndPuncttrue
\mciteSetBstMidEndSepPunct{\mcitedefaultmidpunct}
{\mcitedefaultendpunct}{\mcitedefaultseppunct}\relax
\EndOfBibitem
\bibitem[Hirata \latin{et~al.}(2007)Hirata, Yanai, Harrison, Kamiya, and
  Fan]{2C-CC2007}
Hirata,~S.; Yanai,~T.; Harrison,~R.~J.; Kamiya,~M.; Fan,~P.-D. High-order
  electron-correlation methods with scalar relativistic and spin-orbit
  corrections. \emph{J. Chem. Phys.} \textbf{2007}, \emph{126}, 024104\relax
\mciteBstWouldAddEndPuncttrue
\mciteSetBstMidEndSepPunct{\mcitedefaultmidpunct}
{\mcitedefaultendpunct}{\mcitedefaultseppunct}\relax
\EndOfBibitem
\bibitem[Landau \latin{et~al.}(2000)Landau, Eliav, Ishikawa, and
  Kaldor]{4C-IHFSCC2000}
Landau,~A.; Eliav,~E.; Ishikawa,~Y.; Kaldor,~U. Intermediate Hamiltonian
  Fock-space coupled-cluster method: Excitation energies of barium and radium.
  \emph{J. Chem. Phys.} \textbf{2000}, \emph{113}, 9905--9910\relax
\mciteBstWouldAddEndPuncttrue
\mciteSetBstMidEndSepPunct{\mcitedefaultmidpunct}
{\mcitedefaultendpunct}{\mcitedefaultseppunct}\relax
\EndOfBibitem
\bibitem[Landau \latin{et~al.}(2001)Landau, Eliav, Ishikawa, and
  Kaldor]{4C-IHFSCC2001}
Landau,~A.; Eliav,~E.; Ishikawa,~Y.; Kaldor,~U. Intermediate Hamiltonian
  Fock-space coupled cluster method in the one-hole one-particle sector:
  Excitation energies of xenon and radon. \emph{J. Chem. Phys.} \textbf{2001},
  \emph{115}, 6862--6865\relax
\mciteBstWouldAddEndPuncttrue
\mciteSetBstMidEndSepPunct{\mcitedefaultmidpunct}
{\mcitedefaultendpunct}{\mcitedefaultseppunct}\relax
\EndOfBibitem
\bibitem[Visscher \latin{et~al.}(2001)Visscher, Eliav, and Kaldor]{4C-FSCC2001}
Visscher,~L.; Eliav,~E.; Kaldor,~U. Formulation and implementation of the
  relativistic Fock-space coupled cluster method for molecules. \emph{J. Chem.
  Phys.} \textbf{2001}, \emph{115}, 9720--9726\relax
\mciteBstWouldAddEndPuncttrue
\mciteSetBstMidEndSepPunct{\mcitedefaultmidpunct}
{\mcitedefaultendpunct}{\mcitedefaultseppunct}\relax
\EndOfBibitem
\bibitem[Fleig \latin{et~al.}(2007)Fleig, S{\o}rensen, and Olsen]{4C-CC2007}
Fleig,~T.; S{\o}rensen,~L.~K.; Olsen,~J. A relativistic 4-component
  general-order multi-reference coupled cluster method: initial implementation
  and application to HBr. \emph{Theor. Chem. Acc.} \textbf{2007}, \emph{118},
  347--356\relax
\mciteBstWouldAddEndPuncttrue
\mciteSetBstMidEndSepPunct{\mcitedefaultmidpunct}
{\mcitedefaultendpunct}{\mcitedefaultseppunct}\relax
\EndOfBibitem
\bibitem[Nataraj \latin{et~al.}(2010)Nataraj, K{\'a}llay, and
  Visscher]{4C-CC2010}
Nataraj,~H.~S.; K{\'a}llay,~M.; Visscher,~L. General implementation of the
  relativistic coupled-cluster method. \emph{J. Chem. Phys.} \textbf{2010},
  \emph{133}, 234109\relax
\mciteBstWouldAddEndPuncttrue
\mciteSetBstMidEndSepPunct{\mcitedefaultmidpunct}
{\mcitedefaultendpunct}{\mcitedefaultseppunct}\relax
\EndOfBibitem
\bibitem[Sørensen \latin{et~al.}(2011)Sørensen, Olsen, and
  Fleig]{2C4C-CC2011}
Sørensen,~L.~K.; Olsen,~J.; Fleig,~T. Two- and four-component relativistic
  generalized-active-space coupled cluster method: Implementation and
  application to BiH Two- and four-component relativistic
  generalized-active-space coupled cluster method: Implementation and
  application to BiH. \emph{J. Chem. Phys.} \textbf{2011}, \emph{134},
  214102\relax
\mciteBstWouldAddEndPuncttrue
\mciteSetBstMidEndSepPunct{\mcitedefaultmidpunct}
{\mcitedefaultendpunct}{\mcitedefaultseppunct}\relax
\EndOfBibitem
\bibitem[Pathak \latin{et~al.}(2016)Pathak, Sasmal, Nayak, Vaval, and
  Pal]{4C-CC2016}
Pathak,~H.; Sasmal,~S.; Nayak,~M.~K.; Vaval,~N.; Pal,~S. Relativistic
  equation-of-motion coupled-cluster method using open-shell reference
  wavefunction: Application to ionization potential. \emph{J. Chem. Phys.}
  \textbf{2016}, \emph{145}, 074110\relax
\mciteBstWouldAddEndPuncttrue
\mciteSetBstMidEndSepPunct{\mcitedefaultmidpunct}
{\mcitedefaultendpunct}{\mcitedefaultseppunct}\relax
\EndOfBibitem
\bibitem[Akinaga and Nakajima(2017)Akinaga, and Nakajima]{2C-CC2017}
Akinaga,~Y.; Nakajima,~T. Two-component relativistic equation-of-motion
  coupled-cluster methods for excitation energies and ionization potentials of
  atoms and molecules. \emph{J. Phys. Chem. A} \textbf{2017}, \emph{121},
  827--835\relax
\mciteBstWouldAddEndPuncttrue
\mciteSetBstMidEndSepPunct{\mcitedefaultmidpunct}
{\mcitedefaultendpunct}{\mcitedefaultseppunct}\relax
\EndOfBibitem
\bibitem[Liu \latin{et~al.}(2018)Liu, Shen, Asthana, and Cheng]{Cheng2C-CC2018}
Liu,~J.; Shen,~Y.; Asthana,~A.; Cheng,~L. Two-component relativistic
  coupled-cluster methods using mean-field spin-orbit integrals. \emph{J. Chem.
  Phys.} \textbf{2018}, \emph{148}, 034106\relax
\mciteBstWouldAddEndPuncttrue
\mciteSetBstMidEndSepPunct{\mcitedefaultmidpunct}
{\mcitedefaultendpunct}{\mcitedefaultseppunct}\relax
\EndOfBibitem
\bibitem[Shee \latin{et~al.}(2016)Shee, Visscher, and Saue]{Saue4C-CC2016}
Shee,~A.; Visscher,~L.; Saue,~T. Analytic one-electron properties at the
  4-component relativistic coupled cluster level with inclusion of spin-orbit
  coupling. \emph{J. Chem. Phys.} \textbf{2016}, \emph{145}, 184107\relax
\mciteBstWouldAddEndPuncttrue
\mciteSetBstMidEndSepPunct{\mcitedefaultmidpunct}
{\mcitedefaultendpunct}{\mcitedefaultseppunct}\relax
\EndOfBibitem
\bibitem[Shee \latin{et~al.}(2018)Shee, Saue, Visscher, and Severo
  Pereira~Gomes]{Saue4C-CC2018}
Shee,~A.; Saue,~T.; Visscher,~L.; Severo Pereira~Gomes,~A. Equation-of-motion
  coupled-cluster theory based on the 4-component Dirac--Coulomb (--Gaunt)
  Hamiltonian. Energies for single electron detachment, attachment, and
  electronically excited states. \emph{J. Chem. Phys.} \textbf{2018},
  \emph{149}, 174113\relax
\mciteBstWouldAddEndPuncttrue
\mciteSetBstMidEndSepPunct{\mcitedefaultmidpunct}
{\mcitedefaultendpunct}{\mcitedefaultseppunct}\relax
\EndOfBibitem
\bibitem[Asthana \latin{et~al.}(2019)Asthana, Liu, and Cheng]{2C-EOM-CCSD2019}
Asthana,~A.; Liu,~J.; Cheng,~L. Exact two-component equation-of-motion
  coupled-cluster singles and doubles method using atomic mean-field spin-orbit
  integrals. \emph{J. Chem. Phys.} \textbf{2019}, \emph{150}, 074102\relax
\mciteBstWouldAddEndPuncttrue
\mciteSetBstMidEndSepPunct{\mcitedefaultmidpunct}
{\mcitedefaultendpunct}{\mcitedefaultseppunct}\relax
\EndOfBibitem
\bibitem[Liu and Cheng(2021)Liu, and Cheng]{Cheng2C-CCrev}
Liu,~J.; Cheng,~L. Relativistic coupled-cluster and equation-of-motion
  coupled-cluster methods. \emph{WIREs Comput. Mol. Sci.} \textbf{2021},
  \emph{11}, e1536\relax
\mciteBstWouldAddEndPuncttrue
\mciteSetBstMidEndSepPunct{\mcitedefaultmidpunct}
{\mcitedefaultendpunct}{\mcitedefaultseppunct}\relax
\EndOfBibitem
\bibitem[Halbert \latin{et~al.}(2021)Halbert, Vidal, Shee, Coriani, and Severo
  Pereira~Gomes]{CVS-EOM-CCSD2021}
Halbert,~L.; Vidal,~M.~L.; Shee,~A.; Coriani,~S.; Severo Pereira~Gomes,~A.
  Relativistic EOM-CCSD for Core-Excited and Core-Ionized State Energies Based
  on the Four-Component Dirac--Coulomb (-Gaunt) Hamiltonian. \emph{J. Chem.
  Theory Comput.} \textbf{2021}, \emph{17}, 3583--3598\relax
\mciteBstWouldAddEndPuncttrue
\mciteSetBstMidEndSepPunct{\mcitedefaultmidpunct}
{\mcitedefaultendpunct}{\mcitedefaultseppunct}\relax
\EndOfBibitem
\bibitem[Visscher \latin{et~al.}(1993)Visscher, Saue, Nieuwpoort, Faegri, and
  Gropen]{4C-CISD1993}
Visscher,~L.; Saue,~T.; Nieuwpoort,~W.; Faegri,~K.; Gropen,~O. The electronic
  structure of the PtH molecule: Fully relativistic configuration interaction
  calculations of the ground and excited states. \emph{J. Chem. Phys.}
  \textbf{1993}, \emph{99}, 6704--6715\relax
\mciteBstWouldAddEndPuncttrue
\mciteSetBstMidEndSepPunct{\mcitedefaultmidpunct}
{\mcitedefaultendpunct}{\mcitedefaultseppunct}\relax
\EndOfBibitem
\bibitem[Kim \latin{et~al.}(1996)Kim, Lee, and Lee]{2C-CI1996}
Kim,~M.~C.; Lee,~S.~Y.; Lee,~Y.~S. Spin-orbit effects calculated by a
  configuration interaction method using determinants of two-component
  molecular spinors: test calculations on Rn and T1H. \emph{Chem. Phys. Lett.}
  \textbf{1996}, \emph{253}, 216--222\relax
\mciteBstWouldAddEndPuncttrue
\mciteSetBstMidEndSepPunct{\mcitedefaultmidpunct}
{\mcitedefaultendpunct}{\mcitedefaultseppunct}\relax
\EndOfBibitem
\bibitem[Fleig \latin{et~al.}(2001)Fleig, Olsen, and Marian]{2C-MRCI2001}
Fleig,~T.; Olsen,~J.; Marian,~C.~M. The generalized active space concept for
  the relativistic treatment of electron correlation. I. Kramers-restricted
  two-component configuration interaction. \emph{J. Chem. Phys.} \textbf{2001},
  \emph{114}, 4775--4790\relax
\mciteBstWouldAddEndPuncttrue
\mciteSetBstMidEndSepPunct{\mcitedefaultmidpunct}
{\mcitedefaultendpunct}{\mcitedefaultseppunct}\relax
\EndOfBibitem
\bibitem[Watanabe and Matsuoka(2002)Watanabe, and Matsuoka]{4C-MRCI2002}
Watanabe,~Y.; Matsuoka,~O. Four-component relativistic
  configuration-interaction calculation using the reduced frozen-core
  approximation. \emph{J. Chem. Phys.} \textbf{2002}, \emph{116},
  9585--9590\relax
\mciteBstWouldAddEndPuncttrue
\mciteSetBstMidEndSepPunct{\mcitedefaultmidpunct}
{\mcitedefaultendpunct}{\mcitedefaultseppunct}\relax
\EndOfBibitem
\bibitem[Fleig \latin{et~al.}(2003)Fleig, Olsen, and Visscher]{4C-GASCI2003}
Fleig,~T.; Olsen,~J.; Visscher,~L. The generalized active space concept for the
  relativistic treatment of electron correlation. II. Large-scale configuration
  interaction implementation based on relativistic 2- and 4-spinors and its
  application The generalized active space concept for the. \emph{J. Chem.
  Phys.} \textbf{2003}, \emph{119}, 2963--2971\relax
\mciteBstWouldAddEndPuncttrue
\mciteSetBstMidEndSepPunct{\mcitedefaultmidpunct}
{\mcitedefaultendpunct}{\mcitedefaultseppunct}\relax
\EndOfBibitem
\bibitem[Fleig \latin{et~al.}(2006)Fleig, Jensen, Olsen, and
  Visscher]{4C-CI-MCSCF2006}
Fleig,~T.; Jensen,~H. J.~A.; Olsen,~J.; Visscher,~L. The generalized active
  space concept for the relativistic treatment of electron correlation. III.
  Large-scale configuration interaction and multiconfiguration
  self-consistent-field four-component methods with application to UO$_2$.
  \emph{J. Chem. Phys.} \textbf{2006}, \emph{124}, 104106\relax
\mciteBstWouldAddEndPuncttrue
\mciteSetBstMidEndSepPunct{\mcitedefaultmidpunct}
{\mcitedefaultendpunct}{\mcitedefaultseppunct}\relax
\EndOfBibitem
\bibitem[Bylicki \latin{et~al.}(2008)Bylicki, Pestka, and Karwowski]{4C-CI2008}
Bylicki,~M.; Pestka,~G.; Karwowski,~J. Relativistic Hylleraas
  configuration-interaction method projected into positive-energy space.
  \emph{Phys. Rev. A} \textbf{2008}, \emph{77}, 044501\relax
\mciteBstWouldAddEndPuncttrue
\mciteSetBstMidEndSepPunct{\mcitedefaultmidpunct}
{\mcitedefaultendpunct}{\mcitedefaultseppunct}\relax
\EndOfBibitem
\bibitem[Kim \latin{et~al.}(2012)Kim, Park, Kim, and Lee]{2C-CICC2012}
Kim,~I.; Park,~Y.~C.; Kim,~H.; Lee,~Y.~S. Spin-orbit coupling and electron
  correlation in relativistic configuration interaction and coupled-cluster
  methods. \emph{Chem. Phys.} \textbf{2012}, \emph{395}, 115--121\relax
\mciteBstWouldAddEndPuncttrue
\mciteSetBstMidEndSepPunct{\mcitedefaultmidpunct}
{\mcitedefaultendpunct}{\mcitedefaultseppunct}\relax
\EndOfBibitem
\bibitem[Fleig(2012)]{Fleig2012}
Fleig,~T. Invited review: Relativistic wave-function based electron correlation
  methods. \emph{Chem. Phys.} \textbf{2012}, \emph{395}, 2--15\relax
\mciteBstWouldAddEndPuncttrue
\mciteSetBstMidEndSepPunct{\mcitedefaultmidpunct}
{\mcitedefaultendpunct}{\mcitedefaultseppunct}\relax
\EndOfBibitem
\bibitem[Hu \latin{et~al.}(2020)Hu, Jenkins, Liu, Kasper, Frisch, and
  Li]{2C-MRCI2020}
Hu,~H.; Jenkins,~A.~J.; Liu,~H.; Kasper,~J.~M.; Frisch,~M.~J.; Li,~X.
  Relativistic Two-Component Multireference Configuration Interaction Method
  with Tunable Correlation Space. \emph{J. Chem. Theory Comput.} \textbf{2020},
  \emph{16}, 2975--2984\relax
\mciteBstWouldAddEndPuncttrue
\mciteSetBstMidEndSepPunct{\mcitedefaultmidpunct}
{\mcitedefaultendpunct}{\mcitedefaultseppunct}\relax
\EndOfBibitem
\bibitem[Wang and Sharma(2023)Wang, and Sharma]{4C2C-HCI2023}
Wang,~X.; Sharma,~S. Relativistic semistochastic heat-bath configuration
  interaction. \emph{J. Chem. Theory Comput.} \textbf{2023}, \emph{19},
  848--855\relax
\mciteBstWouldAddEndPuncttrue
\mciteSetBstMidEndSepPunct{\mcitedefaultmidpunct}
{\mcitedefaultendpunct}{\mcitedefaultseppunct}\relax
\EndOfBibitem
\bibitem[Knecht \latin{et~al.}(2014)Knecht, Legeza, and Reiher]{4C-DMRG2014}
Knecht,~S.; Legeza,~{\"O}.; Reiher,~M. Communication: Four-component density
  matrix renormalization group. \emph{J. Chem. Phys.} \textbf{2014},
  \emph{140}, 041101\relax
\mciteBstWouldAddEndPuncttrue
\mciteSetBstMidEndSepPunct{\mcitedefaultmidpunct}
{\mcitedefaultendpunct}{\mcitedefaultseppunct}\relax
\EndOfBibitem
\bibitem[Battaglia \latin{et~al.}(2018)Battaglia, Keller, and
  Knecht]{4C-DMRG2018}
Battaglia,~S.; Keller,~S.; Knecht,~S. Efficient Relativistic Density-Matrix
  Renormalization Group Implementation in a Matrix-Product Formulation.
  \emph{J. Chem. Theory Comput.} \textbf{2018}, \emph{14}, 2353--2369\relax
\mciteBstWouldAddEndPuncttrue
\mciteSetBstMidEndSepPunct{\mcitedefaultmidpunct}
{\mcitedefaultendpunct}{\mcitedefaultseppunct}\relax
\EndOfBibitem
\bibitem[Brandejs \latin{et~al.}(2020)Brandejs, Vi{\v{s}}{\v{n}}{\'a}k, Veis,
  Mat{\'e}, Legeza, and Pittner]{4C-DMRG2020}
Brandejs,~J.; Vi{\v{s}}{\v{n}}{\'a}k,~J.; Veis,~L.; Mat{\'e},~M.;
  Legeza,~{\"O}.; Pittner,~J. Toward DMRG-tailored coupled cluster method in
  the 4c-relativistic domain. \emph{J. Chem. Phys.} \textbf{2020}, \emph{152},
  174107\relax
\mciteBstWouldAddEndPuncttrue
\mciteSetBstMidEndSepPunct{\mcitedefaultmidpunct}
{\mcitedefaultendpunct}{\mcitedefaultseppunct}\relax
\EndOfBibitem
\bibitem[Anderson and Booth(2020)Anderson, and Booth]{4C-FCIQMC}
Anderson,~R.~J.; Booth,~G.~H. Four-component full configuration interaction
  quantum Monte Carlo for relativistic correlated electron problems. \emph{J.
  Chem. Phys.} \textbf{2020}, \emph{153}, 184103\relax
\mciteBstWouldAddEndPuncttrue
\mciteSetBstMidEndSepPunct{\mcitedefaultmidpunct}
{\mcitedefaultendpunct}{\mcitedefaultseppunct}\relax
\EndOfBibitem
\bibitem[Liu(2017)]{eQEDBook2017}
Liu,~W. In \emph{Handbook of Relativistic Quantum Chemistry}; Liu,~W., Ed.;
  Springer-Verlag: Berlin, 2017; pp 345--373\relax
\mciteBstWouldAddEndPuncttrue
\mciteSetBstMidEndSepPunct{\mcitedefaultmidpunct}
{\mcitedefaultendpunct}{\mcitedefaultseppunct}\relax
\EndOfBibitem
\bibitem[Marian(2001)]{Marian2001}
Marian,~C.~M. In \emph{Reviews in Computational Chemistry}; Lipkowitz,~K.~B.,
  Boyd,~D.~B., Eds.; Wiley-VCH: New York, 2001; Vol.~17; pp 99--204\relax
\mciteBstWouldAddEndPuncttrue
\mciteSetBstMidEndSepPunct{\mcitedefaultmidpunct}
{\mcitedefaultendpunct}{\mcitedefaultseppunct}\relax
\EndOfBibitem
\bibitem[Pitzer and Winter(1988)Pitzer, and Winter]{DGCIa}
Pitzer,~R.~M.; Winter,~N.~W. Electronic-structure methods for heavy-atom
  molecules. \emph{J. Phys. Chem.} \textbf{1988}, \emph{92}, 3061--3063\relax
\mciteBstWouldAddEndPuncttrue
\mciteSetBstMidEndSepPunct{\mcitedefaultmidpunct}
{\mcitedefaultendpunct}{\mcitedefaultseppunct}\relax
\EndOfBibitem
\bibitem[Yabushita \latin{et~al.}(1999)Yabushita, Zhang, and Pitzer]{SpinUGA1}
Yabushita,~S.; Zhang,~Z.; Pitzer,~R.~M. Spin-Orbit Configuration Interaction
  Using the Graphical Unitary Group Approach and Relativistic Core Potential
  and Spin-Orbit Operators. \emph{J. Phys. Chem. A} \textbf{1999}, \emph{103},
  5791--5800\relax
\mciteBstWouldAddEndPuncttrue
\mciteSetBstMidEndSepPunct{\mcitedefaultmidpunct}
{\mcitedefaultendpunct}{\mcitedefaultseppunct}\relax
\EndOfBibitem
\bibitem[DiLabio and Christiansen(1997)DiLabio, and Christiansen]{DGCIb}
DiLabio,~G.; Christiansen,~P. Low-lying $0^+$ states of bismuth hydride.
  \emph{Chem. Phys. Lett.} \textbf{1997}, \emph{277}, 473--477\relax
\mciteBstWouldAddEndPuncttrue
\mciteSetBstMidEndSepPunct{\mcitedefaultmidpunct}
{\mcitedefaultendpunct}{\mcitedefaultseppunct}\relax
\EndOfBibitem
\bibitem[Balasubramanian(1988)]{DGCIc}
Balasubramanian,~K. Relativistic configuration interaction calculations for
  polyatomics: Applications to PbH$_2$, SnH$_2$, and GeH$_2$. \emph{J. Chem.
  Phys.} \textbf{1988}, \emph{89}, 5731--5738\relax
\mciteBstWouldAddEndPuncttrue
\mciteSetBstMidEndSepPunct{\mcitedefaultmidpunct}
{\mcitedefaultendpunct}{\mcitedefaultseppunct}\relax
\EndOfBibitem
\bibitem[Sj{\o}voll \latin{et~al.}(1997)Sj{\o}voll, Gropen, and
  Olsen]{DetSOCI1997}
Sj{\o}voll,~M.; Gropen,~O.; Olsen,~J. A determinantal approach to spin-orbit
  configuration interaction. \emph{Theor. Chem. Acc.} \textbf{1997}, \emph{97},
  301--312\relax
\mciteBstWouldAddEndPuncttrue
\mciteSetBstMidEndSepPunct{\mcitedefaultmidpunct}
{\mcitedefaultendpunct}{\mcitedefaultseppunct}\relax
\EndOfBibitem
\bibitem[Buenker \latin{et~al.}(1998)Buenker, Alekseyev, Liebermann, Lingott,
  and Hirsch]{CI-SOC1998}
Buenker,~R.~J.; Alekseyev,~A.~B.; Liebermann,~H.-P.; Lingott,~R.; Hirsch,~G.
  Comparison of spin-orbit configuration interaction methods employing
  relativistic effective core potentials for the calculation of zero-field
  splittings of heavy atoms with a $^2P$ ground state. \emph{J. Chem. Phys.}
  \textbf{1998}, \emph{108}, 3400--3408\relax
\mciteBstWouldAddEndPuncttrue
\mciteSetBstMidEndSepPunct{\mcitedefaultmidpunct}
{\mcitedefaultendpunct}{\mcitedefaultseppunct}\relax
\EndOfBibitem
\bibitem[Kleinschmidt \latin{et~al.}(2006)Kleinschmidt, Tatchen, and
  Marian]{SPOCK2006}
Kleinschmidt,~M.; Tatchen,~J.; Marian,~C.~M. SPOCK. CI: A multireference
  spin-orbit configuration interaction method for large molecules. \emph{J.
  Chem. Phys.} \textbf{2006}, \emph{124}, 124101\relax
\mciteBstWouldAddEndPuncttrue
\mciteSetBstMidEndSepPunct{\mcitedefaultmidpunct}
{\mcitedefaultendpunct}{\mcitedefaultseppunct}\relax
\EndOfBibitem
\bibitem[Wang \latin{et~al.}(2008)Wang, Gauss, and van W{\"u}llen]{WF-CC2008}
Wang,~F.; Gauss,~J.; van W{\"u}llen,~C. Closed-shell coupled-cluster theory
  with spin-orbit coupling. \emph{J. Chem. Phys.} \textbf{2008}, \emph{129},
  064113\relax
\mciteBstWouldAddEndPuncttrue
\mciteSetBstMidEndSepPunct{\mcitedefaultmidpunct}
{\mcitedefaultendpunct}{\mcitedefaultseppunct}\relax
\EndOfBibitem
\bibitem[Tu \latin{et~al.}(2011)Tu, Yang, Wang, and Guo]{WF-CC2011}
Tu,~Z.; Yang,~D.-D.; Wang,~F.; Guo,~J. Symmetry exploitation in closed-shell
  coupled-cluster theory with spin-orbit coupling. \emph{J. Chem. Phys.}
  \textbf{2011}, \emph{135}, 034115\relax
\mciteBstWouldAddEndPuncttrue
\mciteSetBstMidEndSepPunct{\mcitedefaultmidpunct}
{\mcitedefaultendpunct}{\mcitedefaultseppunct}\relax
\EndOfBibitem
\bibitem[Ganyushin and Neese(2013)Ganyushin, and Neese]{CAS-SOC2013}
Ganyushin,~D.; Neese,~F. A fully variational spin-orbit coupled complete active
  space self-consistent field approach: application to electron paramagnetic
  resonance g-tensors. \emph{J. Chem. Phys.} \textbf{2013}, \emph{138},
  104113\relax
\mciteBstWouldAddEndPuncttrue
\mciteSetBstMidEndSepPunct{\mcitedefaultmidpunct}
{\mcitedefaultendpunct}{\mcitedefaultseppunct}\relax
\EndOfBibitem
\bibitem[Cao \latin{et~al.}(2017)Cao, Li, Wang, and Liu]{sf-X2C-EOM-SOC2017}
Cao,~Z.; Li,~Z.; Wang,~F.; Liu,~W. Combining the spin-separated exact
  two-component relativistic Hamiltonian with the equation-of-motion
  coupled-cluster method for the treatment of spin--orbit splittings of light
  and heavy elements. \emph{Phys. Chem. Chem. Phys.} \textbf{2017}, \emph{19},
  3713--3721\relax
\mciteBstWouldAddEndPuncttrue
\mciteSetBstMidEndSepPunct{\mcitedefaultmidpunct}
{\mcitedefaultendpunct}{\mcitedefaultseppunct}\relax
\EndOfBibitem
\bibitem[Mussard and Sharma(2017)Mussard, and Sharma]{HBCISOC2017}
Mussard,~B.; Sharma,~S. One-Step Treatment of Spin--Orbit Coupling and Electron
  Correlation in Large Active Spaces. \emph{J. Chem. Theory Comput.}
  \textbf{2017}, \emph{14}, 154--165\relax
\mciteBstWouldAddEndPuncttrue
\mciteSetBstMidEndSepPunct{\mcitedefaultmidpunct}
{\mcitedefaultendpunct}{\mcitedefaultseppunct}\relax
\EndOfBibitem
\bibitem[Guo \latin{et~al.}(2021)Guo, Wang, Lu, and Wang]{SOC-CCSDT-GPU2021}
Guo,~M.; Wang,~Z.; Lu,~Y.; Wang,~F. Energy correction and analytic energy
  gradients due to triples in CCSD(T) with spin-orbit coupling on graphic
  processing units using single-precision data. \emph{Mol. Phys.}
  \textbf{2021}, \emph{119}, e1974591\relax
\mciteBstWouldAddEndPuncttrue
\mciteSetBstMidEndSepPunct{\mcitedefaultmidpunct}
{\mcitedefaultendpunct}{\mcitedefaultseppunct}\relax
\EndOfBibitem
\bibitem[Nyvang and Olsen(2023)Nyvang, and Olsen]{SO-MRCI2023}
Nyvang,~A.; Olsen,~J. A relativistic configuration interaction method with
  general expansions and initial applications to electronic g-factors. \emph{J.
  Chem. Phys.} \textbf{2023}, \emph{159}, 044102\relax
\mciteBstWouldAddEndPuncttrue
\mciteSetBstMidEndSepPunct{\mcitedefaultmidpunct}
{\mcitedefaultendpunct}{\mcitedefaultseppunct}\relax
\EndOfBibitem
\bibitem[Zhang \latin{et~al.}(2022)Zhang, Xiao, and Liu]{SOiCI}
Zhang,~N.; Xiao,~Y.; Liu,~W. {SOiCI} and {iCISO}: combining iterative
  configuration interaction with spin{\textendash}orbit coupling in two ways.
  \emph{J. Phys.: Condens. Matter} \textbf{2022}, \emph{34}, 224007\relax
\mciteBstWouldAddEndPuncttrue
\mciteSetBstMidEndSepPunct{\mcitedefaultmidpunct}
{\mcitedefaultendpunct}{\mcitedefaultseppunct}\relax
\EndOfBibitem
\bibitem[Hess \latin{et~al.}(1982)Hess, Buenker, Marian, and
  Peyerimhoff]{Hess1982}
Hess,~B.~A.; Buenker,~R.~J.; Marian,~C.~M.; Peyerimhoff,~S.~D. Investigation of
  electron correlation on the theoretical prediction of zero-field splittings
  of $^2\Pi$ molecular states. \emph{Chem. Phys. Lett.} \textbf{1982},
  \emph{89}, 459--462\relax
\mciteBstWouldAddEndPuncttrue
\mciteSetBstMidEndSepPunct{\mcitedefaultmidpunct}
{\mcitedefaultendpunct}{\mcitedefaultseppunct}\relax
\EndOfBibitem
\bibitem[Teichteil \latin{et~al.}(1983)Teichteil, Pelissier, and
  Spiegelmann]{CIPSO1983}
Teichteil,~C.; Pelissier,~M.; Spiegelmann,~F. Ab initio molecular calculations
  including spin-orbit coupling. I. Method and atomic tests. \emph{Chem. Phys.}
  \textbf{1983}, \emph{81}, 273--282\relax
\mciteBstWouldAddEndPuncttrue
\mciteSetBstMidEndSepPunct{\mcitedefaultmidpunct}
{\mcitedefaultendpunct}{\mcitedefaultseppunct}\relax
\EndOfBibitem
\bibitem[Rakowitz and Marian(1997)Rakowitz, and Marian]{CI-SOC1997}
Rakowitz,~F.; Marian,~C.~M. An extrapolation scheme for spin--orbit
  configuration interaction energies applied to the ground and excited
  electronic states of thallium hydride. \emph{Chem. Phys.} \textbf{1997},
  \emph{225}, 223--238\relax
\mciteBstWouldAddEndPuncttrue
\mciteSetBstMidEndSepPunct{\mcitedefaultmidpunct}
{\mcitedefaultendpunct}{\mcitedefaultseppunct}\relax
\EndOfBibitem
\bibitem[Rakowitz and Marian(1996)Rakowitz, and Marian]{SOCsingle-1}
Rakowitz,~F.; Marian,~C.~M. The fine-structure splitting of the thallium atomic
  ground state: LS-versus jj-coupling. \emph{Chem. Phys. Lett.} \textbf{1996},
  \emph{257}, 105--110\relax
\mciteBstWouldAddEndPuncttrue
\mciteSetBstMidEndSepPunct{\mcitedefaultmidpunct}
{\mcitedefaultendpunct}{\mcitedefaultseppunct}\relax
\EndOfBibitem
\bibitem[Danovich \latin{et~al.}(1998)Danovich, Marian, Neuheuser, Peyerimhoff,
  and Shaik]{SOCsingle-2}
Danovich,~D.; Marian,~C.~M.; Neuheuser,~T.; Peyerimhoff,~S.~D.; Shaik,~S.
  Spin-Orbit Coupling Patterns Induced by Twist and Pyramidalization Modes in
  \ce{C2H4}: A Quantitative Study and a Qualitative Analysis. \emph{J. Phys.
  Chem. A} \textbf{1998}, \emph{102}, 5923--5936\relax
\mciteBstWouldAddEndPuncttrue
\mciteSetBstMidEndSepPunct{\mcitedefaultmidpunct}
{\mcitedefaultendpunct}{\mcitedefaultseppunct}\relax
\EndOfBibitem
\bibitem[Tatchen and Marian(1999)Tatchen, and Marian]{SOCsingle-3}
Tatchen,~J.; Marian,~C.~M. On the performance of approximate spin--orbit
  Hamiltonians in light conjugated molecules: the fine-structure splitting of
  HC$_6$H$^+$, NC$_5$H$^+$, and NC$_4$N$^+$. \emph{Chem. Phys. Lett.}
  \textbf{1999}, \emph{313}, 351--357\relax
\mciteBstWouldAddEndPuncttrue
\mciteSetBstMidEndSepPunct{\mcitedefaultmidpunct}
{\mcitedefaultendpunct}{\mcitedefaultseppunct}\relax
\EndOfBibitem
\bibitem[Berning \latin{et~al.}(2000)Berning, Schweizer, Werner, Knowles, and
  Palmieri]{MRCI-SOC2000}
Berning,~A.; Schweizer,~M.; Werner,~H.-J.; Knowles,~P.~J.; Palmieri,~P.
  Spin-orbit matrix elements for internally contracted multireference
  configuration interaction wavefunctions. \emph{Mol. Phys.} \textbf{2000},
  \emph{98}, 1823--1833\relax
\mciteBstWouldAddEndPuncttrue
\mciteSetBstMidEndSepPunct{\mcitedefaultmidpunct}
{\mcitedefaultendpunct}{\mcitedefaultseppunct}\relax
\EndOfBibitem
\bibitem[Vallet \latin{et~al.}(2000)Vallet, Maron, Teichteil, and
  Flament]{Teichteil2000}
Vallet,~V.; Maron,~L.; Teichteil,~C.; Flament,~J.-P. A two-step uncontracted
  determinantal effective hamiltonian-based so--ci method. \emph{J. Chem.
  Phys.} \textbf{2000}, \emph{113}, 1391--1402\relax
\mciteBstWouldAddEndPuncttrue
\mciteSetBstMidEndSepPunct{\mcitedefaultmidpunct}
{\mcitedefaultendpunct}{\mcitedefaultseppunct}\relax
\EndOfBibitem
\bibitem[Kleinschmidt \latin{et~al.}(2002)Kleinschmidt, Tatchen, and
  Marian]{DFTMRCI-SOC2002}
Kleinschmidt,~M.; Tatchen,~J.; Marian,~C.~M. Spin-orbit coupling of DFT/MRCI
  wavefunctions: Method, test calculations, and application to thiophene.
  \emph{J. Comput. Chem.} \textbf{2002}, \emph{23}, 824--833\relax
\mciteBstWouldAddEndPuncttrue
\mciteSetBstMidEndSepPunct{\mcitedefaultmidpunct}
{\mcitedefaultendpunct}{\mcitedefaultseppunct}\relax
\EndOfBibitem
\bibitem[Roos and Malmqvist(2004)Roos, and Malmqvist]{CASPT2-SOC2004}
Roos,~B.~O.; Malmqvist,~P.-{\AA}. Relativistic quantum chemistry: the
  multiconfigurational approach. \emph{Phys. Chem. Chem. Phys.} \textbf{2004},
  \emph{6}, 2919--2927\relax
\mciteBstWouldAddEndPuncttrue
\mciteSetBstMidEndSepPunct{\mcitedefaultmidpunct}
{\mcitedefaultendpunct}{\mcitedefaultseppunct}\relax
\EndOfBibitem
\bibitem[Ganyushin and Neese(2006)Ganyushin, and Neese]{SI-SOC2006}
Ganyushin,~D.; Neese,~F. First-principles calculations of zero-field splitting
  parameters. \emph{J. Chem. Phys.} \textbf{2006}, \emph{125}, 024103\relax
\mciteBstWouldAddEndPuncttrue
\mciteSetBstMidEndSepPunct{\mcitedefaultmidpunct}
{\mcitedefaultendpunct}{\mcitedefaultseppunct}\relax
\EndOfBibitem
\bibitem[Klein and Gauss(2008)Klein, and Gauss]{EOMIPSO2008}
Klein,~K.; Gauss,~J. Perturbative calculation of spin-orbit splittings using
  the equation-of-motion ionization-potential coupled-cluster ansatz. \emph{J.
  Chem. Phys.} \textbf{2008}, \emph{129}, 194106\relax
\mciteBstWouldAddEndPuncttrue
\mciteSetBstMidEndSepPunct{\mcitedefaultmidpunct}
{\mcitedefaultendpunct}{\mcitedefaultseppunct}\relax
\EndOfBibitem
\bibitem[Mai \latin{et~al.}(2014)Mai, M{\"u}ller, Plasser, Marquetand, Lischka,
  and Gonz{\'a}lez]{mai2014perturbational}
Mai,~S.; M{\"u}ller,~T.; Plasser,~F.; Marquetand,~P.; Lischka,~H.;
  Gonz{\'a}lez,~L. Perturbational treatment of spin-orbit coupling for
  generally applicable high-level multi-reference methods. \emph{J. Chem.
  Phys.} \textbf{2014}, \emph{141}, 074105\relax
\mciteBstWouldAddEndPuncttrue
\mciteSetBstMidEndSepPunct{\mcitedefaultmidpunct}
{\mcitedefaultendpunct}{\mcitedefaultseppunct}\relax
\EndOfBibitem
\bibitem[Roemelt(2015)]{DMRGSO2015}
Roemelt,~M. Spin orbit coupling for molecular ab initio density matrix
  renormalization group calculations: Application to g-tensors. \emph{J. Chem.
  Phys.} \textbf{2015}, \emph{143}, 044112\relax
\mciteBstWouldAddEndPuncttrue
\mciteSetBstMidEndSepPunct{\mcitedefaultmidpunct}
{\mcitedefaultendpunct}{\mcitedefaultseppunct}\relax
\EndOfBibitem
\bibitem[Sayfutyarova and Chan(2016)Sayfutyarova, and Chan]{DMRGSO2016}
Sayfutyarova,~E.~R.; Chan,~G. K.-L. A state interaction spin-orbit coupling
  density matrix renormalization group method. \emph{J. Chem. Phys.}
  \textbf{2016}, \emph{144}, 234301\relax
\mciteBstWouldAddEndPuncttrue
\mciteSetBstMidEndSepPunct{\mcitedefaultmidpunct}
{\mcitedefaultendpunct}{\mcitedefaultseppunct}\relax
\EndOfBibitem
\bibitem[Knecht \latin{et~al.}(2016)Knecht, Keller, Autschbach, and
  Reiher]{nosiDMRGSO2016}
Knecht,~S.; Keller,~S.; Autschbach,~J.; Reiher,~M. A Nonorthogonal
  State-Interaction Approach for Matrix Product State Wave Functions. \emph{J.
  Chem. Theory Comput.} \textbf{2016}, \emph{12}, 5881--5894\relax
\mciteBstWouldAddEndPuncttrue
\mciteSetBstMidEndSepPunct{\mcitedefaultmidpunct}
{\mcitedefaultendpunct}{\mcitedefaultseppunct}\relax
\EndOfBibitem
\bibitem[Cheng \latin{et~al.}(2018)Cheng, Wang, Stanton, and
  Gauss]{cheng2018perturbative}
Cheng,~L.; Wang,~F.; Stanton,~J.~F.; Gauss,~J. Perturbative treatment of
  spin-orbit-coupling within spin-free exact two-component theory using
  equation-of-motion coupled-cluster methods. \emph{J. Chem. Phys.}
  \textbf{2018}, \emph{148}, 044108\relax
\mciteBstWouldAddEndPuncttrue
\mciteSetBstMidEndSepPunct{\mcitedefaultmidpunct}
{\mcitedefaultendpunct}{\mcitedefaultseppunct}\relax
\EndOfBibitem
\bibitem[Zhang and Cheng(2020)Zhang, and Cheng]{ChengEOMCCSO2020}
Zhang,~C.; Cheng,~L. Performance of an atomic mean-field spin--orbit approach
  within exact two-component theory for perturbative treatment of spin--orbit
  coupling. \emph{Mol. Phys.} \textbf{2020}, \emph{118}, e1768313\relax
\mciteBstWouldAddEndPuncttrue
\mciteSetBstMidEndSepPunct{\mcitedefaultmidpunct}
{\mcitedefaultendpunct}{\mcitedefaultseppunct}\relax
\EndOfBibitem
\bibitem[Guo \latin{et~al.}(2020)Guo, Wang, and Wang]{WangMP2020}
Guo,~M.; Wang,~Z.; Wang,~F. Treating spin-orbit coupling at different levels in
  equation-of-motion coupled-cluster calculations. \emph{Mol. Phys.}
  \textbf{2020}, \emph{118}, e1785029\relax
\mciteBstWouldAddEndPuncttrue
\mciteSetBstMidEndSepPunct{\mcitedefaultmidpunct}
{\mcitedefaultendpunct}{\mcitedefaultseppunct}\relax
\EndOfBibitem
\bibitem[Zhou and Suo()Zhou, and Suo]{Suo-SOC2021}
Zhou,~Q.; Suo,~B. New implementation of spin-orbit coupling calculation on
  multi-configuration electron correlation theory. \emph{Int. J. Quantum Chem.}
  \emph{121}, e26772\relax
\mciteBstWouldAddEndPuncttrue
\mciteSetBstMidEndSepPunct{\mcitedefaultmidpunct}
{\mcitedefaultendpunct}{\mcitedefaultseppunct}\relax
\EndOfBibitem
\bibitem[Bodenstein \latin{et~al.}(2021)Bodenstein, Fink, Heimermann, and van
  W{\"u}llen]{vanWullen2021}
Bodenstein,~T.; Fink,~K.; Heimermann,~A.; van W{\"u}llen,~C. Development and
  application of a complete active space spin-orbit configuration interaction
  program designed for molecule magnets. \emph{ChemPhysChem} \textbf{2021},
  \emph{22}, 1--13\relax
\mciteBstWouldAddEndPuncttrue
\mciteSetBstMidEndSepPunct{\mcitedefaultmidpunct}
{\mcitedefaultendpunct}{\mcitedefaultseppunct}\relax
\EndOfBibitem
\bibitem[Freitag \latin{et~al.}(2021)Freitag, Baiardi, Knecht, and
  González]{MPSSI2021}
Freitag,~L.; Baiardi,~A.; Knecht,~S.; González,~L. Simplified State
  Interaction for Matrix Product State Wave Functions. \emph{J. Chem. Theory
  Comput.} \textbf{2021}, \emph{17}, 7477--7485\relax
\mciteBstWouldAddEndPuncttrue
\mciteSetBstMidEndSepPunct{\mcitedefaultmidpunct}
{\mcitedefaultendpunct}{\mcitedefaultseppunct}\relax
\EndOfBibitem
\bibitem[Ganyushin and Neese(2013)Ganyushin, and Neese]{SOCASSCF2013}
Ganyushin,~D.; Neese,~F. A fully variational spin-orbit coupled complete active
  space self-consistent field approach: Application to electron paramagnetic
  resonance g-tensors. \emph{J. Chem. Phys.} \textbf{2013}, \emph{138},
  104113\relax
\mciteBstWouldAddEndPuncttrue
\mciteSetBstMidEndSepPunct{\mcitedefaultmidpunct}
{\mcitedefaultendpunct}{\mcitedefaultseppunct}\relax
\EndOfBibitem
\bibitem[Guo \latin{et~al.}(2023)Guo, Zhang, and Liu]{SOiCISCF}
Guo,~Y.; Zhang,~N.; Liu,~W. SOiCISCF: Combining SOiCI and iCISCF for
  Variational Treatment of Spin--Orbit Coupling. \emph{J. Chem. Theory Comput.}
  \textbf{2023}, \emph{19}, 6668--6685\relax
\mciteBstWouldAddEndPuncttrue
\mciteSetBstMidEndSepPunct{\mcitedefaultmidpunct}
{\mcitedefaultendpunct}{\mcitedefaultseppunct}\relax
\EndOfBibitem
\bibitem[Liu and Hoffmann(2016)Liu, and Hoffmann]{iCI}
Liu,~W.; Hoffmann,~M.~R. iCI: Iterative CI toward full CI. \emph{J. Chem.
  Theory Comput.} \textbf{2016}, \emph{12}, 1169--1178, (E) \textbf{2016},
  \emph{12}, 3000\relax
\mciteBstWouldAddEndPuncttrue
\mciteSetBstMidEndSepPunct{\mcitedefaultmidpunct}
{\mcitedefaultendpunct}{\mcitedefaultseppunct}\relax
\EndOfBibitem
\bibitem[Zhang \latin{et~al.}(2020)Zhang, Liu, and Hoffmann]{iCIPT2}
Zhang,~N.; Liu,~W.; Hoffmann,~M.~R. Iterative Configuration Interaction with
  Selection. \emph{J. Chem. Theory Comput.} \textbf{2020}, \emph{16},
  2296--2316\relax
\mciteBstWouldAddEndPuncttrue
\mciteSetBstMidEndSepPunct{\mcitedefaultmidpunct}
{\mcitedefaultendpunct}{\mcitedefaultseppunct}\relax
\EndOfBibitem
\bibitem[Zhang \latin{et~al.}(2021)Zhang, Liu, and Hoffmann]{iCIPT2New}
Zhang,~N.; Liu,~W.; Hoffmann,~M.~R. Further Development of iCIPT2 for Strongly
  Correlated Electrons. \emph{J. Chem. Theory Comput.} \textbf{2021},
  \emph{17}, 949--964\relax
\mciteBstWouldAddEndPuncttrue
\mciteSetBstMidEndSepPunct{\mcitedefaultmidpunct}
{\mcitedefaultendpunct}{\mcitedefaultseppunct}\relax
\EndOfBibitem
\bibitem[Eriksen \latin{et~al.}(2020)Eriksen, Anderson, Deustua, Ghanem, Hait,
  Hoffmann, Lee, Levine, Magoulas, Shen, Tubman, Whaley, Xu, Yao, Zhang, Alavi,
  Chan, Head-Gordon, Liu, Piecuch, Sharma, Ten-no, Umrigar, and
  Gauss]{Blindtest}
Eriksen,~J.~J.; Anderson,~T.~A.; Deustua,~J.~E.; Ghanem,~K.; Hait,~D.;
  Hoffmann,~M.~R.; Lee,~S.; Levine,~D.~S.; Magoulas,~I.; Shen,~J.;
  Tubman,~N.~M.; Whaley,~K.~B.; Xu,~E.; Yao,~Y.; Zhang,~N.; Alavi,~A.; Chan,~G.
  K.-L.; Head-Gordon,~M.; Liu,~W.; Piecuch,~P.; Sharma,~S.; Ten-no,~S.~L.;
  Umrigar,~C.~J.; Gauss,~J. The ground state electronic energy of benzene.
  \emph{J. Phys. Chem. Lett.} \textbf{2020}, \emph{11}, 8922--8929\relax
\mciteBstWouldAddEndPuncttrue
\mciteSetBstMidEndSepPunct{\mcitedefaultmidpunct}
{\mcitedefaultendpunct}{\mcitedefaultseppunct}\relax
\EndOfBibitem
\bibitem[Liu(2024)]{UnifiedH}
Liu,~W. Unified construction of relativistic Hamiltonians. \emph{J. Chem.
  Phys.} \textbf{2024}, \emph{160}, 084111\relax
\mciteBstWouldAddEndPuncttrue
\mciteSetBstMidEndSepPunct{\mcitedefaultmidpunct}
{\mcitedefaultendpunct}{\mcitedefaultseppunct}\relax
\EndOfBibitem
\bibitem[Paldus and Boyle(1980)Paldus, and Boyle]{Paldus1980}
Paldus,~J.; Boyle,~M.~J. Unitary group approach to the many-electron
  correlation problem via graphical methods of spin algebras. \emph{Phys.
  Script.} \textbf{1980}, \emph{21}, 295--311\relax
\mciteBstWouldAddEndPuncttrue
\mciteSetBstMidEndSepPunct{\mcitedefaultmidpunct}
{\mcitedefaultendpunct}{\mcitedefaultseppunct}\relax
\EndOfBibitem
\bibitem[Liu()]{CommentQED}
Liu,~W. Comment on ``Theoretical examination of QED Hamiltonian in relativistic
  molecular orbital theory'' [J. Chem. Phys. 159, 054105 (2023)]. \emph{J.
  Chem. Phys.} \emph{160}\relax
\mciteBstWouldAddEndPuncttrue
\mciteSetBstMidEndSepPunct{\mcitedefaultmidpunct}
{\mcitedefaultendpunct}{\mcitedefaultseppunct}\relax
\EndOfBibitem
\bibitem[Liu(2024)]{ReCommentQED}
Liu,~W. Response to ``Response to `Comment on Theoretical examination of QED
  Hamiltonian in relativistic molecular orbital theory' '' [J. Chem. Phys. 160,
  187102 (2024)]. \textbf{2024}, arXiv:2308.14011\relax
\mciteBstWouldAddEndPuncttrue
\mciteSetBstMidEndSepPunct{\mcitedefaultmidpunct}
{\mcitedefaultendpunct}{\mcitedefaultseppunct}\relax
\EndOfBibitem
\bibitem[Inoue \latin{et~al.}(2023)Inoue, Watanabe, and Nakano]{WrongQED}
Inoue,~N.; Watanabe,~Y.; Nakano,~H. Theoretical examination of QED Hamiltonian
  in relativistic molecular orbital theory. \emph{J. Chem. Phys.}
  \textbf{2023}, \emph{159}, 054105\relax
\mciteBstWouldAddEndPuncttrue
\mciteSetBstMidEndSepPunct{\mcitedefaultmidpunct}
{\mcitedefaultendpunct}{\mcitedefaultseppunct}\relax
\EndOfBibitem
\bibitem[Inoue \latin{et~al.}()Inoue, Watanabe, and Nakano]{RespCommentQED}
Inoue,~N.; Watanabe,~Y.; Nakano,~H. Response to Comment on ``Theoretical
  examination of QED Hamiltonian in relativistic molecular orbital theory'' [J.
  Chem. Phys. 159, 054105 (2023)]. \emph{J. Chem. Phys.} \emph{160}\relax
\mciteBstWouldAddEndPuncttrue
\mciteSetBstMidEndSepPunct{\mcitedefaultmidpunct}
{\mcitedefaultendpunct}{\mcitedefaultseppunct}\relax
\EndOfBibitem
\bibitem[Liu and Peng(2006)Liu, and Peng]{Q4C}
Liu,~W.; Peng,~D. Infinite-order quasirelativistic density functional method
  based on the exact matrix quasirelativistic theory. \emph{J. Chem. Phys.}
  \textbf{2006}, \emph{125}, 044102, (E)\textbf{125}, 149901 (2006).\relax
\mciteBstWouldAddEndPunctfalse
\mciteSetBstMidEndSepPunct{\mcitedefaultmidpunct}
{}{\mcitedefaultseppunct}\relax
\EndOfBibitem
\bibitem[Peng \latin{et~al.}(2007)Peng, Liu, Xiao, and Cheng]{Q4CX2C}
Peng,~D.; Liu,~W.; Xiao,~Y.; Cheng,~L. Making four- and two-component
  relativistic density functional methods fully equivalent based on the idea of
  ``from atoms to molecule''. \emph{J. Chem. Phys.} \textbf{2007}, \emph{127},
  104106\relax
\mciteBstWouldAddEndPuncttrue
\mciteSetBstMidEndSepPunct{\mcitedefaultmidpunct}
{\mcitedefaultendpunct}{\mcitedefaultseppunct}\relax
\EndOfBibitem
\bibitem[X2C()]{X2CName}
The acronym `X2C' (pronounced as `ecstacy') for exact two-component
  Hamiltonians was proposed by W. Liu after intensive discussions with H. J.
  Aa. Jensen, W. Kutzelnigg, T. Saue and L. Visscher during the Twelfth
  International Conference on the Applications of Density Functional Theory
  (DFT-2007), Amsterdam, August 26-30, 2007. Note that the `exact' here
  emphasizes that all the solutions of the matrix Dirac equation can be
  reproduced up to machine accuracy. It is particularly meaningful when
  compared with approximate two-component Hamiltonians.\relax
\mciteBstWouldAddEndPunctfalse
\mciteSetBstMidEndSepPunct{\mcitedefaultmidpunct}
{}{\mcitedefaultseppunct}\relax
\EndOfBibitem
\bibitem[Kutzelnigg and Liu(2005)Kutzelnigg, and Liu]{X2C2005}
Kutzelnigg,~W.; Liu,~W. Quasirelativistic theory equivalent to fully
  relativistic theory. \emph{J. Chem. Phys.} \textbf{2005}, \emph{123},
  241102\relax
\mciteBstWouldAddEndPuncttrue
\mciteSetBstMidEndSepPunct{\mcitedefaultmidpunct}
{\mcitedefaultendpunct}{\mcitedefaultseppunct}\relax
\EndOfBibitem
\bibitem[Liu and Peng(2009)Liu, and Peng]{X2C2009}
Liu,~W.; Peng,~D. Exact two-component Hamiltonians revisited. \emph{J. Chem.
  Phys.} \textbf{2009}, \emph{131}, 031104\relax
\mciteBstWouldAddEndPuncttrue
\mciteSetBstMidEndSepPunct{\mcitedefaultmidpunct}
{\mcitedefaultendpunct}{\mcitedefaultseppunct}\relax
\EndOfBibitem
\bibitem[Kenneth G.~Dyall(2007)]{RQC_BOOK_Dyall}
Kenneth G.~Dyall,~K. F.~J. \emph{Introduction to relativistic quantum
  chemistry}; Oxford University Press, 2007\relax
\mciteBstWouldAddEndPuncttrue
\mciteSetBstMidEndSepPunct{\mcitedefaultmidpunct}
{\mcitedefaultendpunct}{\mcitedefaultseppunct}\relax
\EndOfBibitem
\bibitem[Liu and Hoffmann(2014)Liu, and Hoffmann]{SDS}
Liu,~W.; Hoffmann,~M.~R. SDS: the `static-dynamic-static' framework for
  strongly correlated electrons. \emph{Theor. Chem. Acc.} \textbf{2014},
  \emph{133}, 1481\relax
\mciteBstWouldAddEndPuncttrue
\mciteSetBstMidEndSepPunct{\mcitedefaultmidpunct}
{\mcitedefaultendpunct}{\mcitedefaultseppunct}\relax
\EndOfBibitem
\bibitem[Song \latin{et~al.}(2021)Song, Guo, Lei, Zhang, and Liu]{SDSRev}
Song,~Y.; Guo,~Y.; Lei,~Y.; Zhang,~N.; Liu,~W. The Static--Dynamic--Static
  Family of Methods for Strongly Correlated Electrons: Methodology and
  Benchmarking. \emph{Top. Current Chem.} \textbf{2021}, \emph{379},
  1--56\relax
\mciteBstWouldAddEndPuncttrue
\mciteSetBstMidEndSepPunct{\mcitedefaultmidpunct}
{\mcitedefaultendpunct}{\mcitedefaultseppunct}\relax
\EndOfBibitem
\bibitem[Epstein(1926)]{Epstein}
Epstein,~P.~S. The stark effect from the point of view of Schroedinger's
  quantum theory. \emph{Phys. Rev.} \textbf{1926}, \emph{28}, 695\relax
\mciteBstWouldAddEndPuncttrue
\mciteSetBstMidEndSepPunct{\mcitedefaultmidpunct}
{\mcitedefaultendpunct}{\mcitedefaultseppunct}\relax
\EndOfBibitem
\bibitem[Nesbet(1955)]{Nesbet}
Nesbet,~R.~K. Configuration interaction in orbital theories. \emph{Proc. Roy.
  Soc. of London. Ser. A} \textbf{1955}, \emph{230}, 312--321\relax
\mciteBstWouldAddEndPuncttrue
\mciteSetBstMidEndSepPunct{\mcitedefaultmidpunct}
{\mcitedefaultendpunct}{\mcitedefaultseppunct}\relax
\EndOfBibitem
\bibitem[Tubman \latin{et~al.}(2018)Tubman, Levine, Hait, Head-Gordon, and
  Whaley]{ASCI2018PT2}
Tubman,~N.~M.; Levine,~D.~S.; Hait,~D.; Head-Gordon,~M.; Whaley,~K.~B. An
  efficient deterministic perturbation theory for selected configuration
  interaction methods. \textbf{2018}, arXiv preprint arXiv:1808.02049\relax
\mciteBstWouldAddEndPuncttrue
\mciteSetBstMidEndSepPunct{\mcitedefaultmidpunct}
{\mcitedefaultendpunct}{\mcitedefaultseppunct}\relax
\EndOfBibitem
\bibitem[Zhang \latin{et~al.}(2020)Zhang, Suo, Wang, Zhang, Li, Lei, Zou, Gao,
  Peng, Pu, Xiao, Sun, Wang, Ma, Wang, Guo, and Liu]{BDF2020}
Zhang,~Y.; Suo,~B.; Wang,~Z.; Zhang,~N.; Li,~Z.; Lei,~Y.; Zou,~W.; Gao,~J.;
  Peng,~D.; Pu,~Z.; Xiao,~Y.; Sun,~Q.; Wang,~F.; Ma,~Y.; Wang,~X.; Guo,~Y.;
  Liu,~W. BDF: A relativistic electronic structure program package. \emph{J.
  Chem. Phys.} \textbf{2020}, \emph{152}, 064113\relax
\mciteBstWouldAddEndPuncttrue
\mciteSetBstMidEndSepPunct{\mcitedefaultmidpunct}
{\mcitedefaultendpunct}{\mcitedefaultseppunct}\relax
\EndOfBibitem
\bibitem[Visscher and Dyall(1997)Visscher, and Dyall]{VISSCHER1997207}
Visscher,~L.; Dyall,~K. Dirac–Fock Atomic Electronic Structure Calculations
  Using Different Nuclear Charge Distributions. \emph{At. Data Nucl. Data
  Tables} \textbf{1997}, \emph{67}, 207--224\relax
\mciteBstWouldAddEndPuncttrue
\mciteSetBstMidEndSepPunct{\mcitedefaultmidpunct}
{\mcitedefaultendpunct}{\mcitedefaultseppunct}\relax
\EndOfBibitem
\bibitem[Sun \latin{et~al.}(2018)Sun, Berkelbach, Blunt, Booth, Guo, Li, Liu,
  McClain, Sayfutyarova, Sharma, Wouters, and Chan]{pyscf1}
Sun,~Q.; Berkelbach,~T.~C.; Blunt,~N.~S.; Booth,~G.~H.; Guo,~S.; Li,~Z.;
  Liu,~J.; McClain,~J.~D.; Sayfutyarova,~E.~R.; Sharma,~S.; Wouters,~S.;
  Chan,~G. K.-L. PySCF: the Python-based simulations of chemistry framework.
  \emph{WIREs Computational Molecular Science} \textbf{2018}, \emph{8},
  e1340\relax
\mciteBstWouldAddEndPuncttrue
\mciteSetBstMidEndSepPunct{\mcitedefaultmidpunct}
{\mcitedefaultendpunct}{\mcitedefaultseppunct}\relax
\EndOfBibitem
\bibitem[Sun \latin{et~al.}(2020)Sun, Zhang, Banerjee, Bao, Barbry, Blunt,
  Bogdanov, Booth, Chen, Cui, Eriksen, Gao, Guo, Hermann, Hermes, Koh, Koval,
  Lehtola, Li, Liu, Mardirossian, McClain, Motta, Mussard, Pham, Pulkin,
  Purwanto, Robinson, Ronca, Sayfutyarova, Scheurer, Schurkus, Smith, Sun, Sun,
  Upadhyay, Wagner, Wang, White, Whitfield, Williamson, Wouters, Yang, Yu, Zhu,
  Berkelbach, Sharma, Sokolov, and Chan]{pyscf2}
Sun,~Q.; Zhang,~X.; Banerjee,~S.; Bao,~P.; Barbry,~M.; Blunt,~N.~S.;
  Bogdanov,~N.~A.; Booth,~G.~H.; Chen,~J.; Cui,~Z.-H.; Eriksen,~J.~J.; Gao,~Y.;
  Guo,~S.; Hermann,~J.; Hermes,~M.~R.; Koh,~K.; Koval,~P.; Lehtola,~S.; Li,~Z.;
  Liu,~J.; Mardirossian,~N.; McClain,~J.~D.; Motta,~M.; Mussard,~B.;
  Pham,~H.~Q.; Pulkin,~A.; Purwanto,~W.; Robinson,~P.~J.; Ronca,~E.;
  Sayfutyarova,~E.~R.; Scheurer,~M.; Schurkus,~H.~F.; Smith,~J. E.~T.; Sun,~C.;
  Sun,~S.-N.; Upadhyay,~S.; Wagner,~L.~K.; Wang,~X.; White,~A.;
  Whitfield,~J.~D.; Williamson,~M.~J.; Wouters,~S.; Yang,~J.; Yu,~J.~M.;
  Zhu,~T.; Berkelbach,~T.~C.; Sharma,~S.; Sokolov,~A.~Y.; Chan,~G. K.-L. Recent
  developments in the PySCF program package. \emph{J. Chem. Phys.}
  \textbf{2020}, \emph{153}, 024109\relax
\mciteBstWouldAddEndPuncttrue
\mciteSetBstMidEndSepPunct{\mcitedefaultmidpunct}
{\mcitedefaultendpunct}{\mcitedefaultseppunct}\relax
\EndOfBibitem
\bibitem[Wilson \latin{et~al.}(1999)Wilson, Woon, Peterson, and
  Dunning]{Dunning1999}
Wilson,~A.~K.; Woon,~D.~E.; Peterson,~K.~A.; Dunning,~J.,~Thom~H. {Gaussian
  basis sets for use in correlated molecular calculations. IX. The atoms
  gallium through krypton}. \emph{J. Chem. Phys.} \textbf{1999}, \emph{110},
  7667--7676\relax
\mciteBstWouldAddEndPuncttrue
\mciteSetBstMidEndSepPunct{\mcitedefaultmidpunct}
{\mcitedefaultendpunct}{\mcitedefaultseppunct}\relax
\EndOfBibitem
\bibitem[de~Jong \latin{et~al.}(2001)de~Jong, Harrison, and Dixon]{Dixon2001}
de~Jong,~W.~A.; Harrison,~R.~J.; Dixon,~D.~A. {Parallel Douglas–Kroll energy
  and gradients in NWChem: Estimating scalar relativistic effects using
  Douglas–Kroll contracted basis sets}. \emph{J. Chem. Phys.} \textbf{2001},
  \emph{114}, 48--53\relax
\mciteBstWouldAddEndPuncttrue
\mciteSetBstMidEndSepPunct{\mcitedefaultmidpunct}
{\mcitedefaultendpunct}{\mcitedefaultseppunct}\relax
\EndOfBibitem
\bibitem[Lin \latin{et~al.}(2023)Lin, Zhang, and Cheng]{Cheng-MP-2023}
Lin,~Z.; Zhang,~C.; Cheng,~L. Comparison of state-interaction and
  spinor-representation calculations of spin-orbit coupling within exact
  two-component coupled-cluster theories. \emph{Mol. Phys.} \textbf{2023},
  e2256423\relax
\mciteBstWouldAddEndPuncttrue
\mciteSetBstMidEndSepPunct{\mcitedefaultmidpunct}
{\mcitedefaultendpunct}{\mcitedefaultseppunct}\relax
\EndOfBibitem
\bibitem[Roos \latin{et~al.}(2004)Roos, Lindh, Malmqvist, Veryazov, and
  Widmark]{ANO2}
Roos,~B.~O.; Lindh,~R.; Malmqvist,~P.-A.; Veryazov,~V.; Widmark,~P.-O. Main
  Group Atoms and Dimers Studied with a New Relativistic ANO Basis Set.
  \emph{J. Phys. Chem. A} \textbf{2004}, \emph{108}, 2851--2858\relax
\mciteBstWouldAddEndPuncttrue
\mciteSetBstMidEndSepPunct{\mcitedefaultmidpunct}
{\mcitedefaultendpunct}{\mcitedefaultseppunct}\relax
\EndOfBibitem
\bibitem[Kramida \latin{et~al.}()Kramida, Ralchenko, Reader, and Team]{NIST1}
Kramida,~A.; Ralchenko,~Y.; Reader,~J.; Team,~N. A. S.~D. NIST Atomic Spectra
  Database, version 5.4. \url{http://physics.nist.gov/asd}\relax
\mciteBstWouldAddEndPuncttrue
\mciteSetBstMidEndSepPunct{\mcitedefaultmidpunct}
{\mcitedefaultendpunct}{\mcitedefaultseppunct}\relax
\EndOfBibitem
\bibitem[Marian(2012)]{marian2012SOCISC}
Marian,~C.~M. Spin--orbit coupling and intersystem crossing in molecules.
  \emph{WIRES Comput. Mol. Sci.} \textbf{2012}, \emph{2}, 187--203\relax
\mciteBstWouldAddEndPuncttrue
\mciteSetBstMidEndSepPunct{\mcitedefaultmidpunct}
{\mcitedefaultendpunct}{\mcitedefaultseppunct}\relax
\EndOfBibitem
\bibitem[Kutzelnigg and Liu(2000)Kutzelnigg, and Liu]{MC-DPT1}
Kutzelnigg,~W.; Liu,~W. Relativistic MCSCF by means of quasidegenerate direct
  perturbation theory. I. Theory. \emph{J. Chem. Phys.} \textbf{2000},
  \emph{112}, 3540--3558\relax
\mciteBstWouldAddEndPuncttrue
\mciteSetBstMidEndSepPunct{\mcitedefaultmidpunct}
{\mcitedefaultendpunct}{\mcitedefaultseppunct}\relax
\EndOfBibitem
\bibitem[Li \latin{et~al.}(2014)Li, Xiao, and Liu]{X2CSOC2}
Li,~Z.; Xiao,~Y.; Liu,~W. On the spin separation of algebraic two-component
  relativistic Hamiltonians: Molecular properties. \emph{J. Chem. Phys.}
  \textbf{2014}, \emph{141}, 054111\relax
\mciteBstWouldAddEndPuncttrue
\mciteSetBstMidEndSepPunct{\mcitedefaultmidpunct}
{\mcitedefaultendpunct}{\mcitedefaultseppunct}\relax
\EndOfBibitem
\bibitem[{He{\ss}} \latin{et~al.}(1996){He{\ss}}, {Marian}, {Wahlgren}, and
  {Gropen}]{SOMF1}
{He{\ss}},~B.~A.; {Marian},~C.~M.; {Wahlgren},~U.; {Gropen},~O. A mean-field
  spin-orbit method applicable to correlated wavefunctions. \emph{Chem. Phys.
  Lett.} \textbf{1996}, \emph{251}, 365--371\relax
\mciteBstWouldAddEndPuncttrue
\mciteSetBstMidEndSepPunct{\mcitedefaultmidpunct}
{\mcitedefaultendpunct}{\mcitedefaultseppunct}\relax
\EndOfBibitem
\bibitem[Dunning~Jr(1989)]{Dunning1989}
Dunning~Jr,~T.~H. Gaussian basis sets for use in correlated molecular
  calculations. I. The atoms boron through neon and hydrogen. \emph{J. Chem.
  Phys.} \textbf{1989}, \emph{90}, 1007--1023\relax
\mciteBstWouldAddEndPuncttrue
\mciteSetBstMidEndSepPunct{\mcitedefaultmidpunct}
{\mcitedefaultendpunct}{\mcitedefaultseppunct}\relax
\EndOfBibitem
\bibitem[Woon and Dunning(1993)Woon, and Dunning]{Dunning1993}
Woon,~D.~E.; Dunning,~J.,~Thom~H. {Gaussian basis sets for use in correlated
  molecular calculations. III. The atoms aluminum through argon}. \emph{J.
  Chem. Phys.} \textbf{1993}, \emph{98}, 1358--1371\relax
\mciteBstWouldAddEndPuncttrue
\mciteSetBstMidEndSepPunct{\mcitedefaultmidpunct}
{\mcitedefaultendpunct}{\mcitedefaultseppunct}\relax
\EndOfBibitem
\bibitem[Zhang \latin{et~al.}(2018)Zhang, Vandezande, Reynolds, and
  Schaefer~III]{4C-MR2018}
Zhang,~B.; Vandezande,~J.~E.; Reynolds,~R.~D.; Schaefer~III,~H.~F. Spin-Orbit
  Coupling via Four-Component Multireference Methods: Benchmarking on p-Block
  Elements and Tentative Recommendations. \emph{J. Chem. Theory Comput.}
  \textbf{2018}, \emph{14}, 1235--1246\relax
\mciteBstWouldAddEndPuncttrue
\mciteSetBstMidEndSepPunct{\mcitedefaultmidpunct}
{\mcitedefaultendpunct}{\mcitedefaultseppunct}\relax
\EndOfBibitem
\bibitem[Visscher and Dyall(1996)Visscher, and Dyall]{RelaBasisCorr1}
Visscher,~L.; Dyall,~K.~G. {Relativistic and correlation effects on molecular
  properties. I. The dihalogens F2, Cl2, Br2, I2, and At2}. \emph{J. Chem.
  Phys.} \textbf{1996}, \emph{104}, 9040--9046\relax
\mciteBstWouldAddEndPuncttrue
\mciteSetBstMidEndSepPunct{\mcitedefaultmidpunct}
{\mcitedefaultendpunct}{\mcitedefaultseppunct}\relax
\EndOfBibitem
\bibitem[Guo \latin{et~al.}(2021)Guo, Zhang, Lei, and Liu]{iCISCF}
Guo,~Y.; Zhang,~N.; Lei,~Y.; Liu,~W. iCISCF: An Iterative Configuration
  Interaction-Based Multiconfigurational Self-Consistent Field Theory for Large
  Active Spaces. \emph{J. Chem. Theory Comput.} \textbf{2021}, \emph{17},
  7545--7561\relax
\mciteBstWouldAddEndPuncttrue
\mciteSetBstMidEndSepPunct{\mcitedefaultmidpunct}
{\mcitedefaultendpunct}{\mcitedefaultseppunct}\relax
\EndOfBibitem
\bibitem[Peng \latin{et~al.}(2009)Peng, Ma, and Liu]{Symm2009}
Peng,~D.; Ma,~J.; Liu,~W. On the construction of Kramers paired double group
  symmetry functions. \emph{Int. J. Quantum Chem.} \textbf{2009}, \emph{109},
  2149--2167\relax
\mciteBstWouldAddEndPuncttrue
\mciteSetBstMidEndSepPunct{\mcitedefaultmidpunct}
{\mcitedefaultendpunct}{\mcitedefaultseppunct}\relax
\EndOfBibitem
\bibitem[Saue and Jensen(1999)Saue, and Jensen]{Quaternion1999}
Saue,~T.; Jensen,~H.~A. Quaternion symmetry in relativistic molecular
  calculations: The Dirac--Hartree--Fock method. \emph{J. Chem. Phys.}
  \textbf{1999}, \emph{111}, 6211--6222\relax
\mciteBstWouldAddEndPuncttrue
\mciteSetBstMidEndSepPunct{\mcitedefaultmidpunct}
{\mcitedefaultendpunct}{\mcitedefaultseppunct}\relax
\EndOfBibitem
\end{mcitethebibliography}
